\documentclass[a4paper, 12pt]{book}
\usepackage[dutch,english]{babel}
\usepackage{latexsym}
\usepackage{amsbsy}
\usepackage{amsmath}
\usepackage{fancyhdr}
\usepackage{sectsty}
\usepackage{amssymb}
\usepackage{amstext}
\usepackage{type1cm}
\usepackage{graphicx}
\usepackage{amsfonts}
\usepackage{psfrag}
\usepackage{verbatim}
\usepackage{calc}

\usepackage[active]{srcltx}

\subsubsectionfont{\large}
\subsectionfont{\large}

\renewcommand{\chaptermark}[1]{\markboth{ #1}{}}

\renewcommand{\headrulewidth}{0pt}
\renewcommand{\footrulewidth}{0pt}
\newcommand{\rf}[1]{(\ref{#1})}
\newcommand{\beq}{\begin{equation}}
\newcommand{\eeq}{\end{equation}}
\newcommand{\bea}{\begin{eqnarray}}
\newcommand{\eea}{\end{eqnarray}}

\newcommand{\e}{\mbox{e}}
\renewcommand{\d}{\mbox{d}}
\newcommand{\g}{\gamma}

\renewcommand{\l}{\lambda}
\newcommand{\La}{\Lambda}
\renewcommand{\b}{\beta}

\newcommand{\ointz}{\oint \frac{dz}{2\pi i \, z}\;}

\renewcommand{\th}{\theta}
\newcommand{\ep}{\varepsilon}
\newcommand{\eps}{\epsilon}

\newcommand{\del}{\delta}
\newcommand{\Del}{\Delta}

\renewcommand{\k}{\kappa}

\newcommand{\oh}{\frac{1}{2}}

\newcommand{\ra}{\rangle}
\newcommand{\la}{\langle}
\newcommand{\prt}{\partial}
\newcommand{\mi}{\!-\!}
\newcommand{\equ}{\!=\!}
\newcommand{\pl}{\!+\!}

\newcommand{\cD}{{\cal D}}
\newcommand{\cS}{{\cal S}}

\newcommand{\cT}{{\cal T}}

\newcommand{\cO}{{\cal O}}

\newcommand{\cW}{{W}}

\newcommand{\tL}{{\tilde{\La}}}
\newcommand{\tX}{{\tilde{X}}}

\newcommand{\hcW}{{\hat{W}}}

\newcommand{\bX}{{\bar{X}}}
\newcommand{\bx}{{\bar{x}}}

\newcommand{\no}{\nonumber}
\newcommand{\nn}{\no\\}

\newcommand{\SL}{\sqrt{\La}}
\newcommand{\sL}{\sqrt{\Lambda}}

\newcommand{\genus}{{\mathfrak g}}
\newcommand{\expec}[1]{\left<#1\right>}

\newcommand{\scprod}[2]{ \langle #1 , #2 \rangle}
\newcommand{\Vol}{\mathrm{Vol}}
\newcommand{\Tri}{ \mathcal{T}}

\newcommand{\myref}[1]{(\ref{#1})}
\newcommand{\lin}{l_1}
\newcommand{\lout}{l_2}
\newcommand{\dL}{\frac{\partial}{\partial L}}
\newcommand{\ddL}{\frac{\partial^2}{\partial L^2}}

\begin{document}

\frontmatter \pagestyle{empty}

\begin{center}
\vspace{40pt}

{\large \bf Topology Change and the Emergence of
Geometry in \\Two Dimensional Causal
Quantum Gravity}

\vspace{40pt}

{\large \sl Willem\ Westra}$\ $

\vspace{40pt}

{\footnotesize
Science institute, mathematics division, University of Iceland,\\
Dunhaga 3, 107 Reykjavik,Iceland,\\
{ email: wwestra@raunvis.hi.is}\\
}
\vspace{40pt}

\end{center}

\vspace{2cm}

\begin{center}
{\bf Abstract}
\end{center}
In this thesis we analyze a very simple model of two dimensional quantum gravity based on causal dynamical triangulations (CDT).
We present an exactly solvable model which indicates that it is possible to incorporate spatial topology changes in the nonperturbative path integral.
It is shown that if the change in spatial topology is accompanied by a coupling constant it is possible to evaluate the path integral to all orders in the coupling and that the result can be viewed as a hybrid between causal and Euclidian dynamical triangulation.\\
The second model we describe shows how a classical geometry with constant negative curvature emerges naturally from a path integral over noncompact manifolds. No initial singularity is present, hence the quantum geometry is naturally compatible with the Hartle Hawking boundary condition. Furthermore, we demonstrate that under certain conditions the quantum fluctuations are small! \\
To conclude, we treat the problem of spacetime topology change. Although we are not able to completely solve the path integral over all manifolds with arbitrary topology, we do obtain results that indicate that such a path integral might be consistent, provided suitable causality restrictions are imposed.

\vspace{52pt}

\noindent
\pagebreak

{\bf How to read this thesis?} \vspace{.1cm} \\ 
After completion of this thesis considerable progress has been made and published in \cite{Ambjorn:2008ta} and \cite{Ambjorn:2008jf}. 
People interested in \cite{Ambjorn:2008ta} and \cite{Ambjorn:2008jf} can use this thesis as an introduction to the subject.
The first chapter is written for a very general audience and can be skipped by most readers.
In reference \cite{Ambjorn:2008ta} spatial and spacetime topology change in two dimensional causal quantum gravity are formalized in the form of a string field theory. 
Subsequently, in \cite{Ambjorn:2008jf} we uncovered a matrix model underlying the string field theory. In a forthcoming publication we show why the continuum amplitudes can be obtained from a matrix model. One will see that it emerges from a new continuum limit of the one matrix model. 


\newpage

\makeatletter
\def\thickhrulefill{\leavevmode \leaders \hrule height 1ex \hfill \kern \z@}
\def\@makechapterhead#1{%
  \vspace*{10\p@}%
  {\parindent \z@ \centering \reset@font
        \Huge \bfseries #1\par\nobreak
        \par
        \vspace*{10\p@}%
    \vskip 60\p@
  }}
\def\@makeschapterhead#1{%
  \vspace*{10\p@}%
  {\parindent \z@ \centering \reset@font
        \Huge \bfseries #1\par\nobreak
        \par
        \vspace*{10\p@}%
    \vskip 60\p@
  }}

\pagestyle{fancy}

\tableofcontents \thispagestyle{plain}

{\fancyhead{} \fancyhead[LE, RO]{\thepage}
\fancyhead[CO]{\slshape\leftmark}
\fancyhead[CE]{\slshape\leftmark}

\include{intro}}

\makeatletter
\def\thickhrulefill{\leavevmode \leaders \hrule height 1ex \hfill \kern \z@}
\def\@makechapterhead#1{%
  \vspace*{10\p@}%
  {\parindent \z@ \centering \reset@font
        {\fontsize{35}{15.6pt}\selectfont \bfseries \thechapter \fontsize{13}{15.6pt}\selectfont}
        \par\nobreak
        \vspace*{15\p@}%
        \interlinepenalty\@M
        \vspace*{10\p@}%
        \Huge \bfseries #1\par\nobreak
        \par
        \vspace*{10\p@}%
    \vskip 60\p@
  }}
\def\@makeschapterhead#1{%
  \vspace*{10\p@}%
  {\parindent \z@ \centering \reset@font
        {\fontsize{35}{15.6pt}\selectfont \bfseries \vphantom{\thechapter} \fontsize{13}{15.6pt}\selectfont}
        \par\nobreak
        \vspace*{15\p@}%
        \interlinepenalty\@M
        \vspace*{10\p@}%
        \Huge \bfseries #1\par\nobreak
        \par
        \vspace*{10\p@}%
    \vskip 60\p@
  }}

\renewcommand{\chaptermark}[1]{\markboth{\thechapter\ #1}{}}
\fancyhf{} \fancyhead[LE, RO]{\thepage}
\fancyhead[CO]{\slshape\rightmark}
\fancyhead[CE]{\slshape\leftmark}
\renewcommand{\headrulewidth}{0pt}
\renewcommand{\footrulewidth}{0pt}
\addtolength{\headheight}{15pt} \mainmatter
\chapter{Introduction to quantum gravity}\label{ch1}

\section{Classical gravity, physics of the large}\label{sec:Classical gravity, physics of the large}

Gravity, omnipresent and inescapable...
\\
\\
Unlike the other three fundamental forces of nature its reach is universal.
All objects and substances in the universe are sensitive to the gravitational pull.
Besides being mere slaves to the will of gravity, matter and energy also play a more
proactive role, since everything inside our universe acts as a source for gravity.
\\
\\
In our daily lives we are only confronted with the passive side of gravity.
If we jump, the gravitational pull of the earth will inevitably let us fall back down again.
The only effect by which gravity reveals itself is by dictating the way we move, never
do we experience our role as sources of gravity. More generic, in no microscopic or mesoscopic
experiment does the gravitational pull between the objects play an important role. The reason for this is clear,
gravity is an extremely feeble force when compared to the other fundamental interactions.
Although we usually take this fact for granted, it might strike one as strange that the weakest
force of nature dominates the motion of objects on the scales relevant in our everyday life.
The fundamental reason behind this is that the source of gravity only comes in one flavor, matter
and energy are always positive causing gravity to be always attractive. If gravity would have had both positive and
negative charges similar to electromagnetism, gravity would not have played any role in our everyday lives, since
it would have been overshadowed by the other forces.
\\
\\
One of the key insights that enabled Newton to formulate his theory of gravity is
the realization that gravity is not only important at scales familiar from our experiences,
but it is also the relevant force at solar system scales and even beyond. He realized that
the motion of a falling apple is similar to the trajectory of the moon, both are caused by the tug of gravity.
\\
\\
Even though Newton's theory was tremendously successful, as it explained the motion of the planets
with unprecedented accuracy, Einstein felt uneasy. He embarked on a historic quest to construct a more aesthetic
description of gravity, an endeavor that turned out to be one of the pinnacles of human ingenuity in recent modern history.
One of his motivations to construct a
more elaborate theory was to harmonize Newton's theory with the principles
of his own theory of special relativity. Another motivation that greatly influenced Einstein's work when constructing his
theory of gravity was the equivalence principle, inertial and gravitational mass were measured to be the same
with remarkable accuracy.
From these incentives and a few other rather philosophical arguments he developed
the theory of general relativity. So by combining aesthetic reasoning with known results from experimental physics
he found a geometrical theory that was seen to describe the real world.
In so doing he extended the validity of gravitational theory to
the largest scales possible. In particular, general relativity has allowed us to compute corrections to
Newton's theory that are vital for the study of cosmology. Furthermore, Einstein's description of gravity has
been tested and confirmed to be valid for the largest distances, masses and velocities that we can measure.

\section{Quantum gravity, physics of the small?} \label{sec:Quantum gravity, physics of the small?}

What about the converse regime? Does gravity really become weaker and weaker when we study nature
on increasingly small scales? The simple empirical answer is yes, the gravitational interaction is so
incredibly weak that it has only been tested down to millimeter scales. It was found that at these scales Newton's law
still holds implying that gravity indeed becomes negligible for the extremely small.
The theoretical expectations are more interesting however.
\\
\\
We know that the physics of systems on small distances is well described by the laws of quantum mechanics.
One of the many peculiar features of quantum theory is that it connects small and large scales by
virtue of Heisenberg's uncertainty principle. To probe physics on increasingly small scales one needs progressively
larger momenta. Since a large momentum implies large energy one expects gravity to become very relevant at the
very tiny scales, contrary to the naive extrapolation of the classical theory.
The scale for which the probe gravitational field becomes large is the Planck scale $L_P\simeq 1.6\times10^{-35}m$.
At this scale the energy needed to resolve the microstructure needs to be so concentrated that a black hole would form.
\\
\\
Quantum mechanics is, unlike gravity, a theory of probabilities. We know that all other forces and all matter fields
satisfy its probabilistic laws, so why should gravity be an exception? To avoid the coexistence of classical and
quantum theory, gravity should be quantized too.
\\
\\
Why have we not yet been able to accomplish this? What sets gravity apart from the other forces of nature? What
makes it so hard to unify gravity with the laws of quantum mechanics?
The reasons are plentiful, one essential fact that makes
the analysis of general relativity very hard in general, also on the classical level, is that it is highly
nonlinear. This nonlinearity is much more severe than in the other interactions of the standard model
as the action is not even polynomial in its fundamental field, the metric.
This has dramatic consequences for the quantization of
gravity by perturbative methods. For the most straight forward methods it basically implies that there is
an infinite number of interaction vertices that have to be taken into account
\footnote{With a suitable (non tensorial) field redefinition of the metric it is however possible to write the Einstein action in a polynomial form. This implies that it is possible do perturbative quantum gravity with a finite number of interaction vertices.}.
Therefore we can conclude that the standard formulations of gravity are not easily treated by perturbation theory.
Another glaring problem that exemplifies the tension between gravity and perturbation theory is the absence of a
natural dimensionless
coupling constant to define a perturbative expansion. Instead, the coupling constant of gravity, Newton's constant
$G_N$, has dimensions of inverse energy squared. Subsequently, the natural parameter of the perturbation expansion
is $G_N E^2$ and we see that the coupling of gravitons increases with their energy. At the point where
the gravitons reach the Planck energy the coupling becomes strong, $G_N E^2 \sim 1$, which inevitable leads
to a breakdown of the perturbation expansion. In contemporary terms we say that the dimensionful nature of
Newton's constant makes general relativity nonrenormalizable as a quantum field theory.
\\
\\
Several points of view can be taken regarding the nonrenormalizability of gravity. The most popular stance
is that the problem comes from an inherent mismatch between the principles of general relativity and quantum theory.
According to this attitude, a resolution for a quantum theory of gravity can only be found in a modification of
the physical principles behind either quantum mechanics, general relativity or both. The most popular candidate for
such a scenario is string theory, where gravity is found to be compatible with quantum theory only if it is
accompanied by a plethora of extra fields and dimensions. Gravity by itself is viewed as a mere low energy effective
theory, and the exact harmonization with quantum theory happens only upon considering the dynamics of the
fundamental strings.
 \\
 \\
A perhaps more conservative attitude is to suppose that gravity and quantum mechanics are not fundamentally incompatible
per se, but that the standard perturbation theory simply is an inadequate tool for the quantization of gravity.
Precisely this philosophy is an inspiration for the models we present in the present thesis.
\\
\\
The construction of a nonperturbative formulation of gravity is a far from trivial task however.
For example, even though the other field theories of the standard
model are considerably simpler than gravity, we cannot solve the path integrals exactly.
Often, the path integrals are merely a helpful tool to set up a perturbative
description of the physical problem at hand. Although extremely successful in QED, it is not an adequate scheme
to study physics in strong coupling regimes such as confinement in QCD.
The computation of quantities that go beyond perturbation theory is often
very difficult, but even worse, it is mostly unclear whether there exists a nonperturbative definition of a path
integral at all! In more than two dimensions there are few methods that enable one to address
the nonperturbative existence of path integrals. In most cases the nonperturbative definition of a
path integral in field theory is only possible by defining it as a limit of a discrete theory.
Although mathematically more rigorous, it is in practice not a very convenient tool to compute concrete
amplitudes, analytical methods are largely unavailable. Nonetheless, the enormous growth in computing power over the recent
years has transformed lattice quantum field theory from a mathematically nice idea into a serious competitor in the arena
of theoretical physics. In particular, the study of QCD has benefitted a lot from these developments. Among the successes are
the calculation of realistic values for meson and baryon masses from first principles. Such formidable achievements are
currently beyond reach of other methods.
\\
\\
In this thesis we investigate simple gravitational models that are based on the method known as Causal
Dynamical Triangulations. In spirit the scheme is a succinct gravitational analogue of lattice QCD. It is a natural
method to define the path integral by a lattice regularization. What remains is a finite statistical sum that,
similar to lattice QCD, lends itself perfectly to computer simulations. One distinguishing feature that sets Causal
Dynamical Triangulations apart from other discrete attempts is that a genuine causal structure is imposed on
the quantum geometry from the outset. The results of the simulations are encouraging,
in four dimensions a well behaved continuum limit seems to exist and there is compelling evidence that a
classical spacetime superimposed with small quantum fluctuations emerges from the nonperturbative path integral.
Despite the intriguing results these numerical methods have to offer, the understanding is far from complete
and inherently restricted by computer power. Furthermore, the statistical model is very complicated and
has so far resisted attempts at a solution by analytical methods.
\\
\\
In two dimensions the situation is much better however, the pure gravity model can be explicitly solved and many interesting
results can be obtained. Of course one might contest that two dimensional gravity is an oversimplified model
as it does not possess some of the essential difficulties of four dimensional gravity such as a dimensionful
coupling constant. Nevertheless it still contains some vital characteristics that set gravitational theories
apart from any other. Issues such as diffeomorphism invariance, background independence and the Wick rotation
are as relevant for the two dimensional model as they are for its higher dimensional analogues.
\\
\\
Besides being interesting from a pure quantum gravity point of view, it might also be regarded as a minimal form of string
theory. Particularly, the two dimensional model of Causal Dynamical Triangulations might shed some light on the
role of causality on the worldsheet of a string. A tantalizing indication that this might indeed
be consequential is that the results of two dimensional Causal Dynamical Triangulations
are physically inequivalent to the outcomes of two dimensional Euclidean quantum gravity.
 \\
 \\
For the purposes of this thesis we primarily view two dimensional quantum gravity as
an interesting laboratory were nonperturbative aspects of quantum gravity can be studied in an exactly solvable setting.
Before presenting our original contributions, we first discuss some general remarks and present
the known results of two dimensional Causal Dynamical Triangulations in chapter \ref{ch2}.
 \\
 \\
In chapter \ref{ch3} we start the discussion of our first generalization by reviewing the
previously established relation between Euclidean and Causal dynamical triangulations.
It is discussed that imposing causality has the important consequence that the spatial topology
of the geometries in the path integral is fixed. Additionally, we recall that in Euclidean Dynamical
Triangulations, as for example defined by matrix models,
the quantum geometry is highly degenerate in the sense that the spatial topology fluctuations dominate
the path integral.
\\
\\
In the remaining sections of chapter \ref{ch3} we show that this situation is not as black and white as
is discussed above. In an original contribution we demonstrate that one can allow for spatial topology
change in two dimensional causal quantum gravity in a controlled manner. We argue that the topology fluctuations
are naturally accompanied by a coupling constant reminiscent of the string coupling.
Upon taking a suitable scaling limit we show that the quantum geometry is no longer swamped by the topology
fluctuations. Surprisingly, we are able to compute the relevant amplitudes to all orders in the coupling and
sum the power series uniquely to obtain an exact nonperturbative result!
\\
\\
In chapter \ref{ch4} we return to the ``pure'' model of Causal Dynamical triangulations. In this chapter we
extend the existing formalism by studying boundary conditions that lead to a path integral over
noncompact manifolds. We begin by recalling that a similar mechanism is familiar from non-critical string
theory where the noncompact quantum geometries are known as ``ZZ branes''. Further we show that
a space of constant negative curvature emerges from the background independent sum over noncompact spacetimes.
Fascinatingly, we can compute the quantum fluctuations and are able to show that they are small almost everywhere
on the geometry! The model is a nice example of how a classical background can appear from a background independent
theory of quantum gravity.
 \\
 \\
To conclude, we tackle the problem of \emph{spacetime} topology change in chapter \ref{ch5}.
Although we are not able to
completely solve the path integral over all manifolds with arbitrary topology, we do obtain some
results indicating that such a path integral might be consistent, provided suitable causality
restrictions are imposed. As a first step we generalize the standard amplitudes of causal dynamical
triangulations by a perturbative computation of amplitudes that include manifolds up to genus two.
Furthermore, a toy model is presented where we make the approximation that the holes in the manifold
are infinitesimally small. This simplification allows us to perform an explicit sum over all genera
and analyze the continuum limit exactly. Remarkably, the presence of the infinitesimal wormholes
leads to a decrease in the effective cosmological constant, reminiscent of the suppression
mechanism considered by Coleman and others in the four dimensional Euclidean path integral.

\chapter{2D Causal Dynamical Triangulations}\label{ch2}

As explained in chapter \ref{ch1} there is as yet no satisfactory theory of four dimensional quantum gravity,
even though both quantum mechanics and general relativity have been formulated over eight decades
ago! Many obstructions to the unification of the two theories, both technical and conceptual, still
remain after all this time.

\section{Quantum gravity for $D\leq4$} \label{sec:Quantum gravity for D<4}

Part of the complications of quantum gravity disappear in lower dimensional models for quantum gravity,
making them interesting playgrounds where one is not confronted with all the issues at once.
So one can consider the study of lower dimensional models as an action plan to tackle the problems step by step.
Of course such a simplification comes at a price, some vital features of the real-world four dimensional theory are lost.
The salient property that distinguishes the four dimensional theory from its lower dimensional analogues is that
it possesses two propagating degrees of freedom whilst the lower dimensional theories have none, a fact that
can be shown by a canonical analysis.
Despite missing this essential characteristic, there is a host of problems
that lower dimensional models still share with the four dimensional theory. An example is the
dimensionful nature of the gravitational coupling constant in both three and four dimensional gravity.
Consequently, both theories are perturbatively nonrenormalizable by power counting.
From a canonical analysis however, it is known that there are no local degrees of freedom
so one deduces that there are at most finitely many degrees of freedom. For a comprehensive review of three dimensional
quantum gravity see \cite{Carlip:1998uc}. If four
dimensional quantum gravity shares the same dissimilarity between the perturbative and nonperturbative descriptions,
it might also be a much better behaved theory than the perturbative expansion leads us to believe.

Often the impression is created that since three dimensional quantum gravity only contains finitely many degrees
of freedom by canonical analysis, the theory can be completely solved. This is a deceptive representation
of the state of affairs though. There is no single model that is generally accepted by the theoretical physics
community. Many problems are unsolved and different approaches give different results for important conceptual
problems such as, do space and time come in discrete units or not? Another issue that does not yet have a satisfactory
explanation within three dimensional quantum gravity is the explanation of black hole entropy of the BTZ black hole
\cite{Banados:1992wn}.
An interesting proposal to do this was
recently put forward by E. Witten \cite{Witten:2007kt} where he basically defines the
gravity theory by its two dimensional boundary
conformal field theory and relates the black hole entropy to the degeneracy of states of that conformal field theory.
The article is also a nice example of the fact that three dimensional quantum gravity has not yet been solved in all details
and that points of view keep changing as Witten also personally changed his viewpoint on the subject. In a
seminal work \cite{Witten:1988hc} he showed
that the theory could be written as a Chern-Simons gauge theory which is a fairly simple gauge theory that \emph{can} be
quantized and he was of the opinion that this equivalence should also hold in the quantum theory.
Now on the contrary, he advocates that the equivalence is only valid semiclassically since, amongst other issues,
the Chern-Simons formulation does not require the vielbein to be invertible whereas the metric formulation does.
His current opinion is that Chern-Simons theory is a useful tool, only to be used for perturbative arguments and
it is not rich enough to fully capture all aspects of three dimensional gravity such as the physics of black holes.
Of course a lot of these statements rest on opinions and conjectures and,
although interesting, one should treat them with care. The argumentation with respect to black holes
for example rests on the assertion that three dimensional black holes are a pure gravity phenomenon.
This statement can be called
into doubt for the obvious reason that black holes usually form by collapsing matter distributions.
The situation in higher dimensional gravity is more complicated though since it seems to be possible to form a
black hole by collapsing gravitational radiation. Subsequently,
three dimensional black hole physics might not tell us something about pure gravity but it might describe gravity
coupled to matter.
So we conclude that three dimensional quantum gravity is an interesting arena where a lot more can be said than for
four dimensional gravity. Many of the fundamental issues that it shares with the four dimensional theory remain
unsolved however.

The next step down the ladder is two dimensional quantum gravity. In this step we lose part of the perturbative analogy
to four dimensional theory, in two dimensions Newton's constant is dimensionless and the Einstein Hilbert action
becomes a topological term, making the theory renormalizable by power counting. Even though it is even less similar to four
dimensional gravity than the three dimensional theory in this respect, it still possesses important
conceptual characteristics such as background independence, diffeomorphism invariance and the problem of defining a
Lorentzian theory. A very appealing advantage of working in two dimensions is that there exits a plethora of exactly solvable
models, both discrete combinatorial as continuum, that can be treated with techniques from conformal field theory and
statistical mechanics.

Beyond being a toy model for four dimensional quantum gravity, the two dimensional model is also interesting from the string
theory point of view. To discuss this relation let me give a sketch of some of the basic principles behind string theory.
The most convenient way to define string theory is to start from the Polyakov action.
It essentially describes the string as two dimensional quantum gravity coupled to scalar fields \cite{Polyakov:1981rd},
where the scalar fields
act as the embedding coordinates of the string. In string theory however, the two dimensional world sheet metric is
introduced as a mere auxiliary variable. If one uses the equations of motion the Polyakov action reduces to the
Nambu Goto action which is written purely in terms of the coordinates and metric of the embedding, or equivalently, the
target space. The reason why the Polyakov action is at least locally equivalent to the Nambu-Goto action is that
it is invariant under Weyl rescaling of the world sheet metric. The world sheet metric in general has three
independent components, where two can be seen to be gauge degrees of freedom from diffeomorphism invariance and the third
is unphysical because the action is invariant under Weyl transformations.

Therefore, classically the Polyakov and Nambu-Goto formulations are equivalent but quantum mechanically this is not
true in general since the measure of the path integral over world sheet metrics is not invariant under Weyl transformations.
Commonly this property of the quantum theory is referred to as the conformal anomaly
\cite{Polyakov:1981rd,Polyakov:1987ez,Polyakov:1987zb}.
Subsequently, in general the conformal factor of the world sheet metric is not a pure gauge degree of freedom implying that
the quantum theory based on the Polyakov action is \emph{not} locally equivalent to a, so far unknown, quantum theory
employing the Nambu-Goto action. One can however remedy this situation by coupling precisely $26$ scalar fields to the
world sheet in which case the conformal anomaly is precisely cancelled. This suggests that the bosonic string naturally
``lives'' in a $26$ dimensional target space\footnote{As is well known, bosonic string theory is unstable and possesses a
tachyon. To resolve this problem one needs to add fermions and supersymmetry. In the resulting superstring theory the target
space manifold is 10 dimensional. Upon considering nonperturbative effects it is expected however that the theory should be
described by membranes embedded in a 11 dimensional target space.}.

To gain a deeper understanding of string theory, people also investigate the dynamics of strings in
dimensions different from $26$ where the conformal anomaly
is not cancelled by the target space scalar fields and it has to be interpreted in a different way. The study of these
models is appropriately dubbed non-critical string theory. In this language pure two dimensional quantum gravity
is referred to as $c=0$ non-critical string theory (see for example \cite{Martinec:2003ka}),
where $c$ is a quantity called `the central charge' and is related
to the expectation value of the trace of the energy momentum tensor of the matter fields coupled to two dimensional
quantum gravity.

In this thesis we focus on exactly solvable two dimensional gravity models. We mainly regard these
two dimensional gravity models as a testing ground for higher dimensional models for
quantum gravity. Nevertheless, since spatial sections of two dimensional spacetimes are in fact one dimensional objects
we switch between (non-critical)  string theory and gravity terminology depending on the application.

\section{Problems and solutions in quantum gravity} \label{sec:Problems and solutions in quantum gravity}

Taking two dimensional gravity seriously as a model for quantum gravity one has to deal with some of the same issues that
one faces in the quantization of four dimensional gravity. Let me highlight some of the fundamental questions
that one faces in any background independent approach:

\begin{enumerate}
\item How should one deal with the problem of time in general relativity?
      Can one define a notion of time and define a Wick rotation?
\item As in any gauge theory one is instructed to factor out the volume of the gauge group to avoid divergencies.
      So in the case of gravity one is faced with factoring out the group of diffeomorphisms.
      Can we do this in any practical
      way and is it possible to find a regularization procedure compatible with this symmetry group?
\item Which class of geometries should be included in the path integral? Should the spatial topology be fixed?
      Should one include geometries with arbitrary spacetime topology?
\end{enumerate}

One of the oldest and most influential ideas to deal with question $1$ is due to Hawking who takes the pragmatic
point of view that one should start with a Euclidean formulation from the beginning \cite{Hawking:1980gf}. The hope was that once
the Euclidean theory was solved one would be able to find a natural Wick rotation. Of course ignoring
problem $1$ simplifies the quantization procedure to some extent but still four dimensional Euclidean quantum
gravity shares many of the problems such as $2$ and $3$ with its Lorentzian counterpart. Hence the hope of solving
four dimensional Euclidean quantum gravity and performing a Wick rotation afterwards has so
far not materialized into a concrete theory.

The quest to answer question $2$ has been somewhat more successful. In \cite{Regge:1961px} Regge realized that
if one introduces a specific lattice regularization
one can formulate the dynamics of classical general relativity without explicitly referring to a particular coordinate system
\cite{Sorkin:1975ah}. An intensely studied model in four dimensions that utilizes Regge's ideas is
{\it quantum Regge calculus}.
In this approach the topology of the lattice is fixed and the length of the edges are the fundamental dynamical
degrees of freedom \cite{Rocek:1982fr}. Although an interesting approach it has met with some
technical difficulties that have kept the theory from providing clear cut results on the nonperturbative sector
of the quantum theory.
Particularly, it is not clear how to define the measure, several proposals exist but there does not seem to be a
general consensus.

To retain the benefits of Regge's coordinate invariant geometry but at the same time avoiding
some of the technical issues associated with quantum Regge calculus, the method of dynamical triangulations was developed
(see \cite{Ambjorn:1997di} for a comprehensive review).
In this approach the same lattice regularization is used as in Regge calculus, but instead of fixing the lattice and
promoting the edge lengths to dynamical variables, the lengths of the edges are fixed and the lattice itself becomes
the dynamical object. Unlike Regge calculus, dynamical triangulation methods are not optimally suited to regularize
a given smooth classical geometry but are highly efficient methods for defining a measure for the path integral over geometries.
Dynamical triangulations are particularly effective in two dimensions where the models reduce to systems that can
be exactly solved by methods known from statistical physics. Especially the use of matrix models and their large $N$
limit \cite{tHooft:1973jz} turned out to be particularly fruitful for the construction of two dimensional
Euclidean quantum gravity models, see for example \cite{Ginsparg:1993is,DiFrancesco:1993nw}.

It has been shown that the results of these dynamical triangulation models coincide nicely
with results from continuum calculations in the conformal gauge as introduced by Polyakov \cite{Polyakov:1981rd}. He showed that
the dynamics of two dimensional gravity can be obtained from a nonlocal action, often called the induced action.
In the conformal gauge this action is equivalent to the Liouville action, which is a local action.
About a decade ago the interest in the field of two dimensional Euclidean gravity was revived since it was
shown in two seminal works \cite{Fateev:2000ik,Zamolodchikov:2001ah} that Liouville theory can be quantized using
conformal bootstrap methods.

Note that the induced action does not represent any local propagating degrees of freedom, in
accordance with canonical considerations. Consequently, Liouville theory also does not describe any local metric
degrees of freedom either, since it is a gauge fixed version of the induced action.
Liouville theory does however provide nontrivial relations between global geometric properties of the quantum geometries
such as the volume and the length of its boundaries. We would like to stress that it implies that two dimensional gravity
is not topological, at least not in the strict sense of
the word since it depends on the metric information of the manifolds. The term topological is not used unambiguously
however, one example is Witten's Chern Simons representation of three dimensional gravity. Often this theory is
referred to as a topological theory while in fact it does encode metric information in an explicit fashion. The reason
for this confusion is twofold. Firstly, the Chern Simons theory does not encode any nontrivial \emph{local} metric degrees
of freedom but only describes global characteristics of the manifold related to the metric. Secondly, one does not
need a metric to write Chern Simons actions in general, so a Chern Simons theory where the gauge field does not represent
any metric degrees of freedom \emph{is} a topological theory. A second example where the word topological is not used
in the strictest sense is topological string theory. Here the term topological is used to indicate that
the dynamics of a string is insensitive to the local geometry of the target space but as in gravitational Chern Simons
theory the global metric properties of the manifold are important.

Because of the exact solvability of the two dimensional Euclidean models, they realize the first step
in Hawking's attitude to quantum gravity in the sense that they are explicit solutions of Euclidean path integrals.
Despite the analytical control one has however not been able to take the next step.
An unambiguous continuation to Lorentzian signature has so far not been found.

The successes of dynamical triangulation methods for two dimensional Euclidean quantum gravity inspired J.~Ambj\o rn and
R.~Loll to develop a dynamical triangulation theory that also addresses question $1$, known as
Causal Dynamical Triangulations (CDT) \cite{Ambjorn:1998xu}. The idea behind the CDT
approach is that the path integral should only contain histories that have a built in causal structure. The suggestion
that one should enforce causality on individual geometries in the path integral goes back at least to Teitelboim
\cite{Teitelboim:1983fh,Teitelboim:1983fk}.
In CDT the triangulations are given a definite causal structure by imposing a particular time slicing and
a fixed spatial topology. A fundamental distinction with respect to the Euclidean models is that in CDT one considers
discretizations of spacetimes with a genuine Lorentzian signature. Given the time slicing one can make a clear
distinction between timelike and spacelike edges which allows one to define a Wick rotation that converts the
quantum mechanical sum over probability amplitudes into a weighted statistical mechanical sum.
The statistical model that one obtains after applying the Wick rotation has
been exactly solved for the two dimensional model and encouraging results have been obtained for three and four dimensional
models using computer simulations
\cite{Ambjorn:2004qm,Ambjorn:2004pw,Ambjorn:2005db,Ambjorn:2005qt,Ambjorn:2005jj,Ambjorn:2006jf}.

As discussed above, the results from the Euclidean dynamical triangulation models are corroborated by
continuum conformal gauge calculations. Similarly, the results from the two dimensional CDT model can also be
obtained by a continuum calculation. Even before the advent of CDT, Nakayama showed that in the proper time gauge,
two dimensional quantum gravity reduces to a simple quantum mechanical model \cite{Nakayama:1993we}.
Using this fact he derived the
same amplitudes that one obtains in CDT. So if the Euclidean model is equivalent to quantum gravity in the conformal gauge
and CDT is related to quantum gravity in the proper time gauge, one would expect that the results of the two theories
coincide, since they merely reflect a different choice of gauge. How can it be that calculations in the two different gauges
leads to different results? Is gauge invariance broken?
On the continuum level these questions are not completely understood, but on the dynamical triangulations side
this problem can been analyzed in detail \cite{Ambjorn:1999fp}. From this analysis it is clear
that the Euclidean path integral contains many more geometries than the CDT. One intuitively sees that the Euclidean path
integral contains geometries where the proper time gauge cannot be chosen globally.

Question $3$ on the issue of topology change is a highly debated and controversial topic,
the possible answer seems to vary immensely from approach to approach.
In most conservative approaches to quantum gravity the stance is taken that one should first figure out the quantization
of gravity on a manifold of fixed topology and only a posteriori consider the possibility of topology change.
Even though this statement seems rather unambiguous it creates a bifurcation between methods that are inherently Euclidean
by nature and methods that take Lorentzian aspects of gravity seriously. In most methods incorporating some Lorentzian
aspects one makes the additional assumption that also the spatial topology is fixed. For theories based on Euclidean
geometry there is no a priori distinction between space and time, implying that a fixed topology of just space might
not be very natural.

In more radical theories such as Group Field Theory (GFT) \cite{Oriti:2006se} the point of view is very different
since the change of topology
of space and time are an essential ingredient in its formulation. An even more radical view is taken in for example causal
set theory, a theory where causality is elevated to the main guiding principle \cite{Bombelli:1987aa}.
In this approach the concept of a manifold is abandoned from the beginning and replaced with points that
possess an elementary causal ordering.
Consequently, the configuration space of causal sets is exclusively contains by topological relations.
\\
\\
Recapitulating, the two dimensional CDT model is one of only very few exactly solvable models known to the author
that addresses both questions $1$ and $2$. One of the foremost objectives of this thesis is to build on the success of
this strategy and to present models where we address all three of the questions posed above.
In particular, we take the process of spatial topology change into account in this explicitly Lorentzian setting
by introducing a coupling constant for this
interaction. Luckily we can make a detailed analysis of the sum over spatial topologies,
since we are able to solve the model to all orders in the
coupling constant and sum the series uniquely to obtain a full nonperturbative result for this process! In chapter \ref{ch5}
we also address the issue of spacetime topology change from two different angles. Although we are not able to obtain the
same level of nonperturbative control as for spatial topology changes, interesting results are obtained.

\section{A notion of time}\label{sec:a notion of time}

Before going into the details of the dynamical triangulation approach we discuss a particular continuum aspect of the
path integral that we wish to compute. In particular, we discuss the role of boundaries in a gravitational path integral.

\subsection{Boundaries and preferred frames}\label{subsec:Boundaries and preferred frames}

Although all explicit path integral calculations in this work are performed in the context of $1+1$ dimensional quantum
gravity we start our discussion in the setting of $3+1$ dimensional gravity. By presenting the arguments regarding the time
variable in four dimensions we emphasize that the issues are as relevant for real world $3+1$ dimensional
gravity as for our two dimensional models.

Formally, we define the path integral as,
\beq
G_\La (X,Y;T)  =
\int \cD [g] \; e^{-S[g]},
\eeq
where the $S[g]$ is the standard action for general relativity for manifolds with boundaries. So the approach
we take to quantum gravity is a rather minimal one. No extra degrees of freedom beyond the metric
are introduced as is done for example in string theory and we do not need a non standard action containing a new
undetermined parameter such as the Barbero-Imirzi parameter that appears in the Holst action \cite{Holst:1995pc}. This action
is the classical starting point of present day loop quantum gravity. The most familiar form for the action of
general relativity for manifolds with boundaries was introduced by Gibbons and Hawking \cite{Gibbons:1976ue}
and reads as follows,
\beq
S[g,G_N] = \frac{1}{16 \pi G_N} \int_M d^4x\sqrt{-g}R  + \frac{1}{8 \pi G_N}\int_{\partial M} d^3x \sqrt{h}K.
\eeq
The first term is the standard Einstein-Hilbert action and the second term is the Gibbons-Hawking-York boundary term. The
boundary term is introduced to make the variational principle well defined for manifolds with boundaries. In effect, this
term cancels the second derivatives in the Einstein Hilbert action such that one does not need to specify the derivatives of
the metric, only the metric itself needs to be given. This fact is particularly clear in the first order, or equivalently
Palatini, formalism where we can explicitly write the Gibbons-Hawking-York boundary term as a total derivative,
\beq
S(e,\omega,G_N)  = \frac{1}{16 \pi G_N}\int_M{\epsilon_{abcd}\left(e^a \!\!\wedge e^b\!\! \wedge F^{cd} - \! d \left(e^a \wedge e^b \wedge \omega^{cd}\right) \right) }.
\eeq
Here $F^{cd}$ is the curvature of local Lorentz transformations and is defined in terms of the spin connection,
\beq
F^{ab}= d \omega^{ab}+ \omega^a_{\;\,c}\wedge\omega^{cb}.
\eeq
Combining the bulk and boundary contributions in one term gives
\beq
S(e,\omega,G_N)  = \frac{1}{16 \pi G_N}\int_M{\epsilon_{abcd}\left(e^a \!\!\wedge e^b\!\! \wedge \omega^c_{\;\,e}\wedge\omega^{ed}
+2 e^a\wedge d e^b\wedge \omega^{cd}\right) }.
\eeq
Note that this action does not depend on derivatives of the spin connection! In the first order formalism the vielbein and
the spin connection are treated as independent fields. Without coupling to matter one obtains the following equation
of motion for the spin connection,
\beq \label{eq:eomspinconnection}
\epsilon_{abcd}\left(e^a \!\!\wedge e^b\!\! \wedge \omega^c_{\;\,e}+e^a\wedge d e^b \delta^c_e\right) =0.
\eeq
Using this equation the action can be written concisely as,
\beq \label{eq:firstorderactionsimplified}
S(e,\omega,G_N)  = \frac{1}{16 \pi G_N}\int_M{\epsilon_{abcd}\left(e^a\wedge d e^b \wedge \omega^{cd}\right) }.
\eeq
Contracting equation \rf{eq:eomspinconnection} with an epsilon symbol gives the more familiar form of the equation of motion of the spin connection,
\beq
\epsilon_{abcd}e^a \wedge T^b=0,
\eeq
where $T^b$ is the torsion two form,
\beq
T^b = \left(d e^b + w^b_{\;\,c}\wedge e^c \right).
\eeq
So the equation of motion of the spin connection yields the constraint that the connection is torsion free.
Together with the tetrad postulate,
\beq
\partial_{\mu} e^a_{\;\,\nu} + \omega^{\;\,a}_{\mu\;\,b}e^b_{\;\,\nu}-\Gamma^{\kappa}_{\mu \nu} e^a_{\;\,\kappa}=0,
\eeq
this fully determines the affine connection and the spin connection in terms of the vielbein. The affine connection can be
expressed in terms of the metric only and reduces to the familiar Levi-Civita connection. By again invoking the tetrad postulate
one can derive the following form for the spin connection
\beq \label{eq:spinconnectioneinstein}
\omega_{\mu}^{\;\,ab}(e)  = \tfrac{1}{2}e^{a \kappa}e^{b\lambda}\omega_{\mu \kappa \lambda}=
e^{a \kappa}e^{b\lambda}\left(\Omega_{\mu \kappa \lambda}-\Omega_{\kappa \lambda \mu }+\Omega_{\lambda \mu \kappa} \right).
\eeq
Here $\Omega_{\mu \kappa \lambda}$ is the object of anholonomy
\beq
\Omega_{\mu \kappa \lambda}= e^{a}_{\mu}\partial_{[\kappa}e_{a\lambda]},
\eeq
which one can view as objects that measure how much the tetrad basis deviates from a coordinate basis.
Combining \rf{eq:firstorderactionsimplified} and \rf{eq:spinconnectioneinstein} we can obtain the action defined entirely
in terms of the vielbein,
\beq \label{eq:finalaction}
S(e,G_N)  = \frac{1}{16 \pi G_N}\int_M{\epsilon_{abcd}\left(e^a\wedge d e^b \wedge \omega^{cd}(e) \right)}.
\eeq
Written in this form it is clear that the action only depends on first derivatives of the vielbein
which is what we wanted to show. It should be stressed that this is merely a rewritten form of the
conventional Einstein-Hilbert-Gibbons-Hawking-York action, no new physics is introduced.
\\
\\
An important
point about the gravitational action \emph{including boundary term} \rf{eq:finalaction},
is that it is no longer invariant under local Lorentz transformations! This is evident, since
the spin connection does not transform tensorially under local Lorentz transformations.
So including boundaries in a gravitational path integral necessarily introduces a preferred Lorentz frame.
%
%
The existence of a preferred Lorentz frame should however not
be confused with concepts as diffeomorphism invariance and background independence. These notions
are not incompatible with the existence of a preferred Lorentz frame. Furthermore, one should realize that the
choice of a specific Lorentz frame does not affect the bulk dynamics as the local Lorentz transformations
are still gauge transformations for the bulk geometry, only the boundaries break the symmetry.

\subsection{Summary}

In the previous subsection \rf{subsec:Boundaries and preferred frames} we recalled the role of boundaries in
classical general relativity. It was stressed that the action for a manifold with boundaries is \emph{not} invariant
under local Lorentz transformations. In other words, boundaries naturally introduce a preferred Lorentz frame.
Within the discrete framework of causal dynamical triangulations the preferred frame of the boundary is used
to define a time foliation of the manifolds in the path integral.

\section{Topology change}\label{sec:Topology change}

In this thesis we discuss some models where we lift the constraint on the (spatial) topology and allow for geometries that
have handles and/or baby universes in a constrained way. More concretely, we present models where a coupling constant
is introduced for the splitting of a string. If one includes these more complicated geometries
in the path integral, the spatial sections of the geometry have the topology of several $S^1$'s.
This means that one is in fact considering a multi-particle, or better multi-string, theory.
History has shown that the framework of quantum field theory is the best way to deal
with multi-particle quantum theories. So in our case the best way to deal with the baby universes and the handles would
be to develop a (non-critical)  string-field theory. Although a full-fledged string-field theory based on CDT
is beyond the scope of this thesis, we do show that even nonperturbative results in the coupling constant for the string
interaction vertex can be obtained!

In section \ref{subsec:Lorentzian aspects} we argue that manifolds with spatial topology
change do not admit a Lorentzian metric everywhere. However, in the models we discuss the Lorentzian signature of the metric
only vanishes at a countable number of points. The analysis of the Wick rotation around such points we leave to future
work, for the present purposes we confine the discussion of our models to the Euclidean domain.

In chapter \ref{ch3} we present results where we perform the sum over \emph{all} tree diagrams
of our interacting non-critical string theory based on CDT. In particular we obtain the disc function and propagator that
are nonperturbatively dressed with string interactions. In chapter \ref{ch5} we go beyond tree level and investigate the loop
expansion, enlarging the class of geometries to include manifolds of arbitrary spacetime topology i.e.~arbitrary genus.
The genus expansion is however considerably more complicated than its tree level counterpart, hindering us to find
nonperturbative results in the coupling. So our analysis is limited to perturbation theory, which is used to compute
results up to order two in the genus expansion.

Furthermore, in chapter \ref{ch5} we make an attempt to go beyond perturbation theory even when considering
 the genus expansion.
Even though we are unable to sum the genus expansion in all generality we can study some non-perturbative effects
in a toy model where we constrain the holes to stay at the cutoff scale. Limiting our focus to quantum geometries
that satisfy this constraint one can allow the number of holes to be arbitrary and
we can perform the sum over genera explicitly. An interesting implication of the model is the suppression of the value
for the effective cosmological constant, reminiscent of the suppression mechanism considered by Coleman and others in
the context of the four dimensional Euclidean path integral.

The study of higher genus manifolds and baby universes in the path integral approach to quantum gravity has received
considerable attention in the context of two dimensional Euclidean quantum gravity.
As has been mentioned, two dimensional Euclidean quantum gravity is a different quantum theory of $2D$ gravity
where the metrics in the path integral have Euclidean signature from the outset.
Note that if one applies the Wick rotation as described above to the causal propagator it is also
defined by a path integral over Euclidean geometries. The set of Euclidean geometries in the causal propagator
is however only a very small subset of the geometries included in the path integral for Euclidean quantum gravity.
This can be understood by observing that in Euclidean quantum gravity one does
not enforce the topology of each spatial universe to be an $S^1$. The relation between causal and
Euclidean quantum gravity can be made precise if one defines their respective path integrals by dynamical triangulation
methods. It turns out that the Euclidean theory is precisely related to the causal theory by removing baby universes
\cite{Ambjorn:1999fp}.
One can also show the reverse relation by starting with causal dynamical triangulations and then adding baby universes
\cite{Ambjorn:1998xu}.

The fact that the Euclidean theory contains baby universes and the Causal theory does not, leads to the conclusion that
the Euclidean theory is strictly speaking not a theory of one single string whereas two dimensional causal quantum gravity is,
if one views both as non-critical string theories. Although this is an appealing way to view the dynamics of the
string one must be careful since it is a picture that is purely based on the worldsheet geometry, the relation to
the dynamics in target space is not a priori clear. Interestingly, there is no weight associated with the branching of
baby universes in the Euclidean theory which allows them to proliferate and actually dominate the path integral in
the continuum limit. This leads to the peculiar situation that the dynamics of non-critical string theory defined
through Euclidean quantum gravity is largely independent of the dynamics of an individual string but is dominated by
the multi-particle, or ``multi-string'', nature of the theory\footnote{for more accurate assessment of this statement
see \rf{sec:Euclidean results with causal methods}}. Specifically, it can be shown by direct calculation
that at each point of the quantum geometry there is
an outgrowth, or equivalently a baby universe, at the scale of the cutoff. One of the prime consequences
of this dominance of baby universes is the non canonical dimension of time, exemplifying the fractal nature
of its quantum geometry.

The model presented in chapter \ref{ch3} could be viewed as a theory where the quantum geometry is allowed to form
baby universes arbitrarily as in Euclidean quantum gravity with the essential difference that we introduce
a coupling constant
for each baby universe. It is shown that this weight effectively tames the proliferation of baby universes preventing
the amplitudes to be dominated by cutoff scale outgrowths. The predominant signal that illustrates the mechanism is the
canonical scaling dimension of time, which is intimately related to the fact that the Hausdorff dimension is two as in the
case of ``pure'' CDT \cite{Ambjorn:1998xu,Ambjorn:1998fd} and not four as is the case in the Euclidean
theory \cite{Kawai:1993cj,Gubser:1993vx,Aoki:1995gi}.

\section{Simplicial geometry} \label{sec:Simplicial geometry}

In this section we describe how to nonperturbatively define and compute path integrals in two
dimensional quantum gravity by the method of causal dynamical triangulations.
One of the principles on which the method is based, is the fact that most quantum field theories can only be defined
beyond perturbation theory by implementing a lattice regularization.

\subsection{Quantum particle from simplicial extrinsic geometry}\label{subsec:Quantum particle from simplicial extrinsic geometry}

\begin{figure}[t]
\begin{center}
\includegraphics[width=3in]{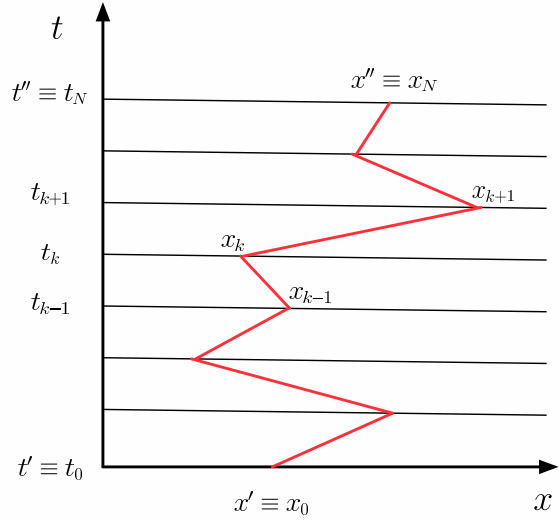}
\caption{Illustration of the path integral for a one-dimensional non-relativistic quantum mechanical problem,
e.g.~a propagating particle. One possible path of the configuration space (path space) is drawn.
The ``virtual'' particle is propagating from $x_0$ to $x_N$ in a piecewise linear path of
N steps of time $\epsilon=(t''-t') /N$ each.}
\label{fig:gravpathqm}
\end{center}
\end{figure}

A classic example where the lattice
regularization plays an important role in the definition of the path integral is the non relativistic free particle.
Although a very simple system, it shares some of its essential features with path integrals for quantum gravity.
Before going into the details of the gravitational path integral we first discuss the quantization of
the non relativistic particle in some detail and highlight the features that are similar to the gravitational quantum theory.
One of these similarities is that both systems are examples of ``random geometry'' i.e.~the individual histories in the
path integral have a geometrical interpretation. A slight difference however is that in quantum gravity
the histories only contain information about their intrinsic geometry whilst the dynamics of the non relativistic particle
is determined by the extrinsic geometry encoded in its velocity as the classical action is given by
\beq
S[x(t) ]=\frac{1}{2} m \int d t \dot{x}^2.
\eeq
Recall that the central object in the path integral formulation of quantum mechanics is the propagator or Feynman kernel
$G(x,t;x',t') $. It describes the time evolution of wavefunctions in quantum mechanics
\beq
\psi(x,t) =\int dx' dt' G(x,t;x',t') \psi(x',t').
\eeq
Physically, it gives the probability amplitude to measure the particle at $x,t$ given the initial location $x',t'$.
In principle the propagator can be found by deriving it from the Schr\"{o}dinger equation. Feynman showed however
that one can instead compute the propagator by computing a weighted integral over all continuous paths between
the initial and final point.
\beq \label{eq:formalpathintegral}
\int {\cal D}[x(t) ]e^{iS[x(t) ]}.
\eeq
Note that \rf{eq:formalpathintegral} merely is a formal expression, one needs to make sense of what it
means to integrate over all trajectories.
To explicitly define the path integral for the free non relativistic particle one is instructed
to first discretize the space of paths to reduce the path integral to a finite dimensional integral.
Commonly this is done by decomposing a general trajectory of the particle into $N$ piecewise linear segments that correspond
to infinitesimal time intervals $\epsilon = (t''-t') /N$ (fig.~\ref{fig:gravpathqm}).
The evaluation of the path integral now amounts to computing $N$ integrals of the following form,
\beq
G(x,t;x',t') =\lim_{\epsilon\rightarrow 0} A^{-N}\prod_{k=1}^{N-1}\int dx_k \exp\left\{ i\sum_{j=0}^{N-1}  S(x_{j+1}-x_{j}) \right\}.
\eeq
Note that this integral is not well defined, since the integrand is a complex valued phase factor.
To be able to compute the integral one has to analytically continue the time variable $t \rightarrow \tau = i t$.
This so-called Wick rotation converts the path integral into a set of real gaussian integrals that can be performed to obtain
the regularized Euclidean amplitude. The continuum amplitude can now be computed by taking the limit where the time
intervals become infinitesimally small, $\epsilon \rightarrow 0$. Notice that in this limit
the velocity and therefore the extrinsic geometry becomes singular at each point of a typical trajectory
in the path integral. In other words, the typical paths that contribute to the path integral are highly non-differentiable.
Similarly, the intrinsic geometry will typically also be singular if one defines a gravitational
path integral by dynamical triangulations. After taking the continuum limit one can obtain the physical
Lorentzian amplitude by applying the inverse Wick rotation $\tau \rightarrow \tau = - i t$.

Notice that in this form the Wick rotation is an analytic continuation in the coordinate $t$ implying that this procedure
is not invariant under coordinate transformations. Taking a gravitational viewpoint, one might ask the question whether it
is possible to define a Wick rotation for point particles that does not involve coordinates?
In the following we suggest that it is indeed possible to construct a Wick rotation for
a point particle that is independent of the particular background and coordinate system.

\subsection{The invariant Wick rotation for particles}\label{subsec:The invariant Wick rotation for particles}

We start by considering the standard action principle for the massive relativistic point particle. The action
is well known and simply proportional to the invariant length of the four dimensional wordline,
\beq
S = -m\int ds,
\eeq
where the invariant length is given as usual by
\beq
ds^2=-g_{\mu \nu}(x) dx^{\mu}dx^{\nu}.
\eeq
Suppose that a classical trajectory is written as $x^{\mu}(\tau) $, where $\tau$ is an arbitrary coordinate that labels the
points along the worldline. Then the action may be written in the familiar form
\beq \label{eq:actionrelativisticparticle}
S = -m\int d\tau \sqrt{-\dot{x}^2}.
\eeq
This action has the very important property that it is invariant under both
four dimensional coordinate transformations and reparameterizations
of the coordinate on the worldline. Thus it really describes the extrinsic geometry
of the worldline as it is embedded in the ambient space and not some particular choice of coordinates.
The action has its shortcomings however, since it has a complicated square root dependence and
is not valid for massless particles. These difficulties can be overcome by introducing a one dimensional einbein
on the worldline. In terms of the einbein the action for a point particle can be written in the following form
\beq \label{eq:einbeinaction}
S=-\frac{1}{2}\int \left(e^{-2}\dot{x}^2 -  m^2\right) e\: d\tau.
\eeq
Solving the equation of motion for the einbein gives
\beq \label{eq:eomeinbein}
\dot{x}^2+e^2m^2=0.
\eeq
Substituting this result back in \rf{eq:einbeinaction} one recovers the action \rf{eq:actionrelativisticparticle}.
The here described relation between the two point particle actions \rf{eq:actionrelativisticparticle} and
\rf{eq:einbeinaction} is completely analogous to the relation
between the Nambu Goto and the Polyakov string actions respectively.
One way to view the einbein is that it is just the lapse function of the one dimensional geometry
along the worldline, since the einbein has just a single component.
Essentially, the action \rf{eq:einbeinaction} describes the physics of a point particle as one dimensional
quantum gravity coupled to the coordinates and metric of the target space.
Now we are in a position to define the coordinate and background independent Wick rotation for the point particle.
If we introduce the signature constant
$\varsigma$ for both the lapse function of the target space and the lapse function of the worldline we obtain the following
action,
\beq \label{eq:einbeinactionvarsigma}
S=\frac{1}{2}\int \left(\frac{\dot{x}^2}{\varsigma e^{2}} +  m^2\right) \sqrt{-\varsigma}e\: d\tau,
\eeq
where
\beq \label{eq:ADMspeedsquaredvarsigma}
\dot{x}^2 = \varsigma N^2 \dot{t}^2 +2 S_{\sigma}\dot{t} \dot{x}^{\sigma}+ h_{\sigma \rho}\dot{x}^{\sigma}\dot{x}^{\rho}.
\eeq
From \rf{eq:einbeinactionvarsigma} and \rf{eq:ADMspeedsquaredvarsigma} it is now easily seen that the Lorentzian
action is rotated to $i$ times the Euclidean action when
$\varsigma$ is analytically continued from $-1$ to $+1$.

\subsection{Quantum gravity from simplicial intrinsic geometry}\label{subsec:Quantum gravity from simplicial intrinsic geometry}

As motivated in section \rf{subsec:Quantum particle from simplicial extrinsic geometry},
regularization by lattice methods can be a powerful tool to
define and evaluate geometrical path integrals. In the example of the point particle the path integral is
evaluated with the help of a discretization of its extrinsic geometry. Similar to the point particle, quantum gravity
is a quantum theory of geometry, yet unlike the quantum particle it is the intrinsic geometry that plays a central role.
So the strategy of causal dynamical triangulations to formulate a path integral for quantum gravity is to employ
a suitable discretization for the intrinsic properties of the space of geometries.

\begin{figure}[t]
\begin{center}
\includegraphics[width=3in]{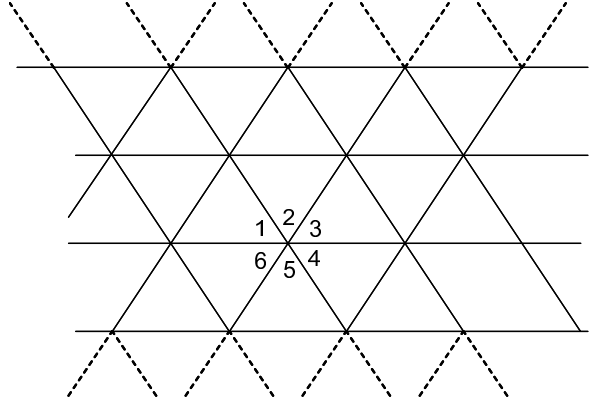}
\caption{A patch of flat space represented by a regular triangulation}
\label{fig:regulartriangulation}
\end{center}
\end{figure}

Discrete methods have had a long tradition in geometry and are a natural tool in the study of the gravitational
dynamics.
Notably, Regge \cite{Regge:1961px} discussed that the intrinsic geometry of a manifold can be discretized by piecewise flat geometries.
In arbitrary dimensions a general piecewise flat geometry consists of flat building blocks called {\it polytopes}.
These polytopes are the natural generalizations of polygons (d=2) and polyhedra (d=3).
Often the piecewise flat geometries are constructed solely from elementary polytopes known as {\it simplices}.
A simplex is the higher dimensional generalization of a triangle, in any dimension it is the polytope with the minimal number
of boundary components. In principle a discrete manifold that is constructed purely from simplices is called a simplicial
manifold. Often however such geometries are simply referred to as {\it triangulations}.
These triangulations can be thought of as a straightforward analogue of the piecewise linear trajectories that
appear in the construction of the path integral for the point particle.

One of the benefits of using a simplicial geometry to approximate an arbitrary manifold
is that the metric properties are completely fixed by specifying all edge lengths.
From the metric information one can extract the local curvature which is the relevant object
when considering gravitational dynamics. The notion of curvature for simplicial manifolds was found initially
by Regge \cite{Regge:1961px} and was later refined in \cite{Cheeger:1983vq}. The prime focus of this work is on two dimensional quantum gravity,
so we explain the Regge curvature in this simple setting.

A simple way to understand the Regge curvature is to first
consider a regular triangulation of equilateral triangles (fig.~\ref{fig:regulartriangulation}). Such a triangulation is a proper simplicial
representation of an everywhere flat manifold. This can be seen by noticing that each vertex is associated with
precisely $6$ triangles. Furthermore, the triangles are equilateral implying that the angle between its sides is
$\tfrac{1}{3} \pi$. So we can conclude that the total angle around each point is equal to $2 \pi$ as should be for a flat manifold.
Introducing local curvature deformations can now be done in two different ways. Historically, the preferred approach is to
deform the triangles by altering the length of the edges. Applying such a deformation to one of the edges of the
flat triangulation causes the total angle around its vertices to be different from $2 \pi$. Since all triangles are still
constrained to be flat this implies that conical singularities are introduced by the deformation.
The amount by which the total angle around a vertex differs from $2\pi$ is called the deficit angle
$\epsilon_v=2\pi-\sum_{i\supset v}\theta_i$. The scalar curvature of a vertex can now be directly related to
this deficit angle by invoking the concept of the so-called {\it dual lattice}, namely

\begin{figure}[t]
\begin{center}
\includegraphics[width=3in]{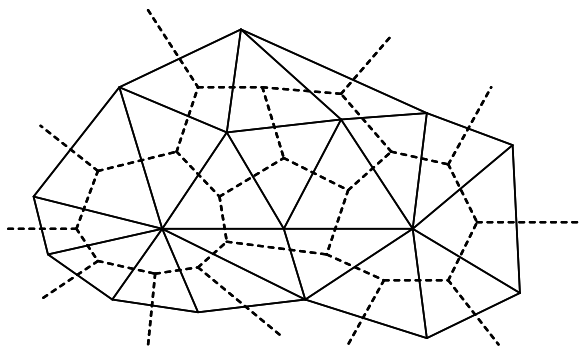}
\caption{A section of a triangulation and its dual graph}
\label{fig:duallattice}
\end{center}
\end{figure}
\beq \label{eq:Reggecurvature}
R_v= 2 \frac{\epsilon_v}{V_v},
\eeq
where $V_v$ is the volume of a cell of the dual lattice. Each triangulation has a unique dual lattice that is easily
constructed by connecting the ``barycenters'' of the triangles with edges of the dual lattice (fig.~\ref{fig:duallattice}).
The relation between the scalar curvature and the deficit angle
is defined by parallel transporting a vector along the edges of the dual lattice that encompass the vertex under
consideration. The curvature is proportional to the angle by which the vector is rotated after one full encircling
of the cell of the dual lattice. The result is given by \rf{eq:Reggecurvature}, the angle by which a vector is rotated after one revolution
along the dual lattice is proportional to the strength of the conical singularity and inversely proportional to
the volume of the cells of the dual lattice.

So the curvature of a geometry which is discretized according to the prescription of Regge
is given by a set of Dirac delta functions located
at the vertices of the triangulation. The distributional character of the curvature is very similar to the non differential
behavior of the histories in the path integral of the point particle and related to the distributional properties
of quantum fields in general.

Given the above notion of curvature it is straightforward to introduce the discrete equivalent of the
Einstein Hilbert action, called the Regge action,
\begin{equation}
S_{\mathrm{Regge}}=  \sum_{v} V_v \left(\lambda-k\,\frac{\epsilon_v}{V_v} \right).
\end{equation}
Observe that this discrete action is manifestly independent of any coordinates as is promised by the
title of \cite{Regge:1961px}. Originally this discrete formulation of the gravitational action was conceived
as a useful tool to study classical aspects of general relativity. It was proposed that the gravitational dynamics
could be conveniently studied by varying the length of the edges of the simplicial manifold.
This approach to simplicial gravity
is called {\it Regge calculus}. In its classical incarnation it has had some success but the interest particularly gained impetus
with the proposition that it might serve as a convenient platform for constructing a quantum theory of gravity
\cite{Rocek:1982fr} \cite{Hamber:1984kx}. The main idea is to use the Regge action to construct a path integral
over the edge lengths of a simplicial manifold with fixed connectivity. The approach is succinctly called
{\it Quantum Regge calculus}, since it is a rather faithful generalization of Regge calculus to the quantum domain.

\begin{figure}[t]
\begin{center}
\includegraphics[width=4in]{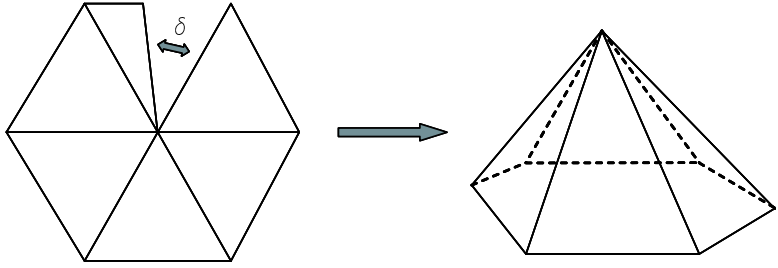}
\caption{A positive deficit angle created by deforming a triangle}
\label{fig:reggedeficit}
\end{center}
\end{figure}

Quantum Regge calculus has not been able give us much understanding beyond semiclassical gravity however.
In the context of two dimensional gravity some doubts have been raised about the
consistency of the approach \cite{Ambjorn:1997ub}. One of the objections is that the approach is not able to
reproduce results from other methods such as Liouville
field theory or dynamical triangulations.

In \cite{Menotti:1996rb} it was proposed that although the formalism of Regge calculus is free of coordinates it
is not completely gauge invariant.
According to their point of view it is possible to perform such a gauge fixing, but
the associated Faddeev-Popov determinants generate a highly non-local measure which makes the theory very hard to handle.

\begin{figure}[t]
\begin{center}
\includegraphics[width=4in]{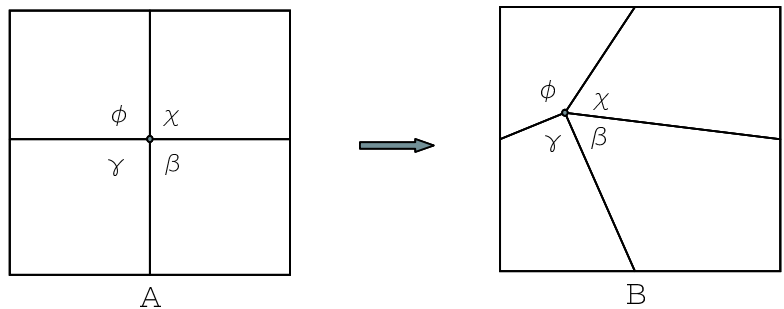}
\caption{All discretizations for which $\phi+\chi+\beta+\gamma = 2 \pi$ represent the same flat geometry.}
\label{fig:reggeovercount}
\end{center}
\end{figure}

An insightful way to visualize the possible overcounting problems in Regge calculus, at least in two dimensions,
is to consider the discretization
of a flat two dimensional manifold in terms of four squares (fig.~\ref{fig:reggeovercount}). If the squares are equilateral
it is clear that
the total angle around the central vertex is $2 \pi$, hence the manifold is flat everywhere also at the central vertex.
If we allow the edges connecting the vertex to fluctuate there still are some possibilities
that keep the manifold flat.
Basically, the fluctuations of the edges that keep the total angle around the central vertex fixed at $2 \pi$
might be considered ``gauge transformations'', since they do not alter the intrinsic geometry of the manifold.
In this context it is interesting to notice that the Regge action only depends on the total angle around a vertex
and not on the angles of individual simplices. One might however argue, that the overcounting problems only appear
when one discretizes manifolds with a high degree of symmetry, such as flat space. Moreover, in the path integral
these special geometries are typically of ``measure zero'' which implies that the overcounting issue is only
a minor problem.

\subsection{Dynamical triangulations} \label{subsec:Dynamical triangulations}

\begin{figure}[t]
\begin{center}
\includegraphics[width=4in]{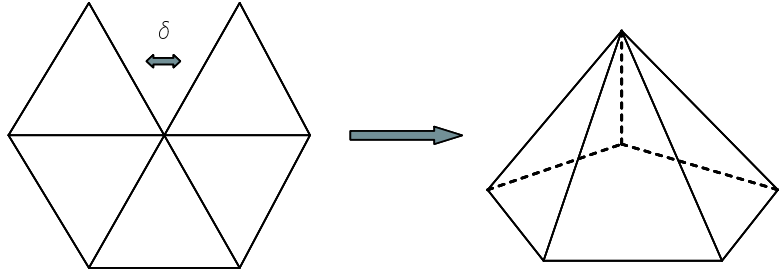}
\caption{A positive deficit angle created by removing a triangle}
\label{fig:dynamicaldeficit}
\end{center}
\end{figure}

To ameliorate the technical issues of the Regge calculus program, the method of dynamical triangulations was developed.
Similar to Regge calculus the scheme is free of coordinates, since both methods are based on the Regge action.
The crucial difference however, is that in dynamical triangulations the length of all edges in the triangulations are
fixed and the geometry is encoded in the nontrivial gluing of the simplices. So instead of altering the edge lengths
one introduces curvature by adding or removing simplices (see fig.~\ref{fig:dynamicaldeficit}). An indication
that this method does not lead to overcounting problems in simple situations
is that the simplicial representation of a given flat manifold is unique.
In addition, adding or removing a triangle always changes the physical curvature and volume of the geometry, implying
that also in more complicated situations the discretization seems to be free of overcounting problems.
Although dynamical triangulation methods are not very efficient to approximate individual classical geometries,
they are ideally suited for a discretization of geometries as histories in a path integral for quantum gravity.

The dynamical triangulation approach to quantum gravity
was initially introduced in the context of two dimensional Euclidean quantum gravity
\cite{Ambjorn:1985az,David:1984tx,Kazakov:1985ea} where it turned out to be a very powerful technique to explicitly
compute the continuum amplitudes for Euclidean quantum gravity. Contrary to
the usual situation, the discrete approach of dynamical triangulations is often more powerful than continuum methods.
Many amplitudes can be obtained with ease and the procedure is considerably more straightforward than the
computation of the amplitudes from the corresponding continuum Liouville field theory.
Actually, the quantization of Liouville theory only gained considerable impetus about $15$ years later in the
seminal work \cite{Fateev:2000ik}. Although the results do not yet cover all aspects of quantum Liouville theory, it
is an important contribution to the quantization of two dimensional gravity.
Another example of the strength of the dynamical triangulation method is that the results for the
amplitudes for Euclidean manifolds of arbitrary genus are exclusive to this method \cite{Ambjorn:1992gw,Eynard:2004mh}.

So the approach of dynamical triangulations is proven to be a useful tool to study Euclidean quantum gravity.
It turns out however that if one studies the quantum geometry of two dimensional quantum gravity
it does not behave as we might expect from a ``realistic'' quantum theory of gravity.
Not even the dimension of geometry is what it is supposed to be, the Hausdorff dimension of the
quantum geometry is four and not two. The principle cause of this
behavior comes from the domination of cutoff scale outgrowths in the path integral. In chapter \ref{ch3} we analyze the
properties of the quantum geometry of two dimensional Euclidean quantum gravity in some detail.

Since the results of the dynamical triangulation scheme can be compared with continuum calculations,
one concludes that the fractal structure of the geometry is not a consequence of the triangulation method,
but an intrinsic property of two dimensional Euclidean quantum gravity.
The initial attitude was that the somewhat degenerate behavior of the quantum geometry is due to
the simplicity of two dimensional gravity.
Therefore, higher dimensional versions of the dynamical triangulation model were developed
and investigated by means of computer simulations, see \cite{Ambjorn:1991wq} and \cite{Agishtein:1991ta} for the three
dimensional model and \cite{Ambjorn:1991pq,Agishtein:1992xx} for the four dimensional model.

The hope that the higher dimensional models might be better behaved than
the two dimensional model did not materialize though.
It was found that in the infinite volume limit these higher dimensional dynamical triangulation models have two phases,
a crumpled phase where the Hausdorff dimension is very large and a tree like phase where the geometry resembles a branched polymer.
Both phases are not satisfactory from a physical point of view, so a suitable continuum limit is not automatically reached
in either phase. Nonetheless, one of the initial ideas was that an appropriate continuum limit perhaps exists at
the critical point separating the crumpled and branched polymer phases. Subsequent analysis revealed those hopes to be
in vain as it was shown that even in the four dimensional model the phase transition is of first order
\cite{Bialas:1996wu,Catterall:1997xj}.

An important lesson that can be learned from these investigations is the following, if one constructs a
random geometry model based on simplices of a certain dimension one is not at all assured that the
quantum geometry behaves anything like a manifold of that dimension.

\section{2D causal dynamical triangulations}

In this section we introduce the concept of causal dynamical triangulations. The motivation behind the inception
of causal dynamical triangulations was twofold:

\begin{enumerate}
\item The quantum geometry of four dimensional Euclidean dynamical triangulation models seem
      incompatible with a well behaved continuum limit.\\
\item The Lorentzian signature of the spacetimes should be taken seriously.
\end{enumerate}

To address the Lorentzian nature of the path integral it is of paramount importance to have an
intrinsically defined notion of time. In causal dynamical triangulations this problem is
addressed by studying piecewise linear geometries that have a layered structure. The layered
structure of the triangulations allows one to globally distinguish timelike and spacelike edges.
Furthermore, the global foliation of the discrete geometries allows one to define a consistent Wick rotation.

The central amplitude one aims to compute in two dimensional causal dynamical triangulations
is the so-called cylinder amplitude or causal propagator. This quantity describes the probability amplitude
for a quantum ensemble of two dimensional geometries of Lorentzian signature with an initial and a final
boundary, where every point of the initial boundary has the same timelike geodesic distance to the final boundary.
In two dimensions the boundaries are one dimensional curves with the topology of a $S^1$ so the only
information that characterizes the intrinsic geometry of the boundaries is their length.

\begin{figure}[t]
\begin{center}
\includegraphics[width=4in]{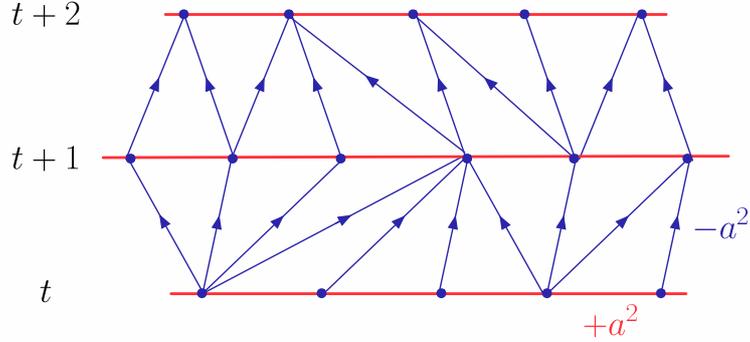}
\caption{Section of a 2d Lorentzian triangulation consisting of spacetime strips of height $\Delta t\equ 1$.
Each spatial slice is periodically identified, such that the simplicial manifold has topology $[0,1]\!\times\! S^1$.
One sees that a single strip with lower boundary length $l_t$ and upper boundary length $l_{t+1}$ consists
exactly of $l_t$ up pointing triangles and $l_{t+1}$ down pointing triangles.}
\label{fig:triangulation}
\end{center}
\end{figure}

The strategy of causal dynamical triangulations is to discretize the manifolds in the path integral with flat triangles.
These Minkowski triangles have one spacelike edge satisfying $L_s^2 = + a^2$
and two timelike edges with $L_{\tau}^2= -a^2$. By construction, one chooses spacetimes which consist of
$T/a$ strips with the topology of $S^1 \times [0,1]$, where the timelike height of the strip
is proportional to $a$. Each strip has two spatial boundaries, one spatial section at time $\tau$ with length
$L(\tau) = a l_t$ and one at time $\tau+a$ with length $L(\tau+a) = a l_{t+1}$.
The geometry of a general circular strip is now determined by the ordering of
$l_{t+1}$ triangles ``pointing up'' and $l_{t}$ triangles ``pointing down'' as illustrated in fig.~\ref{fig:triangulation}.

Since the triangles are genuine patches of flat Minkowski space, they are naturally equipped with a local light cone structure.
Furthermore, from the global distinction between timelike and spacelike edges one sees that
the triangulation equips the manifold with a global causal structure. By virtue of this global causal structure
it is possible to define a Wick rotation for curved manifolds.
In the triangulation context the Wick rotation amounts to changing the
squared length of the timelike edges from negative to positive signature $L_{\tau}^2=-a^2 \mapsto L_{\tau}^2=a^2$. As in the
continuum case this rotation should be treated with some care and one must show that the Lorentzian action
defined with $L_{\tau}^2=-a^2$ and the Euclidean action with $L_{\tau}^2=a^2$ can be connected by a smooth deformation.
In appendix \ref{app:Lorentzian triangles} we discuss the Regge action for these two dimensional Minkowski triangulations. It is shown that the
Lorentzian and the Euclidean Regge action, multiplied with the imaginary unit $i$,
are indeed connected by a smooth analytical continuation of the parameter $\alpha$
from $-1$ to $1$ where $\alpha$ is defined by $L_{\tau}^2= \alpha a^2$.
So the Wick rotation indeed possesses the desired
property that the weight in the path integral is converted from a complex phase factor to a real Boltzmann type weight.
\beq
\mathcal{W}:\quad  e^{i\,S_{\mathrm{Regge}}(T^{lor}) }\mapsto  e^{-\,S_{\mathrm{Regge}}(T^{eu}) }.
\eeq
As we shall see in the forthcoming section the above defined kinematical structures reduce the computation of
the path integral for the causal propagator to a statistical mechanics problem.

\section{The discrete solution} \label{sec:The discrete solution}

The basic ingredients of any statistical model are the entropy and the Boltzmann weight.
For the two dimensional causal dynamical triangulations model the entropy is generated by the number of
geometrically distinct ways one can organize the Minkowski triangles in a layered structure as depicted in
fig.~\ref{fig:triangulation}.
The Boltzmann factor on the other hand is related to the Regge action associated to the triangles. In two
dimensions the Regge action is particularly simple, since the two dimensional Einstein-Hilbert term
is a topological invariant.
\beq
\int_M d^2x \sqrt{|\det g|}\, R(x) =2\pi \chi(M),
\eeq
where $\chi(M) \equ 2\mi 2\genus-b $ is the Euler characteristic of the manifold $M$, $\genus$ is the genus and $b$ is
the number of boundary components of the manifold. For the moment we fix the topology of the manifold to be of the form
$S^1\times[0,1]$. This choice corresponds to what we refer to as ``bare''or equivalently ``pure'' causal dynamical triangulation
model as originally introduced by Ambj\o rn and Loll in \cite{Ambjorn:1998xu}. One of the new contributions of this thesis
is that we go beyond the assumption of fixed topology and in chapter \ref{ch3} we introduce a model where controlled
spatial topology changes are an integral part of the quantum geometry. For such models the Einstein-Hilbert part of the action
is essential as it tames the topology changes by introducing a Boltzmann weight that suppresses manifolds
with complicated topology in the path integral. Since the topology is fixed in the pure model the Einstein-Hilbert action
can act at most as an overall phase factor in the path integral. Furthermore, for the pure
causal dynamical triangulations model we are solely interested in manifolds with
the topology of a cylinder so the Euler characteristic is zero, making Newton's constant irrelevant for two dimensional
quantum gravity without topology change.

The only quantity which is relevant for the dynamics of the pure dynamical triangulation model is the total volume
of the manifold. The Regge action is then simply proportional to the added volume of all triangles of a specific
configuration,
\beq
S_{\mathrm{Regge}}(T) =\tilde{\lambda} \,a^2\, N(T),
\eeq
where $\tilde{\lambda}$ is the bare cosmological constant and $N(T) $ the number of triangles in the triangulation $T$.
Note that an order one factor coming from the volume term has been absorbed into $\tilde{\lambda}$.
The path integral for the propagator can now be written as follows
\beq
G_{\tilde{\lambda}}(l_1,l_2;t) =\sum_{\Tri} \frac{1}{C_{\Tri}}\:\: e^{i\,\tilde{\lambda}\,a^2\,N(T) },
\eeq
where $\Tri$ denotes the causal triangulations with initial boundary length $l_1$ and final boundary length $l_2$, and
$C_{\Tri}$ denotes the volume of the automorphism group of a triangulation. Basically it is the symmetry factor of the
manifold that is still left after factoring out the diffeomorphisms. After a Wick rotation the discrete sum over quantum
amplitudes is converted to a genuine statistical model with a real Boltzmann weight,
\beq
G_{\lambda}(l_1,l_2;t) =\sum_{\Tri} \frac{1}{C_{\Tri}}\:\: e^{-\lambda\,a^2\,N(T) },
\eeq
where it should be noted that $\lambda$ and $\tilde{\lambda}$ differ by an order one constant because of the
different volume of Minkowskian and Euclidean triangles (see appendix \ref{app:Lorentzian triangles}).
The layered structure of the triangulations has the natural implication that the propagator satisfies
the following semi-group property or composition law,
\beq \label{eq:discrcompositionllnonmarked}
G_\l(l_1,l_2;t_1+t_2)  = \sum_{l} l\: G_\l(l_1,l;t_1) \; G_\l(l,l_2;t_2).
\eeq
The measure factor $l$ in the composition law comes from the circular nature of a strip. Writing the composition law
for $t_1=1$ we see that the {\it one step propagator} acts as a transfer matrix,
\beq \label{eq:transfermatrixequation}
G_\l(l_1,l_2;t+1)  = \sum_{l} l\: G_\l(l_1,l;1) \; G_\l(l,l_2;t).
\eeq
In the following we derive $G_\l(l_1,l_2;t) $ by iterating \rf{eq:transfermatrixequation} $t$ times.
The iteration procedure can be conveniently
carried out by introducing the generating function for $G_\l(l_1,l_2;t) $
\beq
G_\l(x,y;t) \equiv \sum_{k,l} x^k\,y^l \;G_\l(k,l;t),
\eeq
where $x$ and $y$ can be naturally interpreted as Boltzmann weights related to the boundary cosmological constants of
individual triangles,

\beq
x=\e^{-\l_ia},~~~~y=\e^{-\l_oa}.
\eeq

Analogously we write the Boltzmann weight related to the bulk cosmological constant as follows,

\beq
g=\e^{-\l a^2}.
\eeq

The above introduced notation implies that the total Boltzmann weight of one strip can be determined
by associating a factor of $g x$ with triangles that have the spacelike edges on the entrance loop and a factor
$g y$ to triangles where the spacelike edges are on the exit loop. The one step propagator is now easily computed
by standard generating function techniques as follows,

\begin{eqnarray} \label{eq:Gxyg1def}
G(x,y;g;1)  & = & \raisebox{-10pt}{\includegraphics[height=25pt]{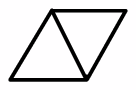}}+
\raisebox{-10pt}{\includegraphics[height=25pt]{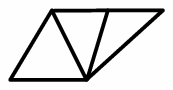}}+
\raisebox{-10pt}{\includegraphics[height=25pt]{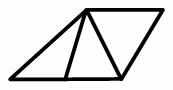}}+... \nonumber\\
 & = &  \sum^{\infty}_{k=1}\frac{1}{k}\left(\sum^{\infty}_{m=1} (gx) ^m\sum^{\infty}_{n=1} (gy) ^n\right) ^k,
\end{eqnarray}

where the factor $\tfrac{1}{k}$ comes from dividing by the volume of the automorphism group for periodic triangulations.
Evaluating the summations in \rf{eq:Gxyg1def} we readily obtain

\beq \label{onesteppropagatorxyZ2}
G(x,y;g;1)  = \sum^{\infty}_{k=1}\frac{1}{k}\left(\frac{gx}{1-gx}\frac{gy}{1-gy} \right) ^k=-\log\left(1-\frac{gx gy}{(1-gx) (1-gy) }\right).
\eeq

From this expression it can be seen that the one step propagator with fixed boundary lengths is given by

\beq \label{eq:onesteppropagatorll}
G(l_1,l_2;g;1)  = g^{l_1+l_2}\frac{1}{l_1+l_2}\binom{l_1+l_2}{l_1},
\eeq

where the division by the volume of the automorphism group now makes its appearance in the guise of
the factor $\tfrac{1}{l_1+l_2}$.

To compute the ``finite time propagator'' $G(x,y;g;t) $, we rewrite the composition law \rf{eq:discrcompositionllnonmarked}
in terms of generating functions and obtain the following,

\beq\label{eq:discrcompositionxynonmarked}
G(x,y;g;t_1+t_2)  = \oint \oint \frac{dz}{2\pi i}\:\frac{dz'}{2\pi i}\: \frac{1}{\left(1-zz'\right) ^2}\:G(x,z;g;t_1)  G(z',y;g;t_2).
\eeq

By setting $t_2=1$ and performing the contour integration over $z$ we obtain
\beq\label{eq:discrcompositionxynonmarkediteration2}
G(x,y;g;t)  = \oint \frac{dz'}{2\pi i z'^2}\: \left[\frac {d}{d z}G_\l(x,z;1) \right]_{z=1/z'} G_\l(z',y;t-1).
\eeq
Inserting the expression for the one step propagator yields the desired iterative equation for $G(x,y;g;t) $,
\beq\label{eq:iterationequationnonmarked}
G(x,y;g;t)  = G\left(\frac{g}{1-gx},y;g;t-1\right) -G_\l\left(g,y;g;t-1\right).
\eeq
The implicit solution of this equation can be written as
\beq\label{eq:discrGxytF}
G(x,y;g;t)  = \log \left(\frac{1}{1-F_t(x)  y}\right) -\log \left(\frac{1}{1-F_{t-1}(g)  y}\right),
\eeq
where $F_t$ is defined iteratively by
\beq \label{eq:iterationF}
F_t(x)  = \frac{g}{1-gF_{t-1}(x) },~~~~F_0(x) =x.
\eeq
The fixed point $F$ as defined by $F_t(x) =F_{t-1}(x) $ is given by
\beq
F=\frac{1-\sqrt{1-4g^2}}{2g},~~~~g=\frac{1}{F+1/F}.
\eeq
By well known methods one can use the fixed point to find the explicit solution to the iterative equation
\rf{eq:iterationF}
\beq\label{eq:FxABC}
F_t(x) = \frac{B_t-x C_t}{A_t-x B_t},~~~~F_{t-1}\left({\textstyle{g}}\right) =\frac{B_t}{A_t},
\eeq
where
\beq \label{eq:defABC}
A_t =1-F^{2t+2},~~~B_t=F-F^{2t+1},~~~C_t=F^2-F^{2t}.
\eeq
The complete finite time propagator is now obtained by substituting \rf{eq:FxABC} in \rf{eq:discrGxytF}, yielding
\beq \label{eq:discrGxytZ}
G(x,y;g;t) = -\log\left(1-Z(x,y;g;t) \right),
\eeq
where we have defined
\beq \label{discrZxyt}
Z(x,y;g;t) ={\textstyle \left(1-\frac{A_t C_t}{B_t^2}\right) }\frac{\frac{B_t}{A_t} x \frac{B_t}{A_t} y}{\left(1- \frac{B_t}{A_t} x\right) \left(1- \frac{B_t}{A_t} y\right) }.
\eeq
The region of convergence of this result as an expansion in powers of $x,y,z$ is
\beq \label{eq:convergencecondition}
|g|< \frac{1}{2}, \quad |x|< 1,\quad|y|<1.
\eeq
%
%
%
%
%
%
\subsection{The continuum limit} \label{subsec:The continuum limit}

One of the central philosophies behind the method of dynamical triangulations is ``universality''.
This concept is well known in statistical mechanics and plays a pivotal role in renormalization theory.
As applied to the case at hand it means that the precise form of the discrete amplitude \rf{eq:discrGxytZ} should not
be important for the physics of the system. The essential physics should for example not depend on the type of
polytopes that are used to regularize the path integral.

One of the prerequisites for such a scenario is that there exists a so-called ``critical'' hypersurface in
the space of parameters of the regularized theory. Near such a region the theory exhibits correlations that are much larger
than the size of the building blocks. If this happens the macroscopic physics is insensitive to the regularization
and one can safely take the limit where the building blocks are infinitesimally small and obtain a more or less unique
continuum theory.

Precisely how the above sketched scenario is realized in two dimensional causal dynamical triangulations
can be most easily understood by analyzing the one step propagator as a function of $g,x,y$. To identify the
critical surface in the parameter space spanned by the three couplings $g$,$x$ and $y$ we compute the average
number of triangles in one strip. This can be readily accomplished by taking the derivative with respect to $g$
of the one step propagator since $g$ can be seen as the fugacity for the number of triangles in the system,
\beq
\langle N_{\bigtriangleup} \rangle =g\frac{d}{d g}G(x,y;g;1)= 1 - \frac{1}{1 - g x} - \frac{1}{1 - g y} + \frac{1}{1 - g (x + y) }.
\eeq
The critical region of the parameter space can now easily be identified by observing that the total number of building blocks
diverges if the couplings satisfy the following relation,
\beq \label{eq:criticalregion}
|g(x+y) |=1,
\eeq
where it should be noted that besides this relation the couplings are also subject to the convergence condition
\rf{eq:convergencecondition}. Near the critical region \rf{eq:criticalregion} the typical length scale
of the system is indeed much larger than the length scale of
the individual building blocks. To define the continuum limit of the theory we assume canonical scaling dimensions for
the bulk and boundary cosmological constants,
\beq
g=g_c e^{-a^2 \La},\quad x=x_c e^{-a X},\quad y=y_c e^{-a Y}.
\eeq
Since we have performed the Wick rotation one is dealing with real valued $g,x,y$. Therefore the natural critical
values for these couplings are determined by \rf{eq:convergencecondition} and \rf{eq:criticalregion} yielding
\beq
g_c = 1/2,~~~~x_c = 1,~~~~y_c=1,
\eeq
leading to the following scaling relations,
\beq \label{eq:scalingrelations}
g=\frac{1}{2} e^{-a^2 \La},\quad x= e^{-a X},\quad y=e^{-a Y}.
\eeq
The continuum limit of the propagator can now be determined uniquely by inserting the scaling relations
\rf{eq:scalingrelations} into the regularized result \rf{eq:discrGxytZ}, giving
\beq \label{eq:contGXYTZ}
G_{\La}(X,Y,T) = -\log\left(1-Z_{\La}(X,Y,T) \right),
\eeq
where
\beq
Z_{\La}(X,Y,T) = \frac{\La}{(X+\SL \coth\SL T) (Y+\SL \coth\SL T) }.
\eeq
Note that in the continuum limit we defined the normalization of the wavefunctions such that
\beq
G_{\La}(X,Y,T) = a G(x,y;g;t).
\eeq
The power of $a$ is uniquely determined by the requirement that the continuum propagator satisfies
a continuum analogue of the composition law \rf{eq:discrcompositionxynonmarked}.
The continuum expression for the propagator where the boundary lengths are fixed instead of the boundary cosmological
constants can now be obtained by an inverse Laplace transformation of \rf{eq:contGXYTZ} with respect to both $X$ and $Y$,
\beq
G_\La(L_1,L_2;T)  = \frac{\e^{- \SL \coth \SL T (L_1+L_2) }}{\sinh \SL T}\; \frac{\sqrt{\La}}{\sqrt{L_1 L_2}}\; \;
I_1\left(\frac{2\sqrt{\La L_1 L_2}}{\sinh \SL T}\right),
\eeq
where $I_1(x) $ is a modified Bessel function of the first kind.
One indication that this object is a good propagator is that it satisfies the expected composition law
\beq \label{eq:compositionlawGLLT}
G_\La (L_1,L_2;T_1+T_2)  = \int_0^\infty \d L  \,L\;G_\La (L_1,L;T_1) \,G_\La(L,L_2,T_2).
\eeq
From the propagator one can naturally define the time dependent disc function by shrinking one of the boundary loops
to zero,
\beq \label{eq:defwavefunctionLT}
W_{\Lambda}(L,T) = G_{\La}(L,L'\equ 0;T),
\eeq
with
\beq
 G_{\La}(L,L'\equ 0;T)=\frac{\Lambda}{\sinh^2\sqrt{\Lambda}T}e^{{-\sqrt{\Lambda}L\coth\sqrt{\La}T }}.
\eeq
This object could be given a cosmological interpretation in that it is the probability amplitude to find a
universe of size $L$ that has existed for a time $T$. Upon doing an integration over time one obtains an object that
has the natural interpretation of a Hartle Hawking wavefunction,
\beq \label{eq:HartleHawkingwavefunctionL}
W_{\Lambda}(L) =\int_0^\infty dT\;W_{\Lambda}(L,T) =\frac{e^{-\sL L}}{L}.
\eeq
Alternatively, this object is often called the disc function.

\subsection{Marking the causal propagator}\label{Marking the causal propagator}

In most expositions on two dimensional causal dynamical triangulations the conventions differ from
the ones used in the previous section. Often, one employs propagators that have a marked vertex on the initial boundary,
\beq
G^{(1;0) }_\l(l_1,l_2;1) = l_1 G_\l(l_1,l_2;1),
\eeq
where the superscript notation $G^{(1;0)}$ denotes that the entrance loop of the propagator is marked,
$G^{(0;1)}$ implies that the exit loop is marked and $G^{(1;1)}$ means that both loops are marked.
In principle all vertices of the triangulation are indistinguishable, the introduction of a mark on the boundary implies
that one vertex on the boundary is distinguishable from all others.
The main virtue of the marking is that it removes the measure factor in the
composition law for the propagator \rf{eq:discrcompositionllnonmarked}
\beq \label{eq:discrcompositionllmarked}
G^{(1;0) }_\l(l_1,l_2;t_1+t_2)  = \sum_{l} G^{(1;0) }_\l(l_1,l;t_1) \; G^{(1;0) }_\l(l,l_2;t_2),
\eeq
which corresponds to the following composition law for the Laplace transformed propagator,
\beq\label{eq:discrcompositionxymarked}
G^{(1;0) }_\l(x,y;t_1+t_2)  = \ointz G^{(1;0) }_\l(x,z^{-1};t_1)  G^{(1;0) }_\l(z,y;t_2).
\eeq
Many formulas simplify with this trick, but the price one pays is that the propagator is no longer symmetric.
Although the physical symmetry of the two boundary components of the causal propagator is not reflected
in the marked expressions we discuss their properties for future convenience.

Since we have already computed the one step propagator without a mark,
we can find the marked version by simply taking a
derivative with respect to the boundary cosmological constant,
\beq \label{eq:G10xyg1}
G^{(1;0)}(x,y;g;1)  = x\frac{d}{d x}G^{(1;0)}_\l(x,y;g;1)  =\frac{g^2 x y}{(1-g x) (1-g(x+y) ) }.
\eeq
To find the marked propagator for finite $t$ one can find an iterative equation analogous to
\rf{eq:iterationequationnonmarked} ,
\beq \label{eq:iterationequationmarked}
G^{(1;0) }(x,y;g;t)  = \frac{gx}{1-gx}\; G^{(1;0) }\left(\frac{g}{1-gx},y;g;t-1\right).
\eeq
Evidently, one can use this relation in a similar iteration procedure as described in section \rf{sec:The discrete solution}
to obtain $G^{(1,0) }(x,y;g;t) $ and then apply the scaling relations to
find the continuum propagator. Instead, one can also apply the scaling relations \rf{eq:scalingrelations}
directly to \rf{eq:iterationequationmarked} and find a differential equation for the continuum marked propagator,
\beq \label{eq:differentialequationGXYT}
\frac{\partial}{\partial T}G^{(1;0)}_{\La}(X,Y,T) =-\frac{\partial}{\partial X}\left[\hcW(X) G^{(1;0) }_{\La}(X,Y,T) \right],
\eeq
where
\beq \label{eq:defhcWpure}
\hcW(X) =X^2-\La.
\eeq
The notation $\hcW(X) $ is introduced to anticipate generalizations, since this differential equation can also be solved
for more general expressions than \rf{eq:defhcWpure}. Equation \rf{eq:differentialequationGXYT}
has the form of a standard first order
differential equation, so to solve it uniquely one needs to supply one boundary condition. The natural condition to
impose on the propagator is that it reduces to the identity for the limit of zero proper time $T$,
\beq
G^{(1;0)}_{\La}(L_1,L_2;T\equ 0)  = \del(L_1-L_2),
\eeq
which has the following analogue in terms of boundary cosmological constants,
\beq
G^{(1;0)}_\La(X,Y;T\equ 0)  = \frac{1}{X+Y}.
\eeq
Solving the differential equation \rf{eq:differentialequationGXYT} with this boundary condition gives
\beq \label{eq:G10XYT}
G^{(1;0) }_{\La}(X,Y,T) =\frac{\hcW(\bX(T) ) }{\hcW(X) }\frac{1}{\bX(T) +Y},
\eeq
where $\bar{X}(T;X) $ is the solution to the characteristic equation
\beq \label{eq:characteristicequation}
\frac{d\bX}{dT}=-\hcW(\bX(T)),\quad \bX(T=0) =X.
\eeq
Since here we are interested in pure CDT, the definite form of this equation is known since $\hcW(X) =X^2-\La$ and
one can solve the characteristic equation explicitly,
\beq \label{eq:xbar}
\bar{X}(T)  = \SL \;
\frac{(\SL+X) -\e^{-2\SL T}(\SL-X) }{(\SL+X) +\e^{-2\SL T}(\SL-X) }.
\eeq
Inserting this expression in \rf{eq:G10XYT} it is easily verified that this form of the propagator is fully compatible
with the previously derived results. If one marks the unmarked propagator \rf{eq:contGXYTZ} by taking a derivative with
respect to the initial boundary cosmology constant $X$, the expression coincides precisely with the above
derived result \rf{eq:G10XYT} .

\subsection{Hamiltonians in causal quantum gravity}\label{subsec:Hamiltonians in causal quantum gravity}

An interesting observation regarding the differential equation for the causal propagator \rf{eq:differentialequationGXYT}
is that it can be viewed as a Wick rotated Schr\"{o}dinger equation,
\beq \label{eq:WickSchrodinger}
-\frac{\partial}{\partial T}G^{(1;0) }_{\La}(X,Y,T) =\hat{H}_X G^{(1;0) }_{\La}(X,Y,T),
\eeq
where the $\hat{H}_X$ is the {\it quantum effective Hamiltonian} and is given by
\beq
\hat{H}_X = (X^2-\La) \frac{\partial}{\partial X}+2X.
\eeq
By inverse Laplace transforming the Schr\"{o}dinger equation \rf{eq:WickSchrodinger} we can find the Hamiltonian in the
``position representation'' or, more appropriately the length representation,
\beq
\hat{H}^{\scriptscriptstyle marked}_L = -L\frac{\partial^2}{\partial L^2}+ \La \: L.
\eeq
This operator is selfadjoint on the Hilbert space $\mathcal{H}\equ \mathcal{L}^2(\mathbb{R}_+,L^{-1}dL) $.
Care should be taken however, since the above defined Hamiltonians are not derived from a symmetric
propagator. The proper Hamiltonian can be obtained by taking the continuum limit of the nonmarked
iteration equation \rf{eq:iterationequationnonmarked}, yielding
\beq
\hat{H}_L = -L\frac{\partial^2}{\partial L^2}- 2 \frac{\partial}{\partial L} + \La \: L,
\eeq
which is selfadjoint on the Hilbert space $\mathcal{H}\equ \mathcal{L}^2(\mathbb{R}_+,LdL) $,
where the measure originates from the basic composition law \rf{eq:compositionlawGLLT}.
Physically, the Hamiltonian is well defined as it is bounded from below
and it is even possible to find its spectrum explicitly.
Solving the eigenvalue equation
\beq
\hat{H}_L \psi_n(L) =E_n\psi_n(L),
\eeq
yields the eigenfunctions
\beq
\psi_n(L) = 2 \sqrt{\tfrac{\La}{n+1}} e^{-\sqrt{\La}L}L^{1}_n(2\sqrt{\La}L),
\eeq
where $L^{1}_n(x) $ is a generalized Laguerre polynomial.
Further, the eigenvalues are given by
\beq
E_n=2\sqrt{\La}(n+1).
\eeq
Observe that the spectrum is equidistant, making the quantum mechanics of the problem rather similar
to that of the harmonic oscillator. The main difference with the harmonic oscillator is that the Hilbert space
of the CDT Hamiltonian is defined by the square integrable functions on the real half line $L\in[0,\infty )$
and not on the whole real line $L\in(-\infty,\infty )$.

To make contact with the continuum treatment of causal quantum gravity by Nakayama \cite{Nakayama:1993we}
we need to absorb the measure factor $L$ in the normalization of the wavefunctions.
The marking procedure described in section \rf{Marking the causal propagator} does precisely this
but the measure is absorbed in the initial state only, leading to a nonsymmetric propagator.
In the continuum theory it is possible to absorb the measure factor in both the initial and final states by
the rescaling $\psi_n(L) =\tfrac{1}{\sqrt{L}} \varphi_n(L) $. Hence, the Hamiltonian with flat measure is defined by
\beq
\hat{H}^{\scriptscriptstyle{flat}}_L \varphi_n(L) =\hat{H}_L\:\tfrac{1}{\sqrt{L}} \varphi_n(L),
\eeq
giving
\beq \label{eq:flathamiltoniancdt}
\hat{H}^{\scriptscriptstyle{flat}}_L = -L\frac{\partial^2}{\partial L^2}- \frac{\partial}{\partial L} + \frac{1}{4L}+ \La \: L.
\eeq
To see how this Hamiltonian appears in the continuum derivation we
consider the non-local ``induced'' action of 2d quantum gravity,
first introduced by Polyakov \cite{Polyakov:1981rd}
\beq \label{eq:inducedaction}
S[g]= \int \d t \d x \sqrt{g} \left( \frac{1}{16} R_g \frac{1}{-\Del_g}R_g +\La \right),
\eeq
where $R$ is the scalar curvature corresponding to the metric $g$, $t$ denotes ``time'' and $x$ the
``spatial'' coordinate.

Nakayama \cite{Nakayama:1993we} analyzed the action \rf{eq:inducedaction} in proper
time gauge assuming the manifold has the topology of a cylinder
with a foliation in proper time $t$, i.e.~the metric was assumed
to be of the form:
\beq\label{eq:metricpropertimegauge}
g = \begin{pmatrix} 1
&  0 \\ 0&
\g(t,x) \end{pmatrix}.
\eeq
It was shown that in this gauge the
classical dynamics is described entirely by the following
one-dimensional action:
\beq \label{eq:nakayamaaction}
S_\k = \int_0^T \d t\left(\frac{\dot{l}^2(t) }{4l(t) }  + \La l(t) + \frac{\k}{l(t) }\right),
\eeq
where
\beq \label{eq:l(t) isint}
l(t)  = \frac{1}{\pi}\int \d x \sqrt{\g},
\eeq
and where $\k$ is an integration constant coming
from solving for the energy-momentum tensor component $T_{01}=0$
and inserting the solution  in \rf{eq:inducedaction}.

Thus $L_{cont}=\pi l(t) $ is precisely the length of the spatial curve
corresponding to a constant value of $t$, calculated in the metric
\rf{eq:metricpropertimegauge}. Nakayama quantized the actions $S_\k$ for $\k = (m+1) ^2$, $m$ a non-negative integer,
and argued that in the quantum theory $\k = (m+\tfrac{1}{2}) ^2$.
The classical Hamiltonian corresponding to the proper time gauge action \rf{eq:nakayamaaction}
was derived and reads as follows
\beq
H_m = \Pi_l l \Pi_l + \left(m+\tfrac{1}{2}\right) ^2\frac{1}{l}+ \La l,
\eeq
where $\Pi_l$ is the canonical momentum conjugate. Subsequently, the quantum Hamiltonian
is found by the straightforward replacement $\Pi_l \rightarrow -i\tfrac{\partial}{\partial l}$
\beq \label{eq:quantumhamiltoniannakayama}
\hat{H}_m = -l\frac{\partial^2}{\partial l^2}- \frac{\partial}{\partial l} + \left(m+\tfrac{1}{2}\right) ^2 \frac{1}{l}+ \La \: l.
\eeq
Rather remarkably, upon the identification $l \leftrightarrow L$ we see that for $m=0$ the result coincides
precisely with the
flat measure Hamiltonian derived by CDT \rf{eq:flathamiltoniancdt}. An interpretation for the higher $m$ quantum numbers
in the context of CDT can be found in \cite{DiFrancesco:1999em}. However, a complete understanding of the significance of
these quantum numbers is in the opinion of the author still lacking.

Recall that the parameter $l$ in Nakayama's Hamiltonian is not the physical length, $L_{cont}$, it differs from $L_{cont}$
by a factor $\pi$. Consequently, it is natural that also in CDT the physical length is defined as $L_{cont}=\pi L$.
This observation will turn out to be important for the emergent geometry discussion in chapter \ref{ch4}.

\chapter{Baby universes} \label{ch3}

Despite recent progress \cite{Ambjorn:2005db,Ambjorn:2005qt},
little is known about the ultimate
configuration space of quantum gravity on which its nonperturbative
dynamics takes place. This makes it difficult to decide
which (auxiliary)  configuration space to choose as starting
point for a quantization. In the context of a path integral
quantization of gravity, the relevant question is which class of
geometries one should be integrating over in the first place.
Setting aside the formidable difficulties in ``doing the integral", there is
a subtle balance between including too many geometries -- such
that the integral will simply fail to exist (nonperturbatively)  in any
meaningful way, even after renormalization -- and including too few
geometries, with the danger of not capturing a physically relevant part
of the configuration space.

A time-honored part of this discussion is the question of whether a
sum over different topologies should be included in the
gravitational path integral. The absence to date of a viable theory of
quantum gravity in four dimensions has not hindered speculation on
the potential physical significance of processes involving topology
change (for reviews, see \cite{Horowitz:1990qb,Dowker:2002hm}). Because such processes
necessarily violate causality, they are usually considered
in a Euclidean setting where the issue does not arise.
In our models for topology change we do wish to capture some Lorentzian aspects though.
In section \rf{subsec:Lorentzian aspects} we therefore address some of the causality issues
and argue that the causal structure of the manifolds is only modified mildly.

The focus of this chapter is devoted to the introduction of spatial topology change in
two dimensional quantum gravity. This might be viewed as a bit of an ironical undertaking since \emph{the} characteristic
that makes the quantum geometry of CDT better behaved than its Euclidean rival is precisely its fixed spatial topology!
We do however show that when a coupling constant is introduced the effect of the spatial topology changes is much less
severe. A natural scaling for this coupling constant is presented, where the dynamics of an individual spacetime
and splitting into baby universes both contribute to the continuum limit of the theory. Consequently, time scales canonically
and the limit where the coupling constant tends to zero is smooth and continuous and gives back the results
of the bare CDT model. This should be
contrasted with Euclidean quantum gravity defined through dynamical triangulations, since in that case
the continuum dynamics of the theory is completely dominated by the baby universes, making the geometry highly fractal.

Rather than going into the details of the theory with the coupling constant straightaway we will
first explain the role that baby universes play in the relation between Euclidean and Causal
dynamical triangulations in section \rf{sec:Euclidean results with causal methods}.
The formalism of CDT can easily be extended to allow for spatial topology
change whereby we admit the geometry to split into baby universes. It turns out that if one allows
for the baby universes the splitting process will dominate the path integral completely, wiping out
all traces of the dynamics of each individual spatial universe \cite{Ambjorn:1998xu}. At the same
time it is shown that this continuum limit corresponds uniquely to the dynamics of Euclidean
quantum gravity. Explicitly it is shown how to rederive the Hartle Hawking wavefunction and
propagator of Euclidean dynamical triangulations in the continuum limit. In this discussion we
closely follow \cite{Ambjorn:1999nc}.

In sections \rf{sec:Introducing the coupling constant},\rf{sec:Dynamics to all orders in the coupling}
and \rf{sec:Relation to random trees} we describe our published work where we generalize the above
described construction \cite{Ambjorn:2007xx}. Particularly, we associate a coupling constant that is reminiscent of
the string coupling constant with the spatial topology fluctuations.
We show that in a suitable scaling limit the spatial topology changes contribute to the path integral in a
controlled manner without dominating the quantum geometry.

\section{Euclidean results with causal methods} \label{sec:Euclidean results with causal methods}

In section \rf{subsec:Spatial topology change} we allow for spatial topology change
and show how to explicitly introduce the baby universes
in the discrete formalism and we compute the Hartle Hawking wavefunction in the continuum limit.
The result is the well known disc amplitude of two dimensional Euclidean quantum gravity.

\subsection{Spatial topology change} \label{subsec:Spatial topology change}

We now address the implementation of spatial topology changes by generalizing the discrete CDT framework.
Although a multitude of constructions that realize topology change exist, they are all equivalent in the continuum limit as
has been checked for some cases by the authors of \cite{Ambjorn:1998xu}.
In the following we introduce one specific realization and show that in the continuum limit the familiar
results from Euclidean quantum gravity are recovered.

\begin{figure}[t]
\begin{center}
\includegraphics[width=4.3in]{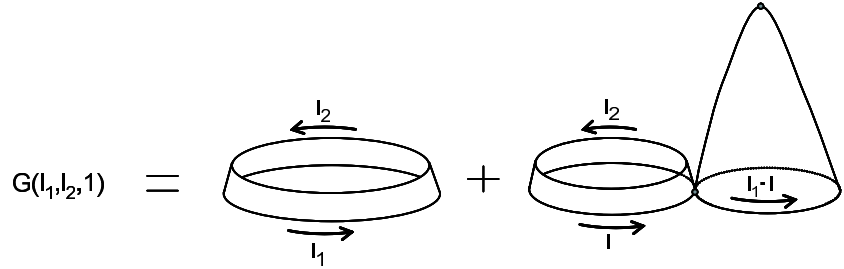}
\caption{Illustration of a one step propagator with a ``baby universe''.}
\label{fig:dressedonesteppropagator}
\end{center}
\end{figure}

The first step in the construction is to generalize the one step propagator, or transfer matrix,
of CDT by alleviating the constraint on the spatial topology of its initial loop.
Particularly, we include strips for which
the entrance loop, say of length $l_1$, has the topology of a ``figure eight''.
A natural procedure to create such a figure eight is to non-locally identify two points of a spatial universe with
topology of an $S^1$. Incorporating this pinching process leads to a factor of $l_1$ in the combinatorics for the one step
propagator since the pinching is allowed to happen at any of the $l_1$ vertices.
A baby universe is now created by associating one of the loops of the figure eight with the boundary of
a disc function whilst the other loop is associated with the initial loop of a regular one step propagator
(fig.~\ref{fig:dressedonesteppropagator}).
Combining all of the above, we can write the new transfer matrix in terms
of the old, or ``bare'',  transfer matrix and the as yet undetermined disc function
\beq \label{eq:onsteppropagatordressed}
G_\l(l_1,l_2;1)  = G_\l^{(b) }(l_1,l_2;1) + \sum_{l=1}^{l_1-1} l_1 w(l_1\mi l,g) \,G_\l^{(b) }(l,l_2;1),
\eeq
where it should be noted that we have dropped the superscript notation for the marking and have
used the notation $G_\l(l_1,l_2;1)=G^{(1,0)}_\l(l_1,l_2;1)$.

This new, or ``dressed'', one step propagator satisfies the same composition law as the bare transfer matrix
\rf{eq:discrcompositionllmarked}, i.e.
\beq
G_\l(l_1,l_2;t_1+t_2)  = \sum_l G_\l(l_1,l;t_1) G_\l(l,l_2;t_2).
\eeq
Particularly this implies that the dressed one step propagator can still be used as a transfer matrix
despite the fact that the disc function contains an infinite sum over time steps
\beq\label{eq:discrcompositionlldressed}
G_\l(l_1,l_2;t)  = \sum_l G_\l(l_1,l;1) \;G_\l(l,l_2;t\mi 1).
\eeq
Performing a (discrete)  Laplace transformation of eq.~\rf{eq:discrcompositionlldressed}
leads to
\bea
{G(x,y;g;t)  =}~~~~~~~~~~~~~~~~~~~~~~~~~~~~~~~~~~~~~~~~~~~~~~~~~~~~~~~
~~~~~~~~~~~~~~~~~~~~~~
 & &  \nonumber \\
\ointz  \left[G_\l^{(b) }(x,z^{-1};1) \pl x \frac{\prt}{\prt x}
\Bigl( w(x;g)  G_\l^{(b) }(x,z^{-1};1) \Bigr)  \right]  G(z,y;g;t \mi 1),
&&\nonumber \\
\!\!\!\!\!\!\!\! \label{eq:GxygtGdressed}
\eea
or, using the explicit form of the transfer matrix $G_{(b)}(x,z;g;1) $,
formula \rf{eq:G10xyg1},
\beq
G(x,y;g;t)  = \Bigl[1+x\frac{\prt w(x,g) }{\prt x}+ x w(x,g) \frac{\prt}{\prt x} \Bigr]
\, \frac{gx}{1-gx} \, G\Bigl( \frac{g}{1\mi gx},y;g;t\mi 1\Bigr) \label{eq:Gxygtdressed}.
\eeq
Note that this is not a closed equation, since so far neither the disc amplitude $w(x,g) $ nor $G(x,y;g;t) $ are known.
Although this means that we cannot derive the discrete expressions, it will be shown by using scaling arguments that
one can uniquely determine the continuum disc amplitude $W_\La(X) $ and propagator $G_\La(X,Y;T) $. As in the case
for CDT without topology change we assume canonical scaling for both the boundaries and the cosmological constant,
\beq
g=\frac{1}{2}e^{-a^2 \La},\quad x=e^{-a X},\quad y=e^{-a Y}.
\eeq
In the following arguments no specific choice for the scaling of the time variable will be assumed as its
scaling will be determined at a later stage. Interestingly, the composition law and the canonical scaling of the boundary lengths
are enough to determine the scaling of the dressed propagator

\beq
G_\l(l_1,l_2,t)  \buildrel{a\rightarrow0}\over\longrightarrow  a\, G_\La(L_1,L_2;T).
\eeq

\begin{figure}[t]
\begin{center}
\includegraphics[width=4in]{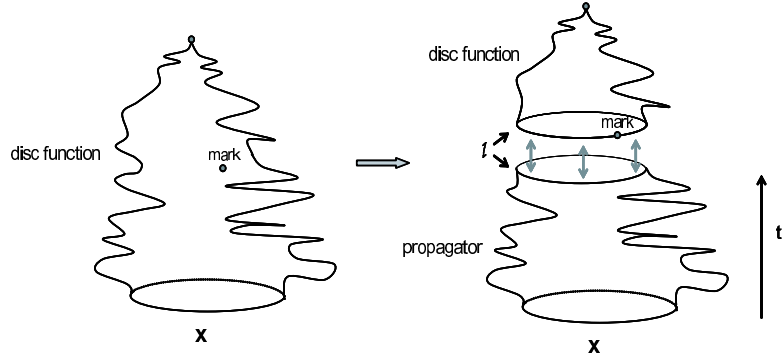}
\caption{Decomposition of the marked CDT disc function in another CDT disc function and a propagator.}
\label{fig:discfunctiondecompositionpure}
\end{center}
\end{figure}

Changing the boundary conditions from fixed boundary length to fixed boundary cosmological constant
amounts to taking a Laplace transformation and implementing the canonical scaling of the boundary cosmological
constant $x=e^{-a X}$, leading to

\beq\label{top8}
G_\l(x,l_2,t)  \buildrel{a\rightarrow 0}\over\longrightarrow G_\La(X,L_2,T)
\eeq

and

\beq\label{top8x}
G_\l(x,y;t)  \buildrel{a\rightarrow 0}\over\longrightarrow a^{-1} G_\La(X,Y;T).
\eeq

\begin{figure}[t]
\begin{center}
\includegraphics[width=4in]{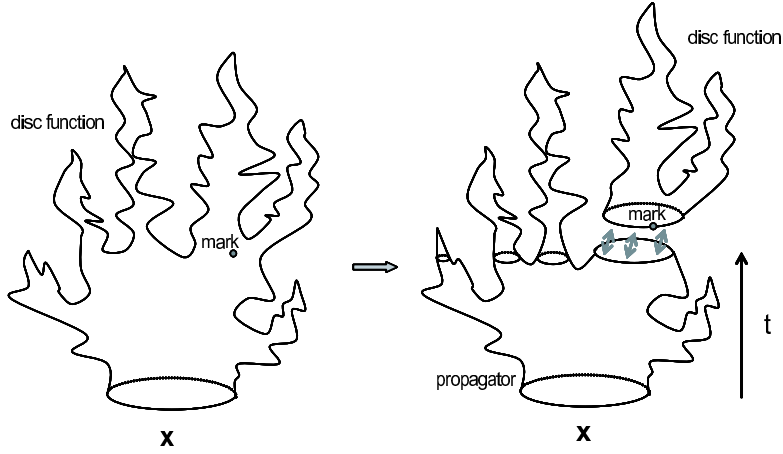}
\caption{Decomposition of the disc function including spatial topology change.}
\label{fig:discfunctiondecompositiontop}
\end{center}
\end{figure}

The scaling relation for the disc function is not as simple to obtain as the scaling for the propagator
and we show that it actually depends on the scaling of time. Where for the propagator we use the composition law
to derive its scaling relation we use the following exact combinatorial identity to determine the scaling relation
for the disc function

\beq\label{an1}
g\ \frac{\prt w(x,g) }{\prt g} = \sum_t \sum_l G(x,l;g;t)  \, l\, w(l,g),
\eeq
or, after the usual Laplace transformation,
\beq\label{an2}
g\ \frac{\prt w(x,g) }{\prt g}=\sum_t \ointz G(x,z^{-1};g;t) \; \frac{\prt w(z,g) }{\prt z}.
\eeq
These identities reflect the fact that if one introduces a mark anywhere in the bulk by taking a derivative
with respect to $g$, the disc function can be decomposed into a propagator and another disc amplitude. The situation
for the bare model is illustrated in fig.~\ref{fig:discfunctiondecompositionpure}.
In the figure on the right we have highlighted the points that have a distance $t$ to the
entrance loop. Since the bare model does not allow for spatial topology change these points constitute one spatial universe
with the topology of an $S^1$. In fig.~\ref{fig:discfunctiondecompositiontop} the situation is illustrated in the case where one does allow for topology change,
in this case all points with equal distance $t$ from the entrance loop form a spatial section with the
topology of several $S^1$'s and the mark is located on one of them.

Since our current objective is to find the scaling for the disc function we assume the following general scaling ansatz
\beq\label{an3}
w(x,g) = w_{ns}(x,g) + a^{\eta}W_\La(X)  + \mbox{less singular terms}.
\eeq
In the case $\eta < 0$ the first term is irrelevant in the continuum limit, it does not appear in the computation
of any continuum quantities. However, if $\eta >0$  a term $w_{ns}$ will generically be present \cite{Ishibashi:1993sv}.
Additionally, the general ansatz for the scaling of the time variable reads as follows
\beq
T = a^{\ep}t,~~~~~\ep >0.
\eeq
As shown in section \rf{subsec:The continuum limit} both space and time scale canonically in the bare model which corresponds to $\ep=1$.
Below we show that allowing the branching into baby universes to contribute to the continuum limit
forces the scaling of time to be anomalous, creating an inherent asymmetry between the time- and space directions.

Inserting the scaling relations \rf{an3} and \rf{eq:scalingrelations} into eq.\ \rf{an2} we obtain
\bea
&&\frac{\prt w_{ns}}{\prt g}- 2a^{\eta-2}\frac{\prt W_\La(X) }{\prt \La} = \nonumber \\
&&\hspace{.6cm} \frac{1}{a^{\ep}}\int \d T \int dZ\;  G_\La(X,-Z;T) \bigg[ \frac{\prt w_{ns}}{\prt z} -a^{\eta-1}\frac{1}{z_{c}} \frac{\prt W_\La(Z) }{\prt Z}\bigg],
\label{an4}
\eea
where $(x,g) =(x_c,g_c) $ in the non-singular part.

From eq.\ \rf{an4} and the requirement $\epsilon >0$ it follows that the
only consistent choices for $\eta$ are
\begin{itemize}
\item[{\bf Scaling 1:}] {\boldmath{${\eta <\ }$}{\bf 0}}

As can be seen from \rf{an3}, this range of values corresponds to the situation where the non scaling part of the disc function
is irrelevant and the physics is completely independent of the cutoff,\
\beq\label{an5c}
a^{\eta-2} \frac{\prt W_\La(X) }{\prt \La} = \frac{a^{\eta-1}}{2 a^{\ep}}
  \int \d T \int dZ\;  G_\La(X,-Z;T)  \;\frac{1}{z_{c}}
  \frac{\prt W_\La(Z) }{\prt Z}.
\eeq
The continuum limit can be taken for any $\eta<0$ since \rf{an5c} does not depend on its explicit value. The value of
$\ep$ on the other hand is fixed and one needs to have $\ep=1$ for the continuum limit to exist.
Summarizing, if the scaling
of the disc function is such that non scaling contributions are negligible in the continuum limit, the time
variable automatically scales canonically. Obviously the bare CDT model falls in this class of scalings since it has
$\eta =-1$ and $\ep=1$.

\item[{\bf Scaling 2:}] {\bf 1}{\boldmath${\ < \eta < \ }$}{\bf 2}.

For this class of scalings for the disc function formula \rf{an4} splits into two equations
\beq\label{an5}
-a^{\eta-2} \frac{\prt W_\La(X) }{\prt \La} = \frac{1}{2 a^{\ep}}\,\frac{\prt w_{ns}}{\prt z}\bigg|_{z=x_c}\;\int \d T \int dZ\;  G_\La(X,-Z;T),
\eeq
and
\beq\label{an6}
\frac{\prt w_{ns}}{\prt g}\bigg|_{g=g_c} = -\frac{a^{\eta-1}}{a^{\ep}} \int \d T \int dZ\;  G_\La(X,-Z;T) \;\frac{1}{z_{c}} \frac{\prt W_\La(Z) }{\prt Z}.
\eeq
From \rf{an5} it follows that $a^{\eta-2}= \frac{1}{a^{\ep}}$ and from  \rf{an6} one sees that
$ \frac{a^{\eta-1}}{a^{\ep}}=1$. Combining these requirements we are led to the conclusion that $\ep= 1/2$ and $\eta=3/2$,
which are precisely the values found in Euclidean $2d$ gravity.
Let us further remark that eq.\ \rf{an5} in this case becomes
\beq\label{an5a}
-\frac{\prt W_\La(X) }{\prt \La} = \mbox{const.}\;G_\La(X,L_2=0).
\eeq
If one rescales the couplings it is possible in general to absorb the constant originating from the non universal terms.
This implies that the continuum dynamics of the theory strongly depends on the fact that there are non universal terms
surviving the continuum limit but their precise value does not play a pertinent role. It should be noted that \rf{an5}
expresses a different relation between the disc function and the propagator than was used for the bare model,
 it differs from \rf{eq:defwavefunctionLT} by a derivative with respect to the cosmological constant.
Finally, inserting $\ep= 1/2$ and $\eta=3/2$ into eq.\ \rf{an6} yields
\beq\label{an7}
\int \d T \int dZ\;  G_\La(X,-Z;T) \;\frac{\prt W_\La(Z) }{\prt Z} = \mbox{const},
\eeq
where as in \rf{an5a} the constant originates from the non scaling terms of the disc function and its value does not
play a significant role in the continuum theory.
\end{itemize}

\begin{figure}[t]
\begin{center}
\includegraphics[width=2in]{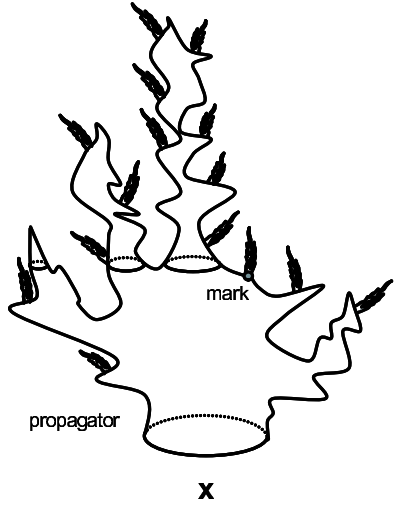}
\caption{At every point in the quantum geometry there is an infinitesimal baby universe.}
\label{fig:mini}
\end{center}
\end{figure}

The relation \rf{an5a} possesses a remarkable interpretation in terms of baby universes. Basically it states that near
any mark in the bulk of the continuum Hartle Hawking wave function there is a baby universe at the scale of the cutoff.
Since the location of the mark is arbitrary it implies that
\emph{near every point there is a baby universe at the cutoff scale} which is illustrated in fig.~\ref{fig:mini}.
This does not mean however that all baby universes are of negligible size. An additional
indication that cutoff size geometries play an important role in the dynamics comes from the Laplace transform of \rf{an7}
\beq
\int \d T\int dL\;  G_\La(L,L';T) \;L'\;W_\La(L')  = \mbox{const}. \times ~\delta(L),
\eeq
which shows that the distribution of geometries is such that it is strongly peaked around universes that have minimal
boundary length. In the following we show that these microscopic artifacts
are an important feature of $2d$ Euclidean quantum gravity since the CDT model with baby universes exactly reproduces
the continuum equations of the Euclidean model in the scaling limit. Furthermore, to make contact with the
Euclidean theory, \rf{an5a} is used in an essential way.

Having derived the scaling relations we can now analyze the scaling limit of \rf{eq:Gxygtdressed} and find an equation
for the dynamics of the propagator. In order for the equation to have a scaling limit at all, $x_c,\, g_c$ and
$w_{ns}(x_c,g_c) $ must satisfy two relations which can be determined
straightforwardly from \rf{eq:Gxygtdressed}. The remaining continuum equation reads
\bea
a^\ep\frac{\prt}{\prt T}\, G_\La(X,Y;T) & =&-a \, \frac{\prt}{\prt X} \Bigl[ (X^2-\La)  G_\La(X,Y;T) \Bigr] \nn
&&-a^{\eta-1}\frac{\prt}{\prt X} \Bigl[W_\La(X)  G_\La(X,Y;T) \Bigr].
\label{top13}
\eea
The first term on the right-hand side of eq.\ \rf{top13} is precisely the one we have already encountered
in the bare model, while the second term is due to the creation of baby universes.
In case where $\eta \leq 1$ the first term in \rf{top13} is subdominant
and the scaling will not be compatible with baby universes in the continuum limit. So from \rf{top13} it can be seen that
for $\eta \leq 1$ we need to have that $\ep=1$ leaving us with the continuum differential equation for the
propagator of the bare model
\beq \label{eq:difequationGXYTpure}
\frac{\prt}{\prt T}\, G_\La(X,Y;T)  = - \frac{\prt}{\prt X} \Bigl[ (X^2-\La)  G_\La(X,Y;T) \Bigr],
\eeq
where we remind the reader that this differential equation can naturally be interpreted as a
Wick rotated Schr\"{o}dinger equation of a single string propagating with respect to its own time on world sheet,
\beq
\frac{\prt}{\prt T}\, G_\La(X,Y;T)  = \hat{H}_{X} G_\La(X,Y;T).
\eeq
From the Laplace transform of this equation one sees that $\hat{H}$ is a simple Hamiltonian containing a kinetic term,
a potential induced by the cosmological constant and no interaction terms,
\beq
\hat{H}^{\scriptscriptstyle marked}_L = -L\frac{\partial^2}{\partial L^2}+ \La \: L.
\eeq
For the second scaling however, where $ 1 < \eta < 2 $, the last term on the right-hand side of \rf{top13} will always
dominate over the first term. Consequently, the first term does \emph{not} survive the continuum limit leaving one with
an equation without a Hamiltonian containing a kinetic term. So for this scaling the dynamics of the world sheet is
governed purely by the interactions of splitting strings. Equivalently, we can say that
{\em once we allow for the creation of baby universes, this process will completely dominate the continuum limit.}
As we have seen from \rf{an5} and \rf{an6}, $(\eta,\ep) =(3/2,1/2) $ is the only consistent scaling that allows
for baby universes. Inserting
this scaling in \rf{top13} we obtain the following continuum equation
\beq\label{top16}
\frac{\prt}{\prt T}\, G_\La(X,Y;T)  = -\frac{\prt}{\prt X} \Bigl[W_\La(X)  G_\La(X,Y;T) \Bigr],
\eeq
%
%
which, combined with eq.\ \rf{an5a}, determines the continuum disc
amplitude $W_\La(X) $.
Integrating \rf{top16} with respect to $T$ and using that
$G_\La(L_1,L_2;T\equ 0) =\delta(L_1\mi L_2) $, i.e.\
\beq\label{top18a}
G_\La(X,L_2\equ 0;T\equ 0) =1,
\eeq
we obtain
\beq\label{top18}
-1 = \frac{\prt}{\prt X}\bigg[ W_\La(X)  \frac{\prt}{\prt\La} W_\La(X)\bigg].
\eeq
From dimensional analysis one can easily see that $W_\La^2(X)  = X^3 F(\SL/X)$. This implies that
the solution of the disc function reads as follows
\beq\label{top19}
W_\La(X)  = \sqrt{-2\La X + b^2 X^3+ c^2 \La^{3/2}}.
\eeq
However, not all values for $b$ and $c$ are physically acceptable.
The inverse Laplace transform of \rf{top19}, $W_\La(L)$, should be bounded for
all $L>0$. This constraint gives the following expression for the disc function
\beq\label{top20}
W_\La(X)  = b \Big(X-\frac{\sqrt{2}}{b\,\sqrt{3}} \,\SL\Big)
\sqrt{X+ \frac{2\sqrt{2}}{b\,\sqrt{3}}\SL},
\eeq
where the constant $b$ is a constant that reflects specific details of the discrete statistical model,
as described by equation \rf{an5a}. We discussed above that this constant does not have an obvious physical significance
since it can be absorbed into the cosmological constant.
Absorbing the irrelevant constant $b$ one obtains the disc function $W^{(eu) }_\La (X) $ of 2d Euclidean quantum gravity,
\beq\label{top21}
W_\La (X)  = (X -\oh \SL )  \sqrt{X+\SL}.
\eeq
Note that the defining equation for the disc function \rf{top16}, can alternatively be derived
by several methods within 2d Euclidean quantum gravity \cite{Kawai:1993cj,Ishibashi:1993sv,Watabiki:1993ym}.
In those computations it is clear that $T$ can be interpreted as the {\em geodesic distance} between
the initial and  final loop.


For a more detailed account on baby universes in
two dimensional Euclidean quantum gravity the reader is referred to \cite{Jain:1992bs}.

%
%
\section{Introducing the new coupling constant}\label{sec:Introducing the coupling constant}

In this section we present a model which can be viewed as a hybrid between Euclidean and causal quantum gravity. Recall
that the pure CDT model possesses a Hamiltonian that governs the dynamics of an individual string, but that there is no interaction term
in the dynamical equation to allow for spatial topology change \rf{eq:difequationGXYTpure}.
In the case of the Euclidean theory the opposite is true, the
equation governing the dynamics of the propagator only contains an interaction term and no kinetic or potential terms
for an individual string \rf{top16}. In the previous section it was shown that both theories can be obtained
by two different scaling
limits of one unifying statistical model based on CDT \rf{eq:onsteppropagatordressed}.
The current objective is to show that there is a natural adaptation of
this unifying model such that its scaling limit acquires single string dynamics while at the same time
allowing interactions in the form of spatial topology change. The essential ingredient in accomplishing this is to
introduce a new coupling in the statistical model \rf{eq:onsteppropagatordressed}
whose scaling is such that both the interaction term and the dynamical term
in \rf{eq:onsteppropagatordressed} contribute in the continuum limit.
\begin{figure}[t]
\begin{center}
\includegraphics[width=3in]{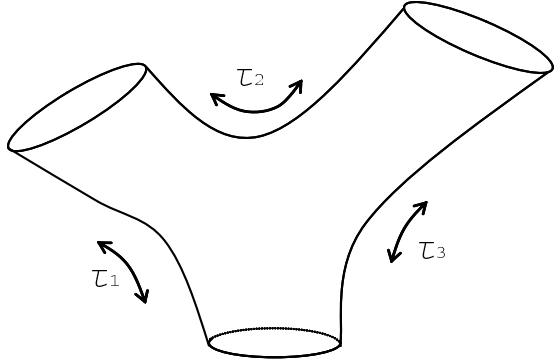}
\caption{A representation of the global aspects of a trouser geometry with modular parameters ($\tau_1, \tau_2, \tau_3$).}
\label{fig:trousermodular}
\end{center}
\end{figure}
Before discussing the construction of the model we examine some aspects of geometries with spatial topology change.
The simplest example of a two dimensional geometry with spatial topology change is the so-called ``trouser'' geometry.
A trouser geometry is a manifold with three boundary components with the topology of an $S^1$. In general the so-called
legs of the trousers can have any length which can be parameterized by three modular parameters as illustrated in
fig.~\ref{fig:trousermodular} .

\begin{figure}[t]
\begin{center}
\includegraphics[width=4in]{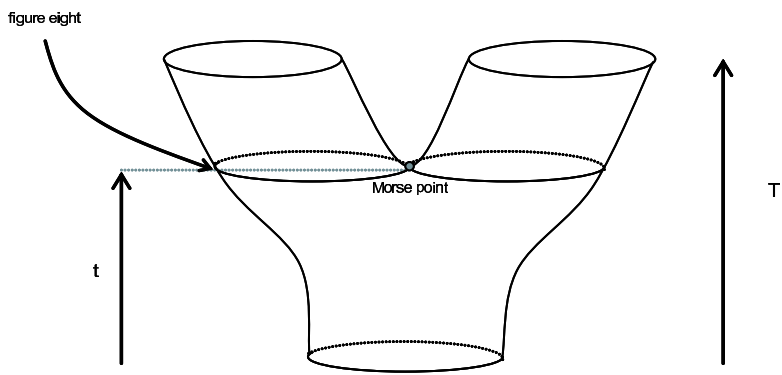}
\caption{A representation of the global geometry of a trouser geometry in causal dynamical triangulations }
\label{fig:trousercdt}
\end{center}
\end{figure}

We are interested however in a more restricted subclass of trouser geometries which are of the form depicted in
fig.~\ref{fig:trousercdt}.
We demand that we have an initial boundary, where every point on this boundary has the same distance to the two final loops.
This restriction means that on top of the length of the boundary components we need just two parameters to describe
the global characteristics of the geometries, one describing the length of the trouser leg belonging to the initial
boundary component and one that specifies the length of the trouser legs related to the final boundary components.
since we required the final boundary components to have the same distance to the initial boundary.

Now we ask ourselves the question, what is the contribution of such a geometry to the path integral? As described in
chapter \ref{ch2} the action for two dimensional gravity contains just a volume term proportional to the cosmological constant
and a topological term coming from the two dimensional Einstein-Hilbert action and the Gibbons-Hawking-York boundary term,
\beq
 S_M= \La V_M + \tfrac{2 \pi}{G_N}\chi(M),
\eeq
where $\chi(M)=\left(2-2\genus-b \right)$ is the Euler characteristic of the manifold.
The weight of an individual geometry in the path integral is therefore proportional to the exponential of the volume and the
exponential of the number of handles and boundary components.
\beq
G_{G_N,\Lambda}(L_1,L_2,T)  = \sum_{topol.} g^{2\genus}_S \int_M D[g_{\mu\nu}] e^{i \La V_M }.
\eeq
%
%
%

%

We see that geometries with the topology of a cylinder are not weighted by $g_S$, since the Euler characteristic of
a cylinder is zero. The Euler characteristic of a trouser geometry is one however, since the
trouser geometry has one more boundary component than the cylinder. Consequently the weight of the trouser geometry
in the path integral is not only determined by the cosmological constant, but also by the coupling $g_S$. For the purposes
of this thesis we are not particularly interested in computing amplitudes with the topology of a trouser, but
we are interested in calculating propagators, i.e.~amplitudes that have two macroscopic boundary components. If one
shrinks one of the final boundary components of a trouser geometry to zero one obtains a geometry with two
boundary components that we interpret as a ``cylinder geometry with one baby universe''.
Shrinking a boundary component does not alter the Euler characteristic, allowing us to conclude that a
geometry with $n$ baby universes contributes to the path integral with a $g_S^n$ topological weight. Note that in the
derivation of the disc amplitude of Euclidean dynamical triangulations of section
\rf{sec:Euclidean results with causal methods} we did not associate
a coupling constant to the baby universes. Subsequently, the formation of baby universe was not suppressed with the
result that the continuum limit is dominated by cutoff scale baby universes. In section
\rf{sec:Dynamics to all orders in the coupling} we show that adding the
coupling constant makes the term \emph{baby} universes a bit misleading since the outgrowths have a certain finite size
in the continuum limit and can be made arbitrarily big or small depending on the value of $g_S$.

\subsection{Lorentzian aspects} \label{subsec:Lorentzian aspects}

\begin{figure}[t]
\begin{center}
\includegraphics[width=4in]{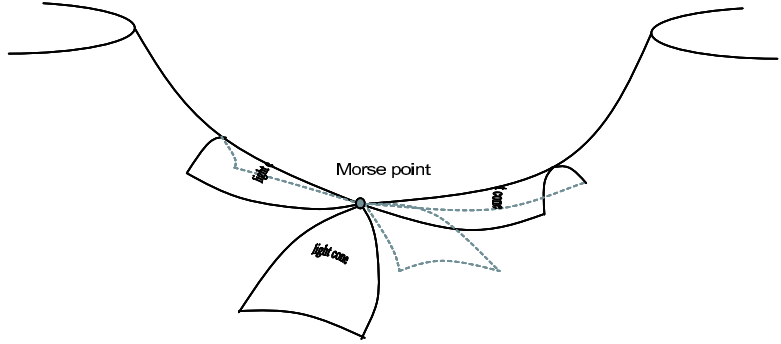}
\caption{Illustration of the double light cone causal structure around a Morse point.}
\label{fig:morsepoint}
\end{center}
\end{figure}

The reasoning in the previous section was based on a picture where the metrics are Wick rotated from Lorentzian to
Euclidean signature. Performing an inverse Wick rotation is not so easy however for geometries with baby universes,
since these geometries do not
admit Lorentzian metric everywhere. The reason for this
comes from the fact that it is not possible to find
a non vanishing vector field everywhere. It is however possible
to find a vector field everywhere except for a finite number of points.
For some values of the time parameter the universe develops a baby universe and the topology of a spatial universe
splits up into two different components. Precisely at
the moment of a split the spatial geometry has the topology of a figure eight and we see that the central point of the
figure eight is a saddle point if one views the geometry as being embedded in $\mathbb{R}^3$.
This point is called a Morse point in the
mathematically oriented literature \cite{Dowker:2002hm}.
Contrary to a generic point on the manifold, such a Morse point does not possess a unique
timelike vector field perpendicular to its spatial slice. In fact it has two future directed normal timelike vectors,
showing that the neighborhood of such a point has an anomalous causal structure. More precisely, such a point has
two future, and two past light cones which we refer to as a double light cone structure \cite{Dowker:2002hm}.
Consequently,
if one universe splits in two, each of the resulting universes carries a light cone belonging to the Morse point as
is illustrated in fig.~\ref{fig:morsepoint}.
Conversely, the universe carries the two past light cones of the Morse point before the split.

Because of the peculiar causal structure around the Morse points we are not quite sure whether the usual definition of the
Wick rotation in CDT is also legitimate here. There are some indications that the Einstein Hilbert action develops complex
valued singularities at these points \cite{Louko:1995jw}. Although the Wick rotation at these points presents one with
an interesting conundrum, we confine ourselves to the Euclidean sector for the remainder of this thesis.

\section{Dynamics to all orders in the coupling}\label{sec:Dynamics to all orders in the coupling}

In this section we discuss the explicit construction of the model with a coupling constant for the spatial topology
changes and show that it can be solved to all orders in the coupling constant in a suitable continuum limit.

Our starting point is equation \rf{eq:onsteppropagatordressed} with the addition of a coupling to the interaction term,
where in section \rf{sec:Introducing the coupling constant} we argued
that the coupling can naturally be denoted as $g_s$,
\beq
G_\l(l_1,l_2;1)  = G_{\l}^{(b) }(l_1,l_2;1) + 2g_s \sum_{l=1}^{l_1-1} l_1 w_{\l,g_s}(l_1\mi l) \,G_\l^{(b) }(l,l_2;1).
\eeq
After a discrete Laplace transform one obtains
\bea
{G(x,y;g;t)  =}~~~~~~~~~~~~~~~~~~~~~~~~~~~~~~~~~~~~~~~~~~~~~~~~~~~~~~~
~~~~~~~~~~~~~~~~~~~~~~
 & &  \nonumber \\
\ointz  \left[G^{(b) }(x,z^{-1};g;1) \pl ~2g_s~x \frac{\prt}{\prt x}
\Bigl( w_{g_s}(x;g)  G^{(b) }(x,z^{-1};g;1) \Bigr)  \right]  G(z,y;g;t \mi 1).
&&\nonumber  \label{eq:Gxygtdressedcomposition}\\
\!\!\!\!\!\!\!\!
\eea
If the baby universes are to survive the continuum limit the coupling constant in a controlled manner we have to scale $g_S$
with the lattice cutoff $a$. To find the appropriate scaling for the coupling we implement the following ansatz
\beq \label{eq:scalinggstr}
g_s = a^{\xi}g_S + \mbox{less singular terms.}
\eeq
Using the explicit form of the transfer matrix $G_\l^{(b) }(x,z^{-1};1) $ and inserting the scaling relations
\rf{eq:scalingrelations}, \rf{eq:scalinggstr} into
expression \rf{eq:Gxygtdressedcomposition} one is led to the following continuum equation,
\bea
a^\ep\frac{\prt}{\prt T}\, G_\La(X,Y;T) & =& -a \, \frac{\prt}{\prt X} \Bigl[ (X^2-\La)  G_\La(X,Y;T) \Bigr] \nn
&&- 2g_S \: a^{\eta+\xi-1}\frac{\prt}{\prt X} \Bigl[\cW_{\La,g_S}(X)  G_\La(X,Y;T) \Bigr].
\label{eq:semicontinuumdifequation}
\eea
If we demand that the first term on the right hand side of eq. \rf{eq:semicontinuumdifequation}
survives the continuum limit
we need $\eps =1$ as in the bare CDT model.
Contrary to \rf{top13} we have the additional freedom to scale the coupling constant $g_S$.
This enables us to adjust the scaling so that
also the second term survives the continuum limit yielding $\eta+\xi=2$. Inserting this relation in
\rf{eq:semicontinuumdifequation} we obtain
the desired dynamical equation that contains both the dynamical term for propagation of single strings and a term governing
string interactions,
\beq \label{eq:difequationGXYTdressed}
\frac{\partial}{\partial T}G^{(1;0) }(X,Y,T) =-\frac{\partial}{\partial X}\left[\left((X^2-\La) + 2 g_S \cW_{\La,g_S}(X) \right) G^{(1;0) }(X,Y,T) \right].
\eeq
We observe that this equation is of the same form as the equations for the CDT models without a coupling constant
\rf{eq:differentialequationGXYT},
\beq
\frac{\partial}{\partial T}G^{(1;0) }(X,Y;T) =-\frac{\partial}{\partial X}\left[\hcW_{\La,g_S}(X)  G^{(1;0) }(X,Y;T) \right],
\eeq
but with a different form of $\hcW(X) $,
\beq \label{eq:hcWintermsofW}
\hcW_{\La,g_S}(X) = (X^2-\La) \;+\;2 g_S\:\cW_{\La,g_S}(X).
\eeq
Note that requiring both terms in \rf{eq:difequationGXYTdressed} to survive the scaling limit does not fix
the scaling uniquely.
However, since we merely introduce a coupling constant to the baby universe model and did not introduce any new
configurations in the path integral, the disc function should still satisfy \rf{an1}.
As before we can use this relation to
further constrain the scaling. The dynamics of our new model requires $\ep=1$, so we are led to the conclusion that the
model falls into the first of the two scaling classes defined in section \rf{sec:Euclidean results with causal methods}
implying $\eta<0$. Consequently, the relation
between the continuum disc function and the continuum propagator is the same as for the bare CDT model,
\beq \label{eq:contmarkinglawW}
\frac{\partial\cW_{\La,g_S}(X) }{\partial \La} = \int dT \int d Z G^{(1;0) }(X,-Z;T) \frac{\partial\cW_{\La,g_S}}{\partial Z}.
\eeq
It seems that we are still left with a predicament since we are unable to specify the exact scaling for the disc function.
In fact it is only an illusory problem, since the disc function only appears in the dynamical equations in combination
with the coupling constant as $g_S \cW$. Therefore, we in principle only need to determine the scaling of this
combination which means that the relevant part of the scaling is already determined by the dynamical equation itself.
In the model we consider however, the scaling is completely fixed by requiring the
disc function to reduce to the disc function of the bare CDT model for zero $g_S$, implying
\beq
w_{\l,g_s}(x) = a^{-1}\cW_{\La,g_S}(X)  , \quad g_s = a^{3}g_S.
\eeq
In the remainder of this section \rf{eq:contmarkinglawW} and \rf{eq:difequationGXYTdressed}
are used to derive a differential equation for $\hcW_{\La,g_S}(X) $.
The equation will be rather implicit however, since the equation does not only depend on $\hcW_{\La,g_S}(X)$, but it also
depends explicitly on the solution of the equation $\hcW_{\La,g_S}(X) =0$. Remarkably, one is
able to solve the equation uniquely provided the disc functions $\cW_{\La,g_S}(L) $ satisfy the natural physical
requirement that they fall of at infinity. As a first step we solve the dynamical equation
\rf{eq:difequationGXYTdressed} to obtain
the propagator in terms of $\hcW_{\La,g_S}(X)$,
\beq
G^{(1;0) }(X,Y,T) =\frac{\hcW_{\La,g_S}(\bX(T) ) }{\hcW_{\La,g_S}(X) }\frac{1}{\bX(T) +Y}.
\eeq
Inserting this into the consistency condition \rf{eq:contmarkinglawW} and performing the integration over $Z$ we obtain
\beq \label{eq:ddlaWisintdT}
\frac{\partial\cW_{\La,g_S}(X) }{\partial \La} = -\int d T \frac{\hcW_{\La,g_S}(\bX(T) ) }{\hcW_{\La,g_S}(X) }\frac{\partial\cW_{\La,g_S}}{\partial \bX(T) }.
\eeq
To evaluate the integral we conveniently use the characteristic equation and convert the integral over time into an integral
over $\bX$,
\beq \label{eq:characteristicequationhcW}
\frac{d\bX}{dT}=-\hcW_{\La,g_S}(\bX(T)).
\eeq
%
%
%
Applying the characteristic equation to \rf{eq:ddlaWisintdT} we obtain
\beq
\frac{\partial\cW_{\La,g_S}(X) }{\partial \La} =\frac{1}{\hcW_{\La,g_S}(X) }
\int\limits^{\bX_{\infty} }_X d\bX \frac{\partial\cW_{\La,g_S}}{\partial \bX},
\eeq
where one can easily evaluate the integral since the integrand is a total derivative,
\beq \frac{\partial\cW_{\La,g_S}(X) }{\partial \La}=\frac{\cW_{\La,g_S}(X)-\cW_{\La,g_S}(\bX_{\infty}) }{\hcW_{\La,g_S}(X) }. \eeq
This equation can be rewritten as an equation for $\hcW_{\La,g_S}(X) $ by reexpressing the disc functions as
\beq
\cW_{\La,g_S}(X) = -\frac{\hcW_{\La,g_S}(X) -(X^2-\La)  }{2 g_S},
\eeq
giving
\beq
\frac{\partial\hcW_{\La,g_S}(X) }{\partial\La}=\frac{\bX_{\infty}^2-X^2}{\hcW_{\La,g_S}(X) },
\eeq
or equivalently,
\beq \label{eq:difequationhcW}
\frac{\partial\hcW_{\La,g_S}(X) ^2}{\partial \La}= 2(\bX_{\infty}^2-X^2).
\eeq
This is the defining equation for $\hcW_{\La,g_S}(X) $ that we intended to derive. The simple form of the equation is deceptive
as $\bX_{\infty}$ is in fact a completely unknown function of $g_S$ and $\La$, the only thing we know is that it is defined as
the solution of $\hcW_{\La,g_S}(X) =0$. Another unknown arises when we try to solve \rf{eq:difequationhcW}
by integrating both sides with
respect to $\La$, the integration ``constant'', $C(g_S,X) $, can in principle be a general function of $g_S$ and $X$. So we
conclude that \rf{eq:difequationhcW} only determines $\hcW_{\La,g_S}(X) $ up to two functions.
Explicitly, the integral of \rf{eq:difequationhcW} can be written as follows,
\beq \label{eq:hcWsqintermsoffandg}
\hcW_{\La,g_S}(X)^2=\La^2 f\left(\textstyle{\frac{g_S}{\sqrt{\La}^3}}\right) -2 X^2\La + X^4h\left(\textstyle{\frac{g_S}{X^3}}\right),
\eeq
where we have used dimensional analysis to parameterize the integration constant and the integral of $\bX_{\infty}$ by two
dimensionless functions,
\beq
\La^2 f\left(\textstyle{\frac{g_S}{\sqrt{\La}^3}}\right)  = 2\int^{\La} d\La' \bX_{\infty}\left(\textstyle{g_S,\sqrt{\La'}^3}\right),
\quad C(g_S,X) = X^4h\left(\textstyle{\frac{g_S}{X^3}}\right).
\eeq
Below we use the following series expansions
\beq
f \left(\textstyle{\frac{g_S}{\sqrt{\La}^3}}\right) =
\sum\limits_{n=0}^{\infty}f_n\left(\textstyle{\frac{g_S}{\sqrt{\La}^3}}\right) ^n,
\quad h \left(\textstyle{\frac{g_S}{X^3}}\right) =
\sum\limits_{n=0}^{\infty}h_n\left(\textstyle{\frac{g_S}{X^3}}\right) ^n.
\eeq
%
%
%
We show that the lowest order coefficients of the expansions are determined by requiring consistency
with the bare CDT model. Amazingly, \emph{all} higher order coefficients are
uniquely determined by demanding that the inverse Laplace transform of the disc function $\cW(X) $ falls of at infinity
at arbitrary order in the $g_S$ expansion. From the definition of $\hcW(X) $ in the pure model \rf{eq:defhcWpure}
one readily sees that to lowest order in $g_S$ we need to have
\beq
\hcW_{\La,g_S}(X)^2=X^4 -2X^2\La +\La^2 + \cO(g_S),
\eeq
which implies $h_0=1,f_0 =1$ when comparing to \rf{eq:hcWsqintermsoffandg}.
Using this result we write the disc function as follows,
\beq \label{cWtaylor}
\cW_{\La,g_S}(X) =\frac{-(X^2-\La) +(X^2-\La) \sqrt{1+\frac{\sum\limits_{n=1}^{\infty}g_S^n\left(h_n X^{4-3n}+f_n \sqrt{\La}^{4-3n}\right) }{(X^2-\La) ^2}}}{2g_S}.
\eeq
If we expand this expression to lowest order in $g_S$ we obtain
\beq \label{cWlowestorder}
\cW_{\La,g_S}(X) =\frac{1}{4}\frac{h_1 X +f_1\sqrt{\La}}{X^2-\La}+\cO(g_S).
\eeq
By equating this result to the marked disc function of the bare model
\beq \label{cWpure}
\cW_{\La,0}(X) = \frac{1}{X+\sqrt{\La}},
\eeq
we conclude that $h_1=4,f_1=-4$.
To obtain the higher order coefficients we proceed by analyzing the $g_S$ expansion of the disc function order by order.
The first order correction to the disc function is obtained by inserting $h_1=4,f_1=-4$ into \rf{cWtaylor} and expanding
to first order in the coupling,
\beq
\cW_{\La,g_S}(X) =\textstyle{\frac{1}{X+\sqrt{\La}}+g_S\left(-\frac{1}{(X^2-\La) (X+\sqrt{\La}) ^2}+\frac{h_2
(4X) ^{-2}+f_2(4\sqrt{\La}) ^{-2}}{X^2-\La}\right) +\cO(g_S^2)}.
\eeq
After an inverse Laplace transform we obtain
\beq
\sqrt{\La}^3\cW_{\La,1}(L) =\frac{1}{8} e^{L \sqrt{\Lambda }}
(f_2+h_2-1) -\frac{h_2 L}{4} +\frac{1}{8} e^{-L\sqrt{\Lambda}}
\left(2 \Lambda  L^2+2 \sqrt{\Lambda } L-f_2-h_2+1\right),
\eeq
where we have introduced the following notation
\beq
\cW_{\La,g_S}(X) =\sum\limits_{n=0}^{\infty}\cW_{\La,n}(X) g_S^n.
\eeq
Demanding the disc function to fall of at infinity implies that the terms proportional to
$L$ and $e^{L \sqrt{\Lambda}}$ must vanish, leading to $h_2=0,f_2=1$. So the result
for the disc function at first order in the coupling reads as follows,
\beq \label{cWpowergsone}
\cW_{\La,1}(L) =\frac{e^{-L \sqrt{\Lambda }} L \left(\sqrt{\Lambda} L+1\right) }{4 \Lambda},
\eeq
which can be confirmed by explicitly computing the ``Feynman diagram'' where the spatial universe
is allowed to split once. Obtaining the higher order coefficients is a bit messy, but the iterative procedure
to compute them is completely analogous to the calculation for $h_2$ and $f_2$. At each order of the $g_S$ expansion
the disc function has the same form as \rf{cWpowergsone}. Every $\cW_{\La,n}(L) $ contains three terms,
a polynomial term in $L$, a term proportional to $e^{\sqrt{\La}L}$ and a term proportional to $e^{-\sqrt{\La}L}$.
Demanding the disc function to be bounded at infinity implies that the polynomial and $ \cO (e^{\sqrt{\La}L})$
terms should vanish.
If one additionally uses the known results for $h_{n-1}$ and $f_{n-1}$ one obtains the $h_n$ and $f_n$ coefficients
uniquely. It turns out that boundedness of the disc function is such a stringent constraint that we need
$h_{n}=0$ for all $n\geq2$. Inserting the nonzero coefficients of $h\left(\textstyle{\frac{g_S}{X^3}}\right) $
into \rf{eq:hcWsqintermsoffandg} we obtain
\beq
\hcW_{\La,g_S}(X) ^2=X^4 -2X^2\La + 4 g_S X + F(g_S,\La),
\eeq
where
\beq
F(g_S,\La) = \La^2 f(q),\quad q=\textstyle{\frac{g_S}{\sqrt{\La}^3}}.
\eeq
As noted above, demanding the disc function to be bounded at infinity also fixes all $f_n$.
Contrary to the $h_n$ coefficients however, the $f_n$ coefficients are rather nontrivial but can be computed by the
algorithm sketched above.
Employing a symbolic computer program such as Mathematica one can readily compute the first coefficients ($\sim 20$).
Making use of Sloan's database of integer sequences it is possible to find a closed analytical expression
for the coefficients $f_n$ in terms of Euler gamma functions,
\beq
f_0 = 1,\quad f_n=\frac{\Gamma(\frac{3}{2}n-2) }{\Gamma(\frac{1}{2}n+1) \Gamma(n) },\quad n\geq1.
\eeq
Given these coefficients we can sum the Taylor expansion and obtain the full non perturbative result for $f\left(q\right) $
\beq
f(q) = {\textstyle \frac{2}{3}+\frac{1}{3}}\, _2F_1\left({\textstyle-\frac{1}{3},-\frac{2}{3};\frac{1}{2}; \frac{2}{3 \sqrt{3}}}q\right) -
{\textstyle 4_2F_1}\left({\textstyle-\frac{1}{6},\frac{1}{6};\frac{3}{2}; \frac{2}{3 \sqrt{3}}}q \right).
\eeq
This means that we have solved our model and we can present the disc function with
a non perturbative sum over spatial topologies,
\beq \label{eq:cWallorders}
\cW_{\La,g_S}(X) =\frac{-(X^2-\La) + \sqrt{X^4 -2X^2\La + 4 q \sqrt{\La}^3  X + \La^2 f\left(q\right) }}{2g_S}.
\eeq
%
%
%

%
%
%
%
%
%
%

\section{Relation to random trees} \label{sec:Relation to random trees}

In this section we give an alternative derivation of the disc function dressed with spatial topology changes
\rf{eq:cWallorders}
to show robustness of the result and to give more insight into the details of the quantum geometry. Particularly,
the derivation we present below highlights the random tree structure of the configurations. To make the connection as clear
as possible we start by deriving the one point function for a random tree model. Sometimes this one point function for
random trees is referred to as a partition function for rooted branched polymers. The random tree model we
consider is a statistical
model consisting of edges that are weighted with a fugacity $z$ and three-valent vertices with a coupling constant
$\lambda$. A convenient way to view the partition function $w(z,g) $ is that it is the generating function for the number of
random trees,
\beq
w(z,\lambda)  = z \sum \limits_{m=0}^{\infty} w_{2m} \left(\lambda z^2\right) ^m,
\eeq
where we use the fact that every vertex except the initial, or marked, vertex is accompanied by two edges.
An easy way to evaluate the partition function $w(z,\lambda) $ is to notice that it should solve the following equation
\beq \label{branchedpolymereq}
w(z,\lambda)  = z + \lambda z w(z,\lambda) ^2.
\eeq
\begin{figure}[t]
\begin{center}
\includegraphics[width=3in]{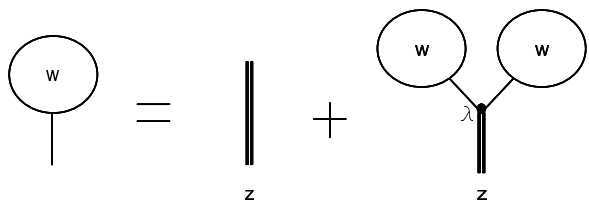}
\caption{Pictorial representation of the iterative equation for the one-point function $w(\lambda,z) $ for branched polymers.}
\label{fig:branchedpolymer}
\end{center}
\end{figure}

\begin{figure}[t]
\begin{center}
\includegraphics[width=3in]{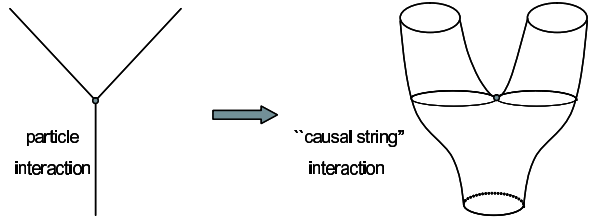}
\caption{Illustration of spatial topology change as a string like generalization of a particle interaction.}
\label{fig:stringinteraction}
\end{center}
\end{figure}

The interpretation of this equation is illustrated in fig.~\ref{fig:branchedpolymer}.
Equation \rf{branchedpolymereq} has two solutions but only
one solution is compatible with the initial condition that there is only one tree with one edge, i.e.~$w_{2m} = 1$.
This solution is given by
\beq
w(z,\lambda)  = \frac{1-\sqrt{1-4 \lambda z^2}}{2 \lambda z},
\eeq
which has the following series expansion,
\beq
w(z,\lambda)  = z \sum \limits_{m=0}^{\infty} \frac{(2m) !}{m!(m+1) !} \left(\lambda z^2\right) ^m.
\eeq
So we see that the number of random trees with $m$ edges, $w_{2m}$, is given by the $m^{th}$ Catalan number.
This random tree model can be regarded as a simple model possessing some features of $\phi^3$ theory. Indeed
it is possible to formulate scalar $\phi^3$ theory in $\mathbb{R}^d$ as a sum over connected diagrams in the same
way as the random tree model. One merely needs to replace the fugacity of the edges $z$ by the standard scalar propagators.
Here we view the random trees not as a model for particles interacting in an ambient space, but as a simple model
for a theory of ``interacting universes'' in two dimensional quantum gravity. The idea is similar to the basic
idea behind string theory, we ``blow up'' the Feynman diagrams by replacing the propagators of an interacting field
theory by the propagators of a string theory as is illustrated in fig.~\ref{fig:stringinteraction}.
In our case this means that we replace
the rather trivial propagators of the branched polymer model $z$ by propagators that we computed from CDT \rf{eq:G10XYT}.
We remind the reader that the resulting model can either be viewed as a toy model for quantum gravity with
topology change or as a string theory without a target space.
To make the analogy as close as possible we write the equation for the generating function of the branched polymer as follows,

\beq \label{eq:branchedpolymereqwG}
w(z,\lambda)  = w_0(z)  + \lambda G_0(z)  w(z,\lambda) ^2,
\eeq
where $w_0(z) $ and $G_0(z) $ denote the one point function and the two point function respectively, to zeroth order
in the coupling $\lambda$. From \rf{branchedpolymereq} we observe that the expressions for these objects coincide
$w_0(z)=G_0(z)=z$. This is an understandable
coincidence, since the branched polymer is a very simple model. The
equation analogous to \rf{eq:branchedpolymereqwG} for our model of interacting spatial universes based on CDT is
\beq \label{eq:iterationequationbabyuniverses}
\cW_{\La,g_S}(L) = \cW_{\La,0}(L)  + g_S \int d T dL_1 dL_2 G^{(1,1) }_{\La,0}(L,L_1+L_2;T) \cW_{\La,g_S}(L_1) \cW_{\La,g_S}(L_2).
\eeq
This is the defining equation for the disc function $\cW(L)$.
It is similar to the way equation \rf{eq:branchedpolymereqwG} defines
the one point function of the branched polymers $w(z,\lambda)$,
since the disc function is the CDT analogue of the branched polymer one point function $w(z,\lambda) $.
Instead of solving \rf{eq:iterationequationbabyuniverses} we work with its Laplace transformed analogue,
\beq \label{eq:iterationequationbabyuniversesXY}
\cW_{\La,g_S}(X) = \cW_{\La,0}(X)  + g_S \int dZ_1 dZ_2 G^{(1,1}_{\La,0}(X;-Z_1,-Z_2) \cW_{\La,g_S}(Z_1) \cW_{\La,g_S}(Z_2),
\eeq
where the propagator is given by
\beq
G^{(1,1) }(X;Y_1,Y_2;T) =
\frac{\hcW_{\La,0}(\bX) }{\hcW_{\La,0}(X) }\left[\frac{1}{(\bX+Y_1) ^2(\bX+Y_2) }+\frac{1}{(\bX+Y_1) (\bX+Y_2) ^2}\right].
\eeq
Performing the integrations over $Z_1$ and $Z_2$ in \rf{eq:iterationequationbabyuniversesXY} gives
\beq
\cW_{\La,g_S}(X) = \cW_{\La,0}(X)  + g_S \int dT  \frac{\hcW_{\La,0}(\bX(T) ) }{\hcW_{\La,0}(X) }\frac{\partial \cW_{\La,g_S}(\bX) ^2}{\partial \bX}.
\eeq
As in \rf{eq:ddlaWisintdT} it is convenient to use the characteristic equation to convert the
integral over $T$ into an integral over $\bX(X,T) $,
\beq
\cW_{\La,g_S}(X) = \cW_{\La,0}(X)  + \frac{g_S}{\hcW_{\La,0}(X) } \int\limits_X^{\bX_{\infty}}d\bX \frac{\partial \cW_{\La,g_S}(\bX) ^2}{\partial \bX}.
\eeq
Seeing that the integrand is a total derivative one easily obtains
\beq
\cW_{\La,g_S}(X) = \cW_{\La,0}(X)  + g_S \frac{\cW_{\La,g_S}(\bX_{\infty}) ^2-\cW_{\La,g_S}(X) ^2}{\hcW_{\La,0}(X)}.
\eeq
Since this is a second order polynomial equation for $\cW(X) $ it is readily solved and we obtain,
\beq
 \cW_{\La,g_S}(X) =\frac{-\hcW_{\La,0}(X) + \hcW_{\La,g_S}(X) }{2 g_S},
\eeq
where
\beq \label{eq:hcWxbinfty}
 \hcW_{\La,g_S}(X) =\sqrt{\hcW_{\La,0}(X) ^2+4 g_S\left(\hcW_{\La,0}(X) \cW_{\La,0}(X) + g_S \cW_{\La,g_S}(\bX_{\infty}) ^2\,\right)}.
\eeq
Note that this equation was derived independently of the particular form of $\cW_{\La,0}(X) $ and $\hcW_{\La,0}(X) $. Currently, however
we are interested in an interacting model based on CDT. So we require the disc function and the propagator to reduce
to the results obtained in the bare CDT theory,
\beq
\cW_{\La,0}(X)  = \frac{1}{X+\sqrt{\La}},\quad \hcW_{\La,0}(X)  = X^2-\La,\quad \bX_{\infty} = \sqrt{\La}.
\eeq
Inserting into \rf{eq:hcWxbinfty} gives,
\beq
\hcW_{\La,g_S}(X) ^2=(X^2-\La) ^2+4 g_S\left((X-\sqrt{\La}) + g_S \cW_{\La,g_S}(\sqrt{\La}) ^2\,\right),
\eeq
which can be conveniently written as
\beq \label{eq:hcWsq}
\hcW_{\La,g_S}(X) ^2=X^4 -2X^2\La + 4 g_S  X + F(g_S,\La),
\eeq
where
\beq
F(g_S,\La) = \La^2-4g_S\sqrt{\La}+4g_S^2\cW_{\La,g_S}(\sqrt{\La}) ^2.
\eeq
Given \rf{eq:hcWsq} and \rf{eq:hcWintermsofW} we can obtain the following result for the disc function,
\beq
\cW_{\La,g_S}(X) =\frac{-(X^2-\La) + \sqrt{X^4 -2X^2\La + 4 g_S  X + F(g_S,\La) }}{2g_S},
\eeq
which is exactly of the same form as derived in the previous section \rf{sec:Dynamics to all orders in the coupling}
but so far we have not yet
determined the precise form of $F(g_S,\La) $. To abbreviate the notation,
we can convert the equation to dimensionless units by
dividing all dimensional quantities by appropriate powers of the cosmological constant
\beq
x = \tfrac{X}{\sqrt{\La}},~~Y=\tfrac{Y}{\sqrt{\La}},~~ q = \tfrac{g_S}{\sqrt{\La}^3},
\eeq
and obtain
\beq
\omega_q(x) =\frac{-(x^2-1) +\hat{\omega}_q(x) }{2q},
\eeq
with
\beq \label{eq:omegaxfq}
\hat{\omega}_q(x) =\sqrt{x^4 -2x^2 + 4 q x  + f(q)}.
\eeq
In the derivation of section \rf{sec:Dynamics to all orders in the coupling} the disc function is expanded
in powers of $q$ and the Taylor coefficients of $F(q)$ are uniquely determined by demanding that the
disc functions decay at infinity at every order of the expansion. Below we take a different route to obtain
the $f(q) $ by again using the characteristic equation in an essential way. Surprisingly, we find $f(q)$
in a form that appears very different from the results of section \rf{sec:Dynamics to all orders in the coupling},
but is in fact exactly the same.
Let us recall that the characteristic equation \rf{eq:characteristicequationhcW}
can be used to define the time variable in terms of $\hat{\omega}_q(x)$ by
\beq
t = \int\limits_{\bx_{\infty} }^{x}\frac{d\bx}{\hat{\omega}(\bx)}.
\eeq
Notice that we can only integrate $t$ all the way to infinity if $\hat{\omega}_q(x)$ has a simple zero since $\bx_{\infty}$ is
defined as the solution of $\hat{\omega}_q(x)=0$. Together with \rf{eq:omegaxfq} it implies that $\hat{\omega}_q(x)=0$
should be of the following form,
\beq \label{eq:omegaxccc}
\hat{\omega}_q(x) =(x-c) \sqrt{(x+c_{+}) (x+c_{-}) },
\eeq
where $c,c_{+}$ and $c_{-}$ are all functions of $q$ that we determine below by taking the square of \rf{eq:omegaxccc}
and equating
it with the square of \rf{eq:omegaxfq}. This gives us four equations for the four unknown functions
$c(q),c_{+}(q),c_{-}(q) $ and $f(q) $,
one equation for each power of $x$, enabling one to solve the system completely.
If one solves the equation belonging to $x^3$
one can eliminate one function and we can write
\beq \label{eq:omegaxcu}
\hat{\omega}_q(x) =(x-c) ^2(x+c+\sqrt{u}) (x+c-\sqrt{u}),
\eeq
where $c_{+}= c + \sqrt{u}$ and  $c_{-}= c-\sqrt{u}$.
If we now expand equation \rf{eq:omegaxcu} in powers of $x$ and than equate it to the square of \rf{eq:omegaxfq}
we obtain
\beq
x^4 - (2c^2+u)  x^2+ 2cu x + c^2(c^2-u) =x^4 - 2x^2 + 4 q x  + f.
\eeq
From this we extract three equations, two simple relations expressing
$f$ and $u$ in terms of $c$,
\beq \label{eq:fintermsofc}
f = c^2(3c^2-2),
\eeq
\beq \label{eq:uintermsofc}
u = 2 - 2c^2,
\eeq
and a third order polynomial equation for $c$
\beq \label{eq:thirorderequationcq}
c^3-c = -q.
\eeq
Inserting the solution of the polynomial equation for $c$ in \rf{eq:fintermsofc} gives $f(q) $ which together
with \rf{eq:omegaxfq} allows us to find the complete solution for $\hat{\omega}(x) $.
The solution of \rf{eq:thirorderequationcq} is most conveniently expressed in terms of $\tilde{q}= \frac{2}{3\sqrt{3}}\,q$
as follows,
\beq \label{eq:cintermsofz}
\sqrt{3}\,c(\tilde{q}) =z_q+\frac{1}{z_q},
\eeq
where
\beq \label{eq:zintermsofq}
z_q=\left(-\tilde{q}+\sqrt{\tilde{q}^2-1}\right) ^{\frac{1}{3}}.
\eeq
Combining \rf{eq:cintermsofz}, \rf{eq:zintermsofq} and \rf{eq:fintermsofc} we obtain,
\beq
f(q) =\frac{1}{3}\left( z_q^4 + 2 z_q^2 + 2 + \frac{2}{z_q^2} +\frac{1}{z_q^4}\right),
\eeq
which appears to be very different from the expression found in the previous section,
\beq
f(q) = {\textstyle \frac{2}{3}+\frac{1}{3}}\, _2F_1\left({\textstyle-\frac{1}{3},-\frac{2}{3};\frac{1}{2}; \frac{2}{3 \sqrt{3}}}q\right) -
{\textstyle 4_2F_1}\left({\textstyle-\frac{1}{6},\frac{1}{6};\frac{3}{2}; \frac{2}{3 \sqrt{3}}}q \right),
\eeq
but when one compares the Taylor expansion of both expressions we see that they are fully equivalent.

\section{Summary}

To set the stage for our model we reviewed the known relation between Euclidean and causal quantum gravity
defined by dynamical triangulations in section \rf{sec:Euclidean results with causal methods}.
Some results of Euclidean quantum gravity can be derived by generalizing
the formalism of causal dynamical triangulations to allow for spatial topology change. No ``energy penalty'' is
associated with these topological fluctuations, manifesting itself by the fact that infinitesimal
baby universes dominate the path integral. The fractal nature of the quantum geometry is reflected by the
non canonical values for both the ``time'' variable and the Hausdorff dimension.

In section \rf{sec:Introducing the coupling constant} we introduced a coupling constant for the spatial topology
changes. From the Einstein Hilbert action of an elementary manifold with a change of spatial topology,
the trouser geometry,
we argued the naturalness of such a coupling. Additionally, we examined the geometry around the point where
the spatial topology change occurs, the Morse point, and recalled that the causal structure around such a
point is non standard. Particularly, the Morse point features a doubling of the light cone structure, it
possesses two past and two future light cones.

In section \rf{sec:Dynamics to all orders in the coupling} we discussed the details of the construction
our new model of two dimensional quantum gravity with spatial topology change.
The discrete kinematical structure of the model is similar to the introductory
section \rf{sec:Euclidean results with causal methods}, the continuum behavior on the other hand
is completely different. The presence of the coupling allows us to define a continuum limit where
manifolds with all spatial topologies contribute but complicated topologies are suppressed by powers
of the coupling constant. Especially, we were able to derive the disc function to all orders in the coupling
constant and sum the series uniquely!

An alternative derivation of the disc function dressed with topology fluctuations was presented in
section \rf{sec:Relation to random trees}. We showed that the disc function of the model can be derived
from an iterative equation that is very similar to the generating function equation that defines the
one point function of rooted random trees, or equivalently branched polymers. Besides providing additional
insight into the structure of the quantum geometry of the model we also found that the hypergeometric functions
that appeared in the result of \rf{sec:Dynamics to all orders in the coupling} can be written in terms of a
solution of a third order polynomial equation.

\chapter{Hyperbolic space} \label{ch4}

In this chapter, which is based on \cite{Ambjorn:2006hu}, we go back to an analysis of pure CDT.
As in most approaches to quantum gravity, the CDT method was originally developed
for the quantization of compact manifolds. In the following however we generalize the boundary conditions
of two dimensional CDT so that the typical geometries in the path integral have an infinite volume.
Particularly, it is shown that given such boundary conditions a classical geometry with constant negative curvature
and superimposed quantum fluctuations emerges from the background independent path integral. Furthermore,
one can choose the boundary conditions such that the relative fluctuations become small in a concrete manner.
To the knowledge of the author this is one of the few cases where a semiclassical geometry emerges from a genuinely
background setup that can at the same time be studied by analytical methods. Another example were a semiclassical
background emerges dynamically from a background independent scheme is four dimensional quantum gravity defined
by causal dynamical triangulations. This model is too complicated to study with analytical methods, nevertheless
it is one of the most promising attempts to formulate a realistic theory of quantum gravity
(for a recent account see \cite{Ambjorn:2006jf}).

\section{Non compact manifolds}\label{sec:non compact manifolds}


As mentioned before, 2d quantum gravity is intimately related with the study of non-critical string theories.
The studies where the quantum gravity aspect has been emphasized mostly consider two dimensional Euclidean quantum
gravity with compact spacetime. The study of 2d Euclidean quantum
gravity with non-compact spacetime was initiated by the
Zamolodchikovs (ZZ)  \cite{Zamolodchikov:2001ah} when they showed how to use
conformal bootstrap and the cluster-decomposition properties to
quantize Liouville theory on the pseudo-sphere (the Poincar\'{e} disc).

Martinec \cite{Martinec:2003ka} and Seiberg et al.\ \cite{Seiberg:2003nm} showed how
the work of ZZ fitted into the framework of non-critical string
theory, where the ZZ-theory could be reinterpreted as special
branes, now called ZZ-branes. Let $W_\tL(\tX) $ be the ordinary
disc amplitude for $2d$ Euclidean gravity on a compact
spacetime. $\tX$ denotes the boundary cosmological constant of
the disc and $\tL$ the cosmological constant. It was found that
the ZZ-brane of 2d Euclidean gravity was associated with the zero
of

\beq\label{0.0}
W_\tL(\tX)  = (\tX -\oh \sqrt{\tL}) \sqrt{\tX + \sqrt{\tL}}.
\eeq

At first sight this is somewhat surprising since
from a world-sheet point of view the disc is compact while the
Poincar\'{e} disc is non-compact. In \cite{Ambjorn:2004my} and \cite{Ambjorn:2007iq} it was shown
how it could be understood in terms of world sheet geometry, i.e.~from a
2d quantum gravity point of view. When the boundary cosmological
constant $\tX$ reaches the value $\tX \equ \sqrt{\tL}/2$ where the
disc amplitude $W_\tL(\tX)  \equ 0$, the geodesic distance from a
generic point on the disc to the boundary diverges, in this way
effectively creating a non-compact spacetime.

Here we show that the same phenomenon occurs in two dimensional
quantum gravity from causal dynamical triangulations.

\section{The hyperbolic plane from CDT}

Recall that the propagator with one mark on the initial boundary is given by
\beq\label{2.a3}
G_\La (X,Y;T)  = \frac{\bX^2(T,X) -\La}{X^2-\La} \; \frac{1}{\bX(T,X) +Y},
\eeq
where $\bX(T,X) $ is the solution of the characteristic equation
\beq\label{2.3}
\frac{\d \bX}{\d T} = -(\bX^2-\La),~~~\bX(0,X) =X,
\eeq
giving
\beq\label{2.4}
\bX(t,X) = \SL \coth \SL(t+t_0),~~~~X=\SL \coth \SL \,t_0.
\eeq
Note that although different in appearance this expression is equivalent to \rf{eq:xbar}.
Viewing $G_\La(X,Y;T) $ as a propagator with X and Y as coupling constants,
$\bX(T) $ can be viewed as a ``running'' boundary cosmological
constant, $T$ being the scale. If $X > - \SL$ then
$\bX(T)  \to \SL$ for $T \to \infty$, $\SL$ being a ``fixed point''
(a zero of the ``$\b$-function'' $-(\bX^2-\La) $ in eq.\ \rf{2.3}).

Let $L_1$ denote the length of the entry boundary and $L_2$ the
length of the exit boundary. Rather than consider a situation
where the boundary cosmological constant $X$ is fixed we can
consider $L_1$ as fixed. We denote the corresponding propagator
$G_\La (L_1,Y;T) $. Similarly we can define $G_\La(X,L_2;T) $
and $G_\La(L_1,L_2;T) $. They are related by Laplace transformations.
For instance:
\beq\label{2.a5}
G_\La(X,Y;T) = \int_0^\infty \d L_2 \int_0^\infty \d L_1\;
G(L_1,L_2;T)  \;\e^{-XL_1-YL_2},
\eeq
and one has the following composition rule for the propagator:
\beq\label{2.a6}
G_\La (X,Y;T_1+T_2)  = \int_0^\infty \d L \;
G_\La (X,L;T_1) \,G(L,Y,T_2).
\eeq

We can now calculate the expectation value of the length of the
spatial slice at  proper time $t \in [0,T]$:

\beq\label{2.a7}
\la L(t) \ra_{X,Y,T} = \frac{1}{G_\La (X,Y;T) } \int_0^\infty \d L\;G_\La (X,L;t)  \;L\; G_\La (L,Y;T-t).
\eeq

In general there is no
reason to expect $\la L(t)  \ra$ to have a classical limit.
Consider for instance the situation where $X$ and $Y$ are larger
than $\SL$ and where $T \gg 1/\SL$. The average boundary lengths
will be of order $1/X$ and $1/Y$. But for $ 0 \ll t \ll T $ the
system has forgotten everything about the boundaries and the
expectation value of $L(t) $ is, up to corrections of order
$e^{-2\SL t}$ or $e^{-2\SL (T-t) }$, determined by the ground state
of the effective Hamiltonian $H_{ef\!f}$ corresponding to
$G_\La(X,Y;T) $. One finds for this ground
state $\la L \ra = 1/\SL$. This picture is confirmed by an
explicit calculation using eq.\ \rf{2.a7} as long as $X,Y > \SL$.
The system is thus, except for boundary effects, entirely
determined by the quantum fluctuations of the ground state of
$H_{ef\!f}$.

We will here be interested in a different and more interesting
situation where a non-compact spacetime is obtained as a limit of
the compact spacetime described by \rf{2.a7}. Thus we want to
take $T \to \infty$ and at the same time also the length of the
boundary corresponding to proper time $T$ to infinity. Since $T
\to \infty$ forces $\bX(T,X)  \to \SL$ it follows from \rf{2.a3}
that the only choice of boundary cosmological constant $Y$
independent of $T$, where the length  $\la L(T) \ra_{X,Y,T}$ goes
to infinity for $T \to \infty$ is $Y\equ \mi \SL$ (fig.~ \ref{fig:criticalboundary}), since we have:

\begin{figure}[t]
\begin{center}
\includegraphics[width=4in]{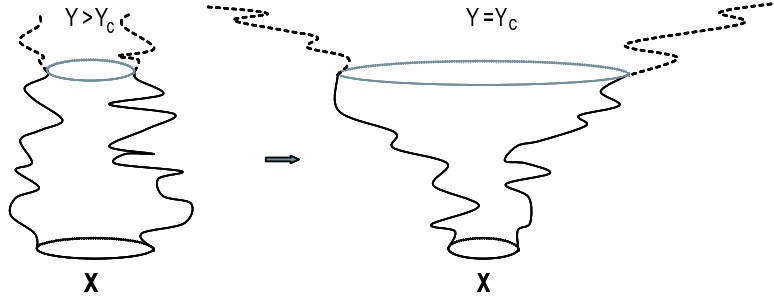}
\caption{For $Y = Y_c =-\sqrt{\La}$ the length of the final boundary diverges as $T \rightarrow \infty$.}
\label{fig:criticalboundary}
\end{center}
\end{figure}

\beq\label{2.a8}
\la L(T) \ra_{X,Y,T} = -\frac{1}{G_\La (X,Y;T) } \, \frac{\prt G_\La (X,Y;T) }{\prt Y} =  \frac{1}{\bX(T,X) +Y}.
\eeq

With the choice $Y \equ - \SL$ one obtains from \rf{2.a7} in the
limit $T \to \infty$:

\beq\label{2.a9}
\la L(t)  \ra_{X} = \frac{1}{\SL} \; \sinh \left(2\SL(t+t_0(X) ) \right),
\eeq

where $t_0(X) $ is defined in eq.\ \rf{2.4}.

We have called $L_2$ the (spatial) length of the boundary
corresponding to $T$ and $\la L(t)  \ra_X$ the spatial length of a
time-slice at time $t$ in order to be in accordance with earlier
notation \cite{Nakayama:1993we,Ambjorn:1998xu}, but starting from a lattice
regularization and taking the continuum limit $L$ is only
determined up to a constant of proportionality which we fix by
comparing with a continuum effective action. In section \rf{subsec:Hamiltonians in causal quantum gravity}
we showed that such a comparison leads to the identification
of $L$ as $L_{cont}/\pi$ and we are led to the following

\beq\label{2.a10}
L_{cont}(t)  \equiv \pi \la L(t) \ra_X = \frac{\pi}{\SL} \; \sinh \left(2\SL(t+t_0(X) )\right).
\eeq

Consider the
classical surface where the intrinsic geometry is defined by
proper time $t$ and spatial length  $L_{cont}(t) $ of the curve
corresponding to constant $t$. It has the line element

\beq\label{2.a11}
\d s^2 = \d t^2 + \frac{L_{cont}^2}{4\pi^2}\; \d \th^2 = \d t^2 + \frac{\sinh^2 \left(2\SL (t+t_0(X) ) \right) }{4 \La} \;\d \th^2,
\eeq

where $t \ge 0$ and $t_0(X) $ is a function of the
boundary cosmological constant $X$ at the boundary corresponding
to $t \equ 0$ (see eq.\ \rf{2.4}). What is remarkable about
formula \rf{2.a11} is that the surfaces for different boundary
cosmological constants $X$ can be viewed as part of the same
surface, the Poincar\'{e} disc with curvature $R= -8\La$, since $t$
can be continued to $t= -t_0 $. The Poincar\'{e} disc itself is
formally obtained in the limit $X \to \infty$ since an infinite
boundary cosmological constant will contract the boundary to a
point.

\section{The classical effective action}

In this section we make a small digression to the ``classical'' theory
and show that the emergence of the hyperbolic plane is natural from this point of view.
In section \rf{subsec:Hamiltonians in causal quantum gravity} we discussed the derivation
of the quantum Hamiltonian of causal quantum gravity from the following classical action.
\beq\label{3.4}
S_\k = \int_0^T \d t \left(\frac{\dot{l}^2(t) }{4l(t) }  +
\La l(t) + \frac{\k}{l}\right).
\eeq
To make contact with the inherently quantum calculation by causal dynamical triangulations
it is interesting to look at the classical behavior corresponding to this action.
The classical solutions corresponding to action \rf{3.4} are
\bea
l(t)  &=& \frac{\sqrt{\k}}{\SL} \; \sinh 2\SL t,~~~~~~~~~~\k>0~~ \mbox{elliptic case}, \label{3.6a}\\
l(t)  &=& \frac{\sqrt{-\k}}{\SL} \; \cosh 2\SL t,~~~~~~~\k<0~~\mbox{hyperbolic case}, \label{3.6b}\\
l(t)  &=& \e^{2\SL t},~~~~~~~~~~~~~~~~~~~~~~~\k=0 ~~\mbox{parabolic case}, \label{3.6c}
\eea
all corresponding to cylinders with constant negative curvature $-8 \La$.
In the elliptic case, where $t$ must be larger than zero, there is
a conical singularity at $t =0$ unless $\k = 1$. For $\k \equ 1$
the geometry is regular at $t=0$ and this value of $\k$
corresponds precisely to the Poincar\'{e} disc, $t=0$ being the
``center'' of the disc. So we see that for $\k \equ 1$ the classical solution
coincides nicely with the emergent geometry derived in the previous section

\section{Quantum fluctuations}

In many ways it is more natural to fix the boundary cosmological
constant than to fix the length of the boundary. However, one pays
the price that the fluctuations of the boundary size are large, in
fact of the order of the average length of the boundary itself
\footnote{This is true also in Liouville quantum theory, the
derivation is essentially the same as that given in \rf{5.1}, as is
clear from \cite{Ambjorn:2004my}.}: from \rf{2.a8} we have

\beq\label{5.1}
\la L^2(T)  \ra_{X,Y;T} - \la L(T)  \ra^2_{X,Y;T} = -\frac{\prt \la L(T)  \ra_{X,Y;T}}{\prt Y} =  \la L(T)  \ra^2_{X,Y;T}.
\eeq

Such large fluctuations are also present around $\la L(t) \ra_{X,Y;T}$
for $t< T$. From this point of view it is even more remarkable that
$\la L(t) \ra_{X,Y=- \SL;T=\infty}$ has such a nice semiclassical
interpretation. Let us now by hand fix the boundary lengths $L_1$
and $L_2$. This is done in the Hartle-Hawking Euclidean path
integral when the geometries $[g]$ are fixed at the boundaries
\cite{Hartle:1983ai}. For our one-dimensional boundaries the geometries at
the boundaries are uniquely fixed by specifying the lengths of the
boundaries, and the relation between the propagator with fixed
boundary cosmological constants and with fixed boundary lengths is
given by a Laplace transformation as shown in eq.\ \rf{2.a5}. Let
us for simplicity analyze the situation where we take the length
$L_1$ of the entrance loop to zero by taking the boundary
cosmological constant $X \to \infty$. Using the decomposition
property \rf{2.a6} one can calculate the connected ``loop-loop''
correlator for fixed $L_2$ and $0< t \leq t+\Del < T$,

\beq\label{5.a1}
\la L(t) L(t+\Del) \ra^{(c) }_{L_2,T} \equiv \la L(t+\Del) L(t) \ra_{L_2,T}-\la L(t) \ra \la L(t+\Del) \ra_{L_2,T}.
\eeq

One finds

\bea\label{5.a2}
\la L(t) L(t\pl\Del) \ra^{(c) }_{L_2,T} &=& \frac{2}{\La} \frac{\sinh^2 \SL t \sinh^2 \SL (T\mi (t\pl\Del) ) }{\sinh^2 \SL T}+
 \\
&& \frac{2L_2}{\SL} \frac{\sinh^2 \SL t \sinh\SL (t\pl\Del)  \sinh\SL(T \mi(t\pl\Del) ) }{\sinh^3 \SL T}.\no
\eea

We also note that

\beq\label{5.b2}
\la L(t) \ra_{L_2,T}= \frac{2}{\SL} \frac{\sinh \SL t \sinh \SL (T\mi t) }{\sinh \SL T}+ L_2\frac{\sinh^2 \SL t}{\sinh^2 \SL T}.
\eeq

For fixed $L_2$ and $T \to \infty$ we obtain

\beq\label{5.a3} \la
L(t) L(t+\Del) \ra^{(c) }_{L_2} = \frac{1}{2\La} \; \e^{-2\SL \Del } \left(1-\e^{-2\SL t} \right) ^2 \eeq and \beq\label{5.b3} \la L(t) \ra_{L_2}=\frac{1}{\SL}\left( 1-\e^{-2\SL t}\right).
\eeq

Eqs.\ \rf{5.a3} and \rf{5.b3}  tell us that except for small $t$
we have   $\la L(t) \ra_{L_2}\equ 1/\SL$. The quantum fluctuations
$\Del L(t) $ of $L(t) $ are defined by $(\Del L(t) ) ^2 = \la
L(t) L(t) \ra^{(c) }$. Thus the spatial extension of the universe is
just quantum size (i.e.~ $1/\SL$, $\La$ being the only coupling
constant)  with fluctuations $\Del L(t) $ of the same size. The time
correlation between $L(t) $ and $L(t+\Del) $ is also dictated by the
scale $1/\SL$, telling us that the correlation between spatial
elements of size $1/\SL$, separated in time by $\Del$ falls of
exponentially as $e^{-2\SL \Del}$ . The above picture is precisely
what one would expect from the classical action, which is proportional
to the area and the boundary cosmological constants only, if we force $T$ to
be large and choose a  $Y$ such that $\la L_2(T) \ra$ is not large,
the universe will be a thin tube, ``classically'' of zero width,
but due to quantum fluctuations of average width $1/\SL$.

A more interesting situation is obtained if we choose $Y = -\SL$,
the special value needed to obtain a non-compact geometry in the
limit $T\to \infty$. To implement this in a setting where $L_2$ is
not allowed to fluctuate we fix $L_2(T) $ to the average value
\rf{2.a8} for $Y\equ \mi \SL$:

\beq\label{5.2}
L_2(T)  = \la L(T)  \ra_{X,Y= -\SL;T} = \frac{1}{\SL} \; \frac{1}{\coth \SL T -1}.
\eeq

From \rf{5.a2} and \rf{5.b2} we have in the limit $T \to \infty$:

\beq\label{5.3}
\la L(t)  \ra = \frac{1}{\SL} \; \sinh 2\SL t,
\eeq

in accordance with \rf{2.a9}, and for the
``loop-loop''-correlator

\beq\label{5.4}
\la L(t+\Del) L(t) \ra^{(c) }= \frac{2}{\La}\;\sinh^2 \SL t= \frac{1}{\SL} \left( \la L(t) \ra -\frac{1}{\SL} \left(1-\e^{-2\SL t}\right) \right).
\eeq

It is seen that the ``loop-loop''-correlator is independent of
$\Del$. In particular we have for $\Del \equ 0$:

\beq\label{5.5}
(\Del L(t) ) ^2 \equiv \la L^2(t) \ra -\la L(t) \ra^2 \sim \frac{1}{\SL} \la L(t) \ra
\eeq

for  $t \gg 1/\SL$. The
interpretation of eq.\ \rf{5.5} is in accordance with the picture
presented below \rf{5.b3}: We can view the curve of length $L(t) $
as consisting of $N(t)  \approx \SL L(t)  \approx e^{2\SL t} $
independently fluctuating parts of size $1/\SL$ and each with a
fluctuation of size $1/\SL$. Thus the total fluctuation $\Del
L(t) $ of $L(t) $ will be of order $1/\SL \times \sqrt{N(t) }$,

\beq\label{5.a5}
\frac{\Del L(t) }{\la L(t) \ra} \sim \frac{1}{\sqrt{\SL \la L(t) \ra}} \sim \e^{-\SL t},
\eeq

i.e.\ the
fluctuation of $L(t) $ around $\la L(t) \ra$ is small for $t \gg
1/\SL$. In the same way the independence of the
``loop-loop''-correlator of $\Del$ can be understood as the
combined result of $L(t+\Del) $ growing exponentially in length
with a factor $e^{2\SL \Del}$ compared to $L(t) $ and, according to
\rf{5.a3}, the correlation of  ``line-elements'' of $L(t) $ and
$L(t+\Del) $ decreasing by a factor $e^{-2\SL \Del}$.

\section{Summary}

We have described how the CDT quantization of 2d gravity for a
special value of the boundary cosmological constant leads to a
non-compact (Euclidean)  AdS-like spacetime of constant negative
curvature dressed with quantum fluctuations. It is possible to
achieve this non-compact geometry as a limit of a compact geometry
as described above. In particular the assignment \rf{5.2} leads to
a simple picture where the fluctuation of $L(t) $ is small compared
to the average value of $L(t) $. In fact the geometry can be viewed
as that of the Poincar\'{e} disc with fluctuations correlated only
over a distance $1/\SL$.

Our construction is similar to the analysis of $ZZ$-branes
appearing as a limit of compact 2d geometries in Liouville quantum
gravity \cite{Ambjorn:2004my}. In the CDT case the non-compactness came when
the running boundary cosmological constant $\bX(T) $ went to the
fixed point $\SL$ for $T \to \infty$. In the case of Liouville
gravity, represented by DT (or equivalently matrix models), the
non-compactness arose when the running (Liouville)  boundary
cosmological constant ${\bX_{Liouville}(T) }$ went to the value
where the disc amplitude $W_\tL(\tX)  \equ 0$, i.e.\ to $\tX \equ
\sqrt{\tL}/2$ (see eq.\ \rf{0.0}). It is the same process in the
two cases, since the relation between Liouville gravity and CDT is
well established and summarized by the mapping \cite{Ambjorn:1999fp}:

\beq\label{6.1}
\frac{X}{\SL} = \sqrt{\frac{2}{3}}\; \sqrt{1+\frac{\tX}{\sqrt{\tL}}},
\eeq

between the coupling
constants of the two theories. The physical interpretation of this
relation is discussed in \cite{Ambjorn:1999fp,Ambjorn:1998xu}: One obtains the CDT model
by chopping away all baby-universes from the Liouville gravity
theory, i.e.\ universes connected to the ``parent-universe'' by a
worm-hole of cut-off scale, and this produces the relation
\rf{6.1} \footnote{The relation \rf{6.1} is similar to the one
encountered in regularized bosonic string theory in dimensions
$d\geq 2$ \cite{Durhuus:1984in,Ambjorn:1985az,Ambjorn:1987wu}: The world sheet degenerates into
so-called branched polymer. The two-point function of these
branched polymers is related to the ordinary two-point function of
the free relativistic particle by chopping off (i.e.\ integrating
out)  the branches, just leaving for each branched polymer
connecting two points in target space one {\it path} connecting
the two points. The mass-parameter of the particle is then related
to the corresponding parameter in the partition function for the
branched polymers as $X/\SL$ to $\tX/\sqrt{\tL}$ in eq.\ \rf{6.1}.
}. It is seen that $X \to \SL$ corresponds precisely to $\tilde{X}
\to \sqrt{\tL}/2$.

While the starting point of the CDT quantization was the desire to
include only Lorentzian, causal geometries in the path integral,
the result \rf{2.a11} shows that after rotation to Euclidean
signature this prescription is in a natural correspondence with
the Euclidean Hartle-Hawking no-boundary condition, since all of
the geometries \rf{2.a11} have a continuation to $t \equ -t_0$,
where the spacetime is regular.  It would be interesting if this
could be promoted to a general principle also in higher
dimensions. The computer simulations reported in
\cite{Ambjorn:2004qm,Ambjorn:2004pw,Ambjorn:2005db,Ambjorn:2005qt,Ambjorn:2005jj,Ambjorn:2006jf}
seem in accordance with this possibility.

\chapter{Topology fluctuations of space and time}\label{ch5}

In this chapter we refocus our attention on the issue of topology change in quantum gravity.
We make some steps to go beyond the results of chapter \ref{ch3} and
incorporate topology fluctuations of space \emph{and time} in the path integral of two dimensional CDT.
In field theory language one would say that in this chapter we go beyond tree
level by computing loop corrections. The incorporation of loop corrections is a notoriously
difficult undertaking however. Even in standard quantum field theory the loop expansion
is not convergent so one can only validly utilize the perturbation theory up to a limited number of
loops. Technically, the expansion merely forms an asymptotic power series instead of a convergent one.
The basic reason behind the divergence of the series expansion is that for large number of loops the
amount of diagrams grows super-exponentially.

The situation for gravity is typically similar, the number of ways in
which one can cut and reglue a manifold to obtain manifolds with
a different topology is very large and leads to uncontrollable divergences
in the path integral. Even in the simplest case of two dimensional
Euclidean geometries, the number of possible configurations
grows faster than exponentially with the volume of the geometry.
A well-known manifestation of this problem is the non-Borel summability of the genus
expansion in string theory. This does not necessarily mean that there is
no underlying well-defined theory, but even in the much-studied case of
two dimensional Euclidean quantum gravity no physically satisfactory,
unambiguous solution has been found \cite{DiFrancesco:1993nw}.

Our contribution to the problem of spacetime topology change is twofold.
In section \rf{Perturbation theory} we take a modest point of view and analyze
the issue of spacetime topology change perturbatively in the loop expansion.
Particularly, we expand the standard formalism of two dimensional
causal quantum gravity and compute the Hartle Hawking wavefunction
up to second order in the genus expansion. The statistical mechanics
of loop diagrams is considerably more complicated than the tree diagrams
that were analyzed in chapter \ref{ch3} henceforh we are not be able to address the summability
of the expansion. 

On a more positive and perhaps more speculative note we introduce
a toy model in section \rf{sec:Nonperturbative sum over topologies?}
where we show that it might be possible to define a nonperturbative sum
over topologies provided suitable causality restrictions are imposed
\footnote{Also in the context of two dimensional Euclidean quantum gravity a nonperturbative sum over genera has been
performed in a simplified model \cite{Ambjorn:1990wp}.}.
This section is based on \cite{Loll:2005dr}.


\section{Perturbation theory} \label{Perturbation theory}

\begin{figure}
\begin{center}
\includegraphics[width=2.8in]{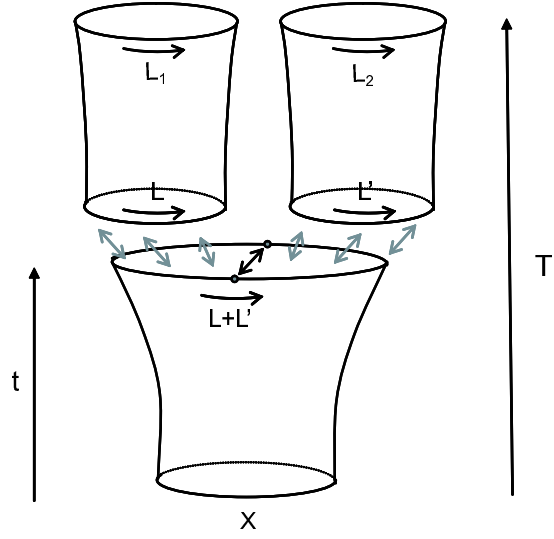}
\caption{Elementary trouser amplitude by gluing propagators}\label{fig:detailtrouser}
\end{center}
\end{figure}

Recall that the time dependent disc amplitude of pure CDT is given by

\beq \label{wavefunctionLT}
W_{\Lambda}(L,T) =\frac{\Lambda}{\sinh^2\sqrt{\Lambda}T}e^{{-\sqrt{\Lambda}L\coth\sqrt{\La}T }}
\eeq
which is obtained from the propagator by shrinking the initial boundary to a point, i.e.~
$W_{\Lambda}(L,T) \equ G_\La(L_1\equ0,L;T) $. Upon integrating over time
one obtains the Hartle Hawking wavefunction,
\beq \label{wavefunctionL}
W_{\Lambda}(L) =\int_0^\infty dT\;W_{\Lambda}(L,T) =\frac{e^{-\sL L}}{L}.
\eeq
To go beyond this genus zero result we need to extend the existing literature on two dimensional
CDT by computing amplitudes where the spatial topology is allowed to change. A simple example of
such an object is the before mentioned trouser amplitude (fig.~\ref{fig:detailtrouser}). It can be
obtained by ``gluing'' three cylinder amplitudes. In this gluing procedure one takes a propagator
with a final loop length $(L+L') $ that is equal to the sum of the lengths of the initial loops $L$
and $L'$ of two other propagators and integrates over $L$ and $L'$. As in the case of the
composition rule \eqref{eq:discrcompositionllnonmarked}
one has to include a measure factor to obtain the correct
amplitudes. Furthermore, from a continuum perspective one can say that the gluing is defined such that
Dehn twists around spatial slices are not present.
Instead of writing out the measure factors
explicitly we absorb them into the propagators by marking the
loops of the amplitudes that are being glued. For convenience we
also use a mixed representation for the propagators where one of
the boundaries has fixed cosmological constant and the other has
fixed boundary length,

\beq \label{propagatorXL}
G_\Lambda(X,L,T) =\frac{e^{-\bar{X}_X(T) L}}{L}-\frac{e^{-L \bar{X}_{\infty}(T) }}{L}
\eeq
where $\bar{X}_X(T) $ is given by

\beq \label{Xbar}
\bar{X}_X(T)  = \bar{X}_{\infty}(T)-\frac{\Lambda}{\sinh^2\sqrt{\Lambda }T\left( X+\bar{X}_{\infty}(T)\right) }
\eeq
and $\bar{X}_{\infty}(T) \!=\!\sqrt{\Lambda } \coth \sqrt{\Lambda }T$. 

We can now calculate the trouser amplitude from an initial boundary with boundary cosmological constant $X$ to two final
boundaries of length $L_1$ and $L_2$ in time $T$, where the
splitting occurs after a fixed time $t$ (fig.~\ref{fig:detailtrouser}),

\bea \label{deftrouserXLLTt}
 \cT_{\Lambda}(X,L_1,L_2;T,t) \!&=&\!2\!\int\limits^{\infty}_0 \int\limits^{\infty}_0 dL dL' G_\Lambda^{(0;1) }\!(X,L \!+\!L',t)\times \nonumber \\
 && G_\Lambda^{(1;0) }\!(L,L_1,T\!-\!t) G_\Lambda^{(1;0) }\!(L',L_2,T\!-\!t),
\eea
where the superscript notation of the propagators denotes the marking of its loops, as introduced in section \rf{Marking the causal propagator}.
Performing the integrations yields
\bea
 \cT_{\Lambda}(X,L_1,L_2;T,t) &=&  2~e^{-\bar{X}_{X}(T) (L_1+L_2) } \left(\frac{\sqrt{\La} \cosh \sqrt{\Lambda}t + X \sinh \sqrt{\Lambda}t}{\sqrt{\La}\cosh \sqrt{\Lambda}T + X \sinh \sqrt{\Lambda}T}\right) ^4 \nonumber \\
 && \!+~2~  e^{-\bar{X}_{\infty}(t) (L_1+L_2) }\! \left(\frac{\sinh\sqrt{\Lambda}t}{\sinh\sqrt{\Lambda}T}\right) ^4.\label{trouserXLLTt}
\eea
%
In the rest of the section we only present results where the length of the initial loop is zero to
keep the presentation as transparent as possible.
The trouser amplitude with zero initial length is dubbed the splitting amplitude which we denote by $\cS$.
In particular we are interested in the splitting amplitude 
to two final loops of length $L_1$ and $L_2$,
where the time before the splitting is arbitrary and the time after the splitting is fixed to be $t'$
 \bea
\cS_{\Lambda}(L_1,L_2;t') &=&  2 \int\limits^{\infty}_0 dL \int\limits^{\infty}_0  dL' \, W_{\Lambda}^{(1) }(L+L') G_\Lambda^{(1;0) }(L,L_1;t') G_\Lambda^{(1;0) }(L',L_2;t') \nonumber\\
&=& 2\,e^{-(L1 + L2)  \sqrt{\Lambda }}e^{-4 t' \sqrt{\Lambda }} \label{defsplitT}.
\eea
This amplitude might seem a bit unnatural, since it treats the time intervals before and after splitting differently.
It is however useful for studying averages of the duration of a hole (section \rf{subsec:Higher genus Hartle Hawking wave functions}).
If we integrate over the time after the split in spatial topology as well we obtain
\beq \label{split}
\cS_{\Lambda}(L_1,L_2) = \int\limits^{\infty}_0 dt' \cS_{\Lambda}(L_1,L_2;t') = \frac{e^{-(L1 + L2)  \sqrt{\Lambda }}}{2 \sL}.
\eeq
In the following we use the results obtained in this subsection to calculate higher genus Hartle Hawking wavefunctions within our model.

\subsection{Higher genus Hartle Hawking wavefunctions}\label{subsec:Higher genus Hartle Hawking wave functions}

\begin{figure}
\begin{center}
\includegraphics[width=4in]{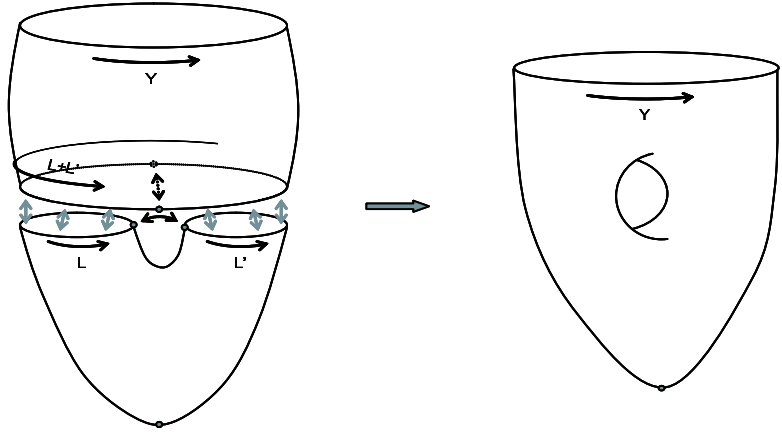}
\caption{Construction of the genus one disc function from the splitting amplitude and a propagator.}\label{fig:genusonediscfunction}
\end{center}
\end{figure}

In this section we present genus one and genus two generalizations of the Hartle Hawking wave function where we integrate
over the time intervals. It is possible to obtain results where the time intervals are fixed, but we choose to omit them for
sake of clarity. Given the splitting amplitude \eqref{split} it is straightforward to compute the genus one generalization
of the Hartle Hawking wavefunction \eqref{wavefunctionL}. To simplify calculations we first obtain the genus one wave function with fixed boundary cosmological
constant $Y$ and a mark on the boundary (see fig.~\ref{fig:genusonediscfunction}),

\beq \label{defwavefunctiongenusoneYmark}
W^{(1) }_{\Lambda,\genus=1}(Y) = e^{-2 \kappa}\int\limits^{\infty}_0 dL \int\limits^{\infty}_0  dL' \cS_{\Lambda}^{(1,1) }(L,L') G^{(1;1) }_{\Lambda}(L+L',Y),
\eeq
where $\kappa \!=\! 2\pi/G^b_N$ is proportional to the inverse \emph{bare} Newton's constant and
$\cS^{(1,1)}$ denotes that both loops of the splitting amplitude possess a mark. Since 2D quantum gravity can
be viewed as string theory with a zero dimensional target space, $e^{-\kappa}$ can be identified with a bare string coupling $g_s$.
The renormalization of this coupling is addressed in the next section. Performing the integrations in \eqref{defwavefunctiongenusoneYmark} one obtains
\beq \label{wavefunctiongenusoneYmark}
W^{(1) }_{\Lambda,\genus=1}(Y) = g^2_s \frac{Y^3+5 \sqrt{\Lambda } Y^2+11 \Lambda  Y+15 \Lambda ^{3/2}}{32 \left(Y+\sqrt{\Lambda }\right) ^5 \Lambda ^{5/2}}.
\eeq
If we now do an inverse Laplace transformation from $Y$ to $L$ and divide by $L$ to remove the mark, we obtain the desired genus one
Hartle Hawking wave function,
\beq \label{wavefunctiongenusoneL}
W_{\Lambda,\genus=1}(L) = g^2_s\frac{e^{-L \sqrt{\Lambda }} \left(\Lambda ^{3/2} L^3+2 \Lambda  L^2+3 \sqrt{\Lambda } L+3\right) }{96 \Lambda ^{5/2}}.
\eeq

One can use the previous results to compute simple observables. For example, using \eqref{defsplitT} we can obtain the average duration of the hole in the genus one disc amplitude:

\bea
\expec{t}_{\text{hole} }&=&\frac{1}{W^{(1) }_{\Lambda,\genus=1}(Y) } \int\limits^{\infty}_0 dt' \, t'  \int\limits^{\infty}_0 dL \int\limits^{\infty}_0  dL' \cS_{\Lambda}^{(1,1) }(L,L';t') G^{(1;1) }_{\Lambda}(L+L',Y) \nonumber\\
&=&\frac{1}{4 \sL}.
\eea

The size of the hole is determined by
$1/\sL$ which is the only length scale in the model. Similarly we
get for the fluctuations

\beq
\expec{\Delta t}_{\text{hole}}=\sqrt{ \expec{t^2}_{\text{hole}}-\expec{t}_{\text{hole}}^2}=\frac{1}{4 \sL}.
\eeq

It is
interesting to observe that $\expec{\Delta
t}_{\text{hole}}\!=\!\expec{t}_{\text{hole} }$. This reflects the
fact that as for the spatial geometry, where we have
$\expec{\Delta L}\!\sim\!\expec{L}$, the geometry of the holes is
also purely governed by quantum fluctuations.

The procedure to obtain the genus two wave function is analogous
to the one outlined above. In practice it involves doing much more
tedious calculations however, since one has to glue more and more
propagators. This does not introduce any fundamental complications
and does not give one much more physical insight so we only
present the result of the computations. To obtain the genus two
wave function one needs to add contributions from three different
diagrams,
\beq
W_{\Lambda,\genus=2}(L) = W^a_{\Lambda,\genus=2}(L)  +W^b_{\Lambda,\genus=2}(L) +W^c_{\Lambda,\genus=2}(L) =
\raisebox{-20pt}{\includegraphics[height=45pt]{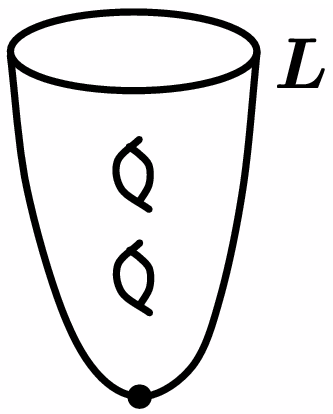}}+
\raisebox{-20pt}{\includegraphics[height=45pt]{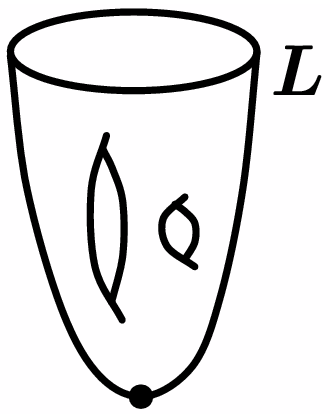}}+
\raisebox{-20pt}{\includegraphics[height=45pt]{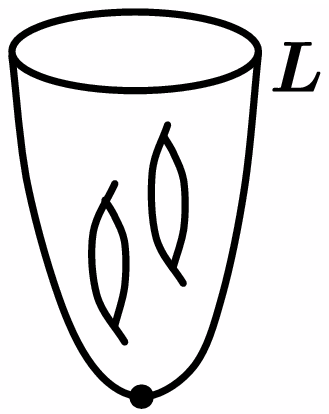}}
\eeq
which evaluates to,
\beq \label{wavefunctiongenustwoL}
W_{\Lambda,\genus=2}(l) =g^4_s{\textstyle \frac{e^{-l} \left(3 l^7+28 l^6+150 l^5+570 l^4+1575 l^3+3150 l^2+4725 l+4725\right) }{92160 \Lambda ^{11/2}}},
\eeq
where the dimensionless variable $l=L\sqrt{\Lambda}$ is the length of the boundary in units of $\sqrt{\Lambda}$.

\subsubsection{Double scaling limit}

Looking at the Hartle Hawking wave functions for genus $\genus=0,1,2$ , i.e.~ Eqs.~\eqref{wavefunctionL}, \eqref{wavefunctiongenusoneL} and \eqref{wavefunctiongenustwoL}, one can see that they all have different dimensions. Specifically,
the $\genus=0,1,2$ wave functions 
have dimension of
$\Lambda^{1/2}$, $\Lambda^{-5/2}$ and $\Lambda^{-11/2}$ respectively. This would
imply that the different genus contributions cannot be added to a single wave function. In fact one \emph{can} add them if one
takes into account their wave function renormalization factors that appear when taking the continuum limit of the discrete
sums. The reason 
why those factors did not appear in the previous sections is
that for the case of fixed topology one absorbs these factors in the boundary states
of the theory. 
If one would have kept those factors from the outset the $\genus=0,1,2$ wave functions would all be dimensionless and would behave as
$a \sqrt{\Lambda}$, $(a \sqrt{\Lambda}) ^{-5}$ and $(a\sqrt{\Lambda}) ^{-11}$ respectively, where $a$ is the cutoff of the theory induced by the discrete lattice spacing.

To remove this
cutoff dependence on the lattice spacing $a$ one can do a renormalization of the string coupling such that all higher genus
diagrams will
contribute in the continuum limit. The simultaneous scaling of the string coupling and the cosmological constant is
called the ``double scaling'' limit. In particular, we obtain
\beq
g_S=g_s (a\sqrt{\Lambda}) ^3.
\eeq
Observe that this scaling limit is the same as the scaling we considered in chapter \ref{ch3}.
This implies that the spatial and spacetime topology changes are in fact part of one topological
expansion. Using this scaling limit we can now derive the Hartle Hawking wave function up to order
two in the genus expansion,
\bea
\!\!\!\!\!\!\!\!\!W_{\Lambda,g_S}(l) \!\!\!&=&\!\!\!{\textstyle \sL\, e^{-l}\left(\frac{1}{l}+ g^2_S\frac{ l^3+2 l^2+3 l +3}{96}+\right.}
\nonumber\\
&&\,\,\,\,\,\,\,\,{\textstyle\left. +g_S^4 {\textstyle \frac{e^{-l} \left(3 l^7+28 l^6+150 l^5+570 l^4+1575 l^3+3150 l^2+4725 l+4725\right) } {92160}}+\mathcal{O}(g_S^6) \right),}\nonumber\\
\eea
where we again used the dimensionless variable $l=L\sqrt{\Lambda}$.

\subsection{Summary}

We have described how to include manifolds of higher genus in the path integral
of two dimensional CDT. One of the main results is the computation of the Hartle Hawking
wave function up to two loops in the genus expansion.
Generalization to higher genus is in principle straightforward, however,
the calculations become more and more cumbersome.
It is interesting to note that the Hartle Hawking wave functions in the framework of CDT
are very similar to the Hartle Hawking amplitudes in Euclidian
Dynamical Triangulations \cite{DiFrancesco:1993nw}. 
Using the method of loop equations it is possible to also obtain higher
genus results for 2D Euclidean quantum gravity \cite{Ambjorn:1992gw,Eynard:2004mh}.
A detailed comparison of these results might lead to a better understanding of the relationship
between Euclidean and Lorentzian quantum gravity.

\newpage

\section{Nonperturbative sum over topologies?}\label{sec:Nonperturbative sum over topologies?}

As shown in previous work, there is a well-defined nonperturbative gravitational path integral
including an explicit sum over topologies in the setting of causal dynamical triangulations in two
dimensions. In this section we derive a complete analytical solution of the quantum continuum
dynamics of this model, obtained uniquely by means of a double-scaling limit. We show that the
presence of infinitesimal wormholes leads to a decrease of the effective cosmological constant,
reminiscent of the suppression mechanism considered by Coleman and others in the four-dimensional
Euclidean path integral. Remarkably, in the continuum limit we obtain a finite spacetime density of
microscopic wormholes without assuming fundamental discreteness. This shows that one can in
principle make sense of a gravitational path integral which includes a sum over spacetime topologies,
provided suitable causality restrictions are imposed on the path integral histories.

\subsection{Outline}

A new idea to tame the divergences associated with spacetime topology
changes in the path integral was advanced in \cite{Loll:2003rn}
and implemented in a model of two dimensional nonperturbative Lorentzian quantum
gravity. The idea is to include a sum over topologies, or over some
subclass of topologies, in the state sum, but to restrict this class further
by certain {\it geometric} (as opposed to topological) constraints.
These constraints involve the causal (and therefore Lorentzian)
structure of the spacetimes and thus would have no analogue in a purely
Euclidean formulation. In the concrete two dimensional model
considered in \cite{Loll:2003rn}, the path integral is taken over a
geometrically distinguished class of spacetimes with arbitrary numbers
of ``wormholes", which violate causality only relatively mildly (see
also \cite{Loll:2003yu}). As a consequence, the nonperturbative
path integral turns out to be well defined.
This is an extension of the central idea of the approach of causal
dynamical triangulations, namely, to use physically motivated
causality restrictions to make the gravitational path integral better behaved
(see \cite{Loll:2002xb} for a review).

In section \rf{discretesection}, we will present a complete analytical solution of the
statistical model of two dimensional Lorentzian random geometries
introduced in \cite{Loll:2003rn}, whose starting point is a regularized
sum over causal triangulated geometries {\it including} a sum over
topologies. For a given genus (i.e.~number of (worm)holes in the spacetime)
not all possible triangulated geometries are included in the sum,
but only those which satisfy certain causality constraints. As shown
in \cite{Loll:2003rn}, this makes the statistical model well defined,
and an unambiguous continuum limit is obtained by taking a
suitable double-scaling limit of the two coupling constants of the model,
the gravitational or Newton's constant and the cosmological
constant. The double-scaling limit presented here differs from the
one found in \cite{Loll:2003rn,Loll:2003yu}, where only the partition
function for a single spacetime strip was evaluated. We will show that
when one includes the boundary lengths of the strip explicitly --
as is necessary to obtain the full spacetime dynamics -- the natural
renormalization of Newton's constant involves the boundary
``cosmological" coupling constants conjugate to the boundary lengths.
Although the holes we include exist only for an infinitesimal time, and we
do not keep track of them explicitly in the states of the Hilbert
space, their integrated effect is manifest in the continuum Hamiltonian
of the resulting gravity theory. As we will see, their presence leads
to an effective lowering of the cosmological constant and therefore
represents a concrete and nonperturbative implementation of an idea
much discussed in the late eighties in the context of the ill-defined
continuum path integral formulation of Euclidean quantum gravity
(see, for example, \cite{Coleman:1988tj,Klebanov:1988eh}).

The remainder of this section is structured as follows.
In the next subsection,
we briefly describe how a nonperturbative theory of two dimensional
Lorentzian quantum gravity can be obtained by the method of causal
dynamical triangulation (CDT), and how a sum over topologies can be
included. For a more detailed account of the construction of
topology-changing spacetimes and the
geometric reasoning behind the causality constraints
we refer the reader to \cite{Loll:2003rn,Loll:2003yu}.
The main result of Sec.\ \rf{discretesection} is the computation
of the Laplace transform of the one-step propagator of the discrete
model for arbitrary boundary geometries.
In Sec.\ \rf{continuumsection} we make a scaling ansatz for the
coupling constants and show that just one of the choices for
the scaling of Newton's constant leads to a new and physically
sensible continuum theory. We calculate the corresponding quantum
Hamiltonian and its spectrum, as well as the full propagator of
the theory. Using these results, we compute
several observables of the continuum theory in Sec.\ \rf{sec_observ},
most importantly, the expectation
value of the number of holes and its spacetime density.
In Sec.\ \rf{conclusions}, we summarize
our results and draw a number of conclusions. In Appendix \ref{app2}, we
discuss the properties of alternative scalings for Newton's constant
which were discarded in the main text. This also establishes a
connection with previous attempts \cite{DiFrancesco:2000nn,Durhuus:2001sp}
to generalize the original Lorentzian model without topology
changes. In Appendix \ref{app3}, we calculate the spacetime density
of holes from a single infinitesimal spacetime strip.

\subsection{Implementing the sum over topologies}

Our aim is to calculate the (1+1) -dimensional
gravitational path integral
\begin{equation}\label{contPI}
    Z(G_N,\Lambda)  = \sum_{topol.} \int D[g_{\mu\nu}] e^{i
    S(g_{\mu\nu}) }
\end{equation}
nonperturbatively by using the method of Causal Dynamical
Triangulations (CDT).
The sum in \rf{contPI} denotes the inclusion in the path integral
of a specific, causally preferred class of
fluctuations of the manifold topology. The action $S(g_{\mu\nu}) $
consists of the usual Einstein-Hilbert curvature term and
a cosmological constant term. Since we work in two dimensions, we recall that
integrated curvature term is proportional to the Euler characteristic
$\chi = 2\mi 2\genus\mi b$ of the spacetime manifold, where $\genus$
denotes the genus (i.e.~the
number of handles or holes)  and $b$ the number of boundary components.
Explicitly, the action reads
\begin{equation}\label{action}
    S = 2 \pi \chi K - \Lambda \int d^2x\sqrt{|\det g_{\mu\nu}|},
\end{equation}
where $K=1/G_N$ is the inverse Newton's constant and $\Lambda$ the
cosmological constant (with dimension of inverse length squared).

Just like in the original CDT model \cite{Ambjorn:1998xu}
we will first regularize the path integral \myref{contPI} by a sum
over piecewise flat two dimensional spacetimes, whose flat
building blocks are identical Minkowskian triangles, all with one
spacelike edge of squared length $+a^2$ and two timelike edges
of squared length $-\alpha a^2$, where $\alpha$ is a real positive
constant. The CDT path integral takes the form of a sum over
triangulations, with each triangulation consisting of a sequence
of spacetime strips of height $\Delta t =1$ in the time direction.
A single such strip is a set of  $l_1$ triangles pointing up
and $l_2$ triangles pointing down (fig.~\ref{fig:triangulation}).
Because the geometry has a sliced structure, one can easily
Wick-rotate it to a triangulated manifold of Euclidean signature
by analytically continuing the parameter $\alpha$ to a real
negative value \cite{Ambjorn:2001cv}. For simplicity, we will set
$\alpha\equ -1$ in evaluating the regularized, real and
Wick-rotated version of the path integral \rf{contPI}.

\begin{figure}
\begin{center}
\includegraphics[width=3.5in]{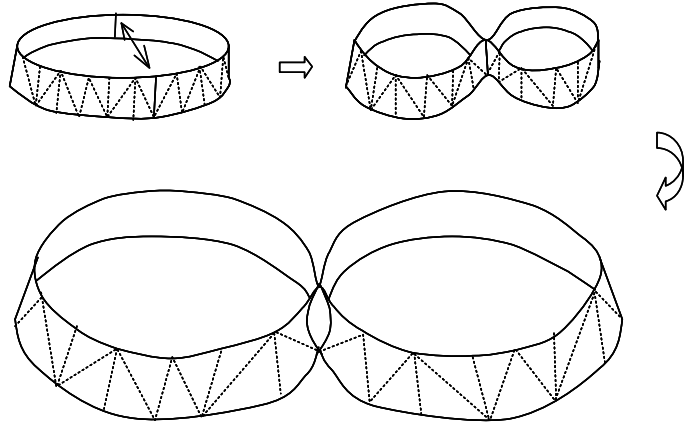}
\caption{Construction of a wormhole by identifying two timelike
edges of a spacetime strip and cutting open the geometry along the
edge.}\label{figure02}
\end{center}
\end{figure}

In the pure CDT model the one-dimensional spatial slices of constant
proper time $t$ are usually chosen as circles, resulting
in cylindrical spacetime geometries. For our present purposes, we
will enlarge this class of geometries by allowing the genus to be variable.
We define the sum over topologies by
performing surgery moves directly on the triangulations to obtain
regularized versions of higher-genus manifolds \cite{Loll:2003rn,Loll:2003yu}.
They are generated by adding tiny wormholes that connect two regions of
the same spacetime strip. Starting from a regular strip of
topology $[0,1]\times S^1$ and height $\Delta t =1$, one can construct a hole by
identifying two of the strip's timelike edges and subsequently cutting open the
geometry along this edge (fig.~\ref{figure02}). By applying this procedure
repeatedly (obeying certain causality constraints \cite{Loll:2003rn,Loll:2003yu}),
more and more wormholes can be created.
Once the regularized path integral has been performed, including a sum over
geometries with wormholes, one takes a continuum limit by letting $a\rightarrow 0$
and renormalizing the coupling constants appropriately, as will be described
in the following sections.

Note that our wormholes are minimally causality- and locality-violating
in that they are located within a single proper-time step (the smallest time
unit available in the discretized theory)  and the associated baby universes
which are born at time $t$ are reglued at time $t+1$ ``without twist" \cite{Loll:2003rn,Loll:2003yu}.
In a macroscopic interpretation one could describe them as wormholes
which are instantaneous in the proper-time frame of an ensemble of
freely falling observers. Note that this is invariantly defined (on Minkowski
space, say)  once an initial surface has been chosen. Such a restriction is
necessary if one wants to arrive at a well-defined unitary evolution
via a transfer matrix formalism, as we are doing. To include wormholes
whose ends lie on different proper-time slices, one would have to
invoke a third-quantized formulation, which would very likely result in
{\it macroscopic} violations of causality, locality and therefore unitarity,
something we are trying to avoid in the present model.

One could wonder whether the effect in the continuum theory of our choice
of wormholes is to single out a preferred coordinate system. Our
final result will show that this is not the case, at least not over and above
that of the pure gravitational model without topology changes. The effect
of the inclusion of wormholes turns out to be a rather mild ``dressing" of the
original theory without holes. We believe that the essence of our model lies
not so much in how the wormholes are connected, because they do not
themselves acquire a nontrivial dynamics in the continuum limit.
Rather, it is important that their number is sufficiently large to have an effect
on the underlying geometry, but on the other hand sufficiently
controlled so as to render the model computable.

A similar type of wormhole has played a prominent role in past attempts to
devise a mechanism to explain the smallness of the cosmological constant
in the Euclidean path integral formulation of {\it four}-dimensional quantum gravity
in the continuum \cite{Coleman:1988tj,Klebanov:1988eh}.
The wormholes considered there resemble those of our toy model
in that both are non-local identifications of the spacetime geometry of
infinitesimal size. The counting of our wormholes is of course different since
we are working in a genuinely Lorentzian setup where certain causality
conditions have to be fulfilled. This enables us to do the sum over
topologies completely explicitly. Whether a similar construction is
possible also in higher dimensions is an interesting, but at this stage
open question.

\subsection{Discrete solution: the one-step propagator}\label{discretesection}

For the (1+1) -dimensional Lorentzian gravity model including a sum over
topologies, the partition function of a single spacetime strip of infinitesimal
duration with summed-over boundaries was evaluated in
\cite{Loll:2003rn} and \cite{Loll:2003yu}.
In the present section, we will extend this treatment by calculating the
full one-step propagator, or, equivalently, the generating function for the
partition function of a single strip with given, fixed boundary lengths.
This opens the way for investigating the full dynamics of the model.

The discrete set-up described above leads to the Wick-rotated one-step
propagator
\begin{equation}\label{Transferlaplacedef}
G_{\lambda,\kappa}(l_1,l_2,t=1) =
e^{-\lambda (\lin+\lout) }\sum_{T|l_1,l_2}e^{-2\kappa\genus},
\end{equation}
where $\kappa$ is the bare inverse Newton's constant and $\lambda$ the bare (dimensionless)
cosmological constant, and we have omitted an overall constant coming from the
Gauss-Bonnet integration. The sum in \myref{Transferlaplacedef} is to be taken over all
triangulations with $l_1$ spacelike links in the initial and $l_2$ spacelike links in the
final boundary.
Note that the number of holes does not appear as one of the arguments of the one-step
propagator since we only consider holes that exist within one strip.
Consequently, the number of holes does not appear explicitly as label for the quantum
states, and the Hilbert space coincides with that of the pure CDT model.
Nevertheless, the integrated effect of the topologically non-trivial
configurations changes the dynamics and the quantum Hamiltonian, as we shall see.

The one-step propagator \myref{Transferlaplacedef} defines a transfer matrix $\hat T$
by
\begin{equation}
\label{transfer}
G_{\lambda,\kappa}(l_1,l_2,1) =\langle l_2|\hat{T}| l_1\rangle,
\end{equation}
from which we obtain the propagator for $t$ time steps as usual by
iteration,
\begin{equation}
G_{\lambda,\kappa}(l_1,l_2,t) =\langle l_2 | \hat{T}^t | l_1 \rangle.
\label{fullprop}
\end{equation}
For simplicity we perform the sum in \myref{Transferlaplacedef} over triangulated
strips with periodically identified boundaries in the spatial direction and one
marked timelike edge. By virtue of the latter,
$G_{\lambda,\kappa}(l_1,l_2,t) $ satisfies the desired composition
property of a propagator,
\begin{eqnarray}
\label{composition}
G_{\lambda,\kappa}(l_1,l_2,t_1+t_2) &=& \sum_{l}G_{\lambda,\kappa}(l_1,l,t_1)
G_{\lambda,\kappa}(l,l_2,t_2),\\
G_{\lambda,\kappa}(l_1,l_2,t+1) &=& \sum_{l}G_{\lambda,\kappa}(l_1,l,1)
G_{\lambda,\kappa}(l,l_2,t),
\end{eqnarray}
where the sums on the right-hand sides are performed over an intermediate
constant-time slice of arbitrary discrete length $l$.

Performing the fixed-genus part of the sum over triangulations in
\myref{Transferlaplacedef} yields
\begin{equation}
\label{sumfixedg}
G_{\lambda,\kappa}(l_1,l_2,1) =e^{-\lambda
N}\sum_{\genus=0}^{[N/2]}
\binom{N}{\lin}\binom{N}{2\genus}\frac{(2\genus) !}{\genus !
(\genus+1) !}e^{-2\kappa\genus},
\end{equation}
with $N \equ \lin \pl \lout$. To simplify calculations we will use
the generating function for\-malism with
\begin{equation}
\label{gen}
G(x,y,g,h,1) =\sum_{\lin,\lout=0}^\infty
G_{\lambda,\kappa}(l_1,l_2,1)  x^{\lin} y^{\lout},
\end{equation}
where we have defined $g\equ e^{-\lambda }$ and
$h\equ e^{-\kappa}$. The quantities $x$ and $y$ can be seen
as purely technical devices, or alternatively as exponentiated
bare boundary cosmological constants
\begin{equation}
x=e^{- \lambda_{in}},\quad y=e^{- \lambda_{out}}.
\end{equation}
Upon evaluating the
sum over $\lin$ and $\lout$ one obtains the generating
function of the one-step propagator
\begin{equation}
\label{Glaplace}
 G(x,y,g,h,1) =    \frac{1}{1 - g\,\left( x + y \right)  }
 \frac{2}{1 + {\sqrt{1 - 4 u^2}}},
\end{equation}
with
\begin{equation} \label{z}
  u = \frac{h}{ \frac{1}{g\,\left( x + y \right)  }-1}.
\end{equation}
Note that in order to arrive at the final result \rf{Glaplace},
we have performed an explicit sum over all topologies!
The fact that this infinite sum converges for appropriate values of
the bare couplings has to do with the causality constraints
imposed on the model, which were geometrically motivated
in \cite{Loll:2003rn}, and which effectively reduce the number of
geometries in the genus expansion.

In \myref{Glaplace} one recognizes the generating function
${\rm Cat}(u^2) $ for the Catalan numbers,
\begin{equation}
\label{catalan}
{\rm Cat}(u^2) =\frac{2}{1 + {\sqrt{1 - 4 u^2}}}.
\end{equation}
For $h\equ 0$ one has ${\rm Cat}(u^2) \equ 1$ and expression
\rf{Glaplace}  reduces to the
one-step propagator without topology changes,
\begin{equation}
\label{transfernog}
G(x,y,g,h=0,1) =\frac{1}{1 - g\,\left( x + y \right)  }.
\end{equation}
Furthermore, one recovers the one-step partition function with
summed-over boundaries of  \cite{Loll:2003rn,Loll:2003yu} by
setting $x\equ y\equ 1$,
\begin{equation}
\label{Znog}
Z(g,h,1)  = \frac{1}{1 - 2 g }
\frac{2}{1 + {\sqrt{1 - 4 (\frac{2g h}{1-2g}) ^2}}}.
\end{equation}

\subsection{Taking the continuum limit}\label{continuumsection}

Taking the continuum limit in the case without topology changes is
fairly straightforward \cite{Ambjorn:1998xu}. The joint region of
convergence of \myref{transfernog} is given by
\begin{equation}
\label{convergence }
|x|<1,\quad |y|<1,\quad |g|<\frac{1}{2}.
\end{equation}
One then tunes the couplings to their critical values according to
the scaling relations
\begin{eqnarray}
g = \frac{1}{2}(1-&\!\!\!\!\!\!\!\!\! &a^2\,\Lambda)  + \mathcal{O}(a^{3}) \label{scalinga}, \\
x  =  1-a\,X + \mathcal{O}(a^{2}),& \!\!\!\!\!\!\!\!\!\!\! &\quad 
y  =  1-a\,Y + \mathcal{O}(a^{2}) \label{scalingb}.
\end{eqnarray}
Up to additive renormalizations, $x$, $y$ and $\lambda$ scale canonically,
with corresponding renormalized couplings $X$, $Y$ and $\Lambda$.
In the case with
topology change we have to introduce an additional scaling
relation for $h$. Since Newton's constant is dimensionless in two
dimensions,
there is no preferred canonical scaling for $h$.
We make the multiplicative ansatz\footnote{Here the factor
$\frac{1}{\sqrt{2}}$ is chosen to give a proper parametrization of
the number of holes in terms of Newton's constant (see Section
\rf{sec_observ}).}
\begin{equation}
\label{scalingh} h =\frac{1}{\sqrt{2}} h_{ren} (a d) ^\beta,
\end{equation}
where $h_{ren}$ depends on the renormalized Newton's constant
$G_N$ according to
\begin{equation}
\label{hrendef}
h_{ren}=e^{- 2\pi / G_N}.
\end{equation}
In order to compensate the powers of the cut-off $a$ in \rf{scalingh},
$d$ must have dimensions of inverse length.
The most natural ansatz in terms of the dimensionful quantities
available is
\begin{equation}
\label{d}
d = (\sqrt{\Lambda}^{\alpha} (X+Y) ^{1-\alpha}).
\end{equation}
The constants $\beta$ and $\alpha$ in relations \rf{scalingh}
and \rf{d}  must be chosen such as to
obtain a physically sensible continuum theory. By
this we mean that the one-step propagator should
yield the Dirac delta-function to lowest order in $a$, and that the
Hamiltonian should be bounded below and not
depend on higher-order terms in \myref{scalinga}, \myref{scalingb},
in a way that would introduce a dependence on new couplings without
an obvious physical interpretation.

To calculate the Hamiltonian operator $\hat H$ we use the analogue of the
composition law \myref{composition} for the Laplace
transform of the one-step propagator \cite{Ambjorn:1998xu},
\begin{equation}
\label{laplacecomposition}
G(x,y,t+1) = \oint \frac{dz}{2 \pi i z} G(x,z^{-1};1) G(z,y,t).
\end{equation}
In a similar manner we can write the time evolution of the wave function as
\begin{equation}
\label{timeevolution1}
\psi(x,t+1) = \oint \frac{dz}{2 \pi i z} G(x,z^{-1};1) \psi(z,t).
\end{equation}
When inserting the scaling relations \myref{scalinga}, \myref{scalingb}
and $t\equ \frac{T}{a}$ into this equation it is convenient to treat separately the
first factor in the one-step propagator \myref{Glaplace}, which is nothing but the
one-step propagator without topology changes \myref{transfernog}, and
the second factor, the Catalan generating function \myref{catalan}.
Expanding both sides of \myref{timeevolution1} to order $a$ gives
\begin{equation}
\label{scaledtransfernog}
\left(1-a \hat{H}+\mathcal{O}(a^2) \right) \psi(X)  = \int^{i\infty}_{-i\infty}
\frac{dZ}{2\pi i} \left\{\left(\frac{1}{Z-X} + a\,\frac{2\Lambda - X Z}{(Z-X) ^2}\right)
{\rm Cat}(u^2)  \right\} \psi(Z),
\end{equation}
where we have used
\begin{equation}
\psi(X,T+a) =e^{-a\,\hat{H}}\psi(X,T),
\end{equation}
with $\psi(X) \equiv\psi(x\equ 1- a X) $.
Note that the first term on the right-hand side of \myref{scaledtransfernog}, $\frac{1}{Z-X}$,
is the Laplace-transformed delta-function. The interesting new behaviour of the Hamiltonian
is contained in the expansion of the Catalan generating function.
Combining \myref{catalan} and \myref{z}, and inserting the scalings
\myref{scalinga}, \myref{scalingb}, yields
\begin{equation}
\label{scalingcatalan}
{\rm Cat}(u^2)  = 1 + { \frac{2\,{d }^{2\beta }\,h_{ren}^2}{\,{\left( Z-X \right)  }^2}
a^{2\beta-2} + {\rm h.o.} },
\end{equation}
where h.o. refers to terms of higher order in $a$.
In order to preserve the delta-function and have a
non-vanishing contribution to the Hamiltonian one is thus naturally led to
$\beta \equ 3/2$. For suitable choices of $\alpha$ it is also possible to obtain
the delta-function by setting $\beta\equ 1$, but
the resulting Hamiltonians turn out to be unphysical or at least do not
have an interpretation as gravitational models with wormholes,
as we will
discuss in Appendix A.\footnote{One might also consider scalings of the form
$ h \rightarrow  c_1 h_{ren} (a d) +c_2 h_{ren} (a d) ^{3/2}$, but
they can be discarded by arguments similar to those of Appendix A.
}

For $\beta = 3/2$ the right-hand side of \myref{scaledtransfernog} becomes
\begin{equation}
\label{scaledtransfernognew}
 \int^{i\infty}_{-i\infty} \frac{dZ}{2\pi i} \left\{\frac{1}{Z-X} + a\,\left( \frac{2\Lambda - X Z}
{(X-Z) ^2}-\frac{2\,\sqrt{\Lambda}^{3 \alpha} \,h_{ren}^2}
{\,{\left( X-Z \right)  }^{3 \alpha}} \right) \right\} \psi(Z).
\end{equation}
We observe that for $\alpha\leqslant 0$ the last term in \myref{scaledtransfernognew}
does not contribute to the Hamiltonian.
Performing the integration for $\alpha>0$ and discarding the possibility of fractional poles
the Hamiltonian reads
\begin{equation}
\label{HamiltonianX}
\hat{H}(X,\frac{\partial}{\partial X}) =X^2 \frac{\partial}{\partial X} +
X -2 \Lambda \frac{\partial}{\partial X}+
2\Lambda^{\frac{3 \alpha}{2}}h_{ren}^2
\frac{(-1) ^{3\alpha}}{ \Gamma(3\alpha) } \frac{\partial^{3\alpha-1}}{\partial X^{3\alpha-1}},\quad \alpha=\frac{1}{3},\frac{2}{3},1,...\, .
\end{equation}
For all $\alpha$'s, these Hamiltonians do not depend on higher-order terms in the
scaling of the coupling constants. One
can check this by explicitly introducing a term quadratic in $a$ (which can
potentially contribute to $\hat H$)  in the
scaling relations \myref{scalingb}, namely,
\begin{eqnarray}
x &=& 1-a\,X +\frac{1}{2}\gamma\,a^2\,X^2+\mathcal{O}(a^3),\nonumber\\
y &= & 1-a\,Y
+\frac{1}{2}\gamma\,a^2\,Y^2+\mathcal{O}(a^3),\label{scalingbnew}
\end{eqnarray}
and noticing that \myref{HamiltonianX} does not depend on $\gamma$.
After making an inverse Laplace transformation $\psi(L) =\int_0^\infty d X e^{X\,L}\psi(X) $
to obtain a wave function in the ``position" representation (where it depends on the
spatial length $L$ of the universe),
and introducing $m=3\alpha-1$ the Hamiltonian reads
\begin{equation}
\label{Hamiltonian}
\hat{H}(L,\dL) =-L\ddL-\dL+2\,\Lambda\,L- \frac{2\,\Lambda^\frac{m+1}{2}h_{ren}^2 }
{ \Gamma(m+1) } L^m,\quad m=0,1,2,...\, .
\end{equation}
Since $\hat H$ is unbounded below for $m \geqslant 2$, we are left
with $m\equ 0$ and $m\equ 1$ as possible choices for the scaling.
However, setting $m\equ 0$ merely has the effect of adding a constant term
to the Hamiltonian, leading to a trivial phase factor for the wave function.
We conclude that the only new and potentially interesting model corresponds
to the scaling with $m\equ 1$ and
\begin{equation}
\label{hm1}
h^2=\frac{1}{2} h_{ren}^2\,\Lambda\,(X+Y) \,a^3,
\end{equation}
with the Hamiltonian given by
\begin{equation}
\label{Hamiltonian1}
\hat{H}(L,\dL) =-L\ddL-\dL+  \left(1- h_{ren}^2 \right) \,2\,\Lambda\,L.
\end{equation}
Note that for all values $G_N\geq 0$ of the renormalized
Newton's constant \myref{scalingh} the Hamiltonian is bounded from
below and therefore well defined. It is self-adjoint
with respect to the natural measure $d\mu(L) \equ dL$ and has a discrete
spectrum, with eigenfunctions
\begin{equation}
\label{eigenfunc}
\psi_n(L) =\mathcal{A}_n e^{ -\sqrt{2\Lambda(1- h_{ren}^2) }L}
L_n(\, 2\sqrt{2\Lambda(1-h_{ren}^2) }\, ),\qquad n=0,1,2,...\, ,
\end{equation}
where $L_n$ denotes the $n$'th Laguerre polynomial. Choosing the normalization
constants as
\begin{equation}
\mathcal{A}_n=\sqrt[4]{8\Lambda(1- h_{ren}^2) },
\end{equation}
the eigenvectors $\{\psi_n(L) $, $n\equ 0,1,2,...\}$
form an orthonormal basis, and
the corresponding eigenvalues are given by
\begin{equation}
E_n=\sqrt{2\Lambda(1- h_{ren}^2) }\left(2n+1\right),\quad n=0,1,2,...\, .
\end{equation}
Having obtained the eigenvalues one can easily calculate the
Euclidean partition function for finite time $T$ (with time periodically
identified)
\begin{equation}
\label{continuumpartfunc} Z_T(G_N,\Lambda) =\sum_{n=0}^\infty e^{-T\,E_n} =
\frac{e^{-\sqrt{2\Lambda(1- h_{ren}^2) }
T}}{1-e^{-2\,\sqrt{2\Lambda(1- h_{ren}^2) }T}},\quad h_{ren}=e^{- 2\pi / G_N}.
\end{equation}
For completeness we also compute the finite-time propagator
\begin{equation}
    G_{\Lambda,G_N}(L_1,L_2,T)  \equiv \langle{L_2} | e^{-T\hat{H}}| L_1\rangle =
\sum_{n=0}^{\infty}e^{-TE_n}\psi_n^*(L_2) \psi_n(L_1) \label{FTPropdef}.
\end{equation}
Inserting \myref{eigenfunc} into \myref{FTPropdef} and using known relations
for summing over Laguerre polynomials \cite{Prudnikov} yields
\begin{eqnarray}\label{calogero20}
    G_{\Lambda,G_N}(L_1,L_2,T) =\omega\,
    \frac{e^{-\omega(L_1+L_2) \coth(\omega T) }}{\sinh(\omega T) }
     I_0 \left(\frac{2 \omega \, \sqrt{L_1 L_2} }{\sinh(\omega T) } \right),
\end{eqnarray}
where we have used the shorthand notation $\omega=\sqrt{2\Lambda(1-
h_{ren}^2) }$. As expected, for $h_{ren}\rightarrow 0$ the results
reduce to those of the pure two dimensional CDT model.

\subsection{Observables} \label{sec_observ}

Due to the low dimensionality of our quantum-gravitational model,
it has only a few observables which characterize its physical properties.
Given the eigenfunctions \myref{eigenfunc} of the Hamiltonian \myref{Hamiltonian1}
one can readily calculate the average spatial extension $\langle L\rangle$ of the
universe and all higher moments
\begin{equation}
\label{moments}
\expec{L^m}_n = \int_0^\infty dL\,L^m |\psi_n(L) |^2.
\end{equation}
Using integral relations for the Laguerre polynomials \cite{Prudnikov}
one obtains\footnote{Note that the poles of $\Gamma(-m) $ cancel with
those of the hypergeometric function.}
\begin{eqnarray}
\expec{L^m}_n
= \left(\frac{1}{8\Lambda (1- h_{ren}^2) }\right) ^\frac{m}{2}\frac{\Gamma(n-m)
\Gamma(m+1) }{\Gamma(n+1) \Gamma(-m) }\times\nonumber\\\nonumber\\
 \times\,\, {}_3F_2 (-n,1+m,1+m;1,1+m-n;1),
\end{eqnarray}
where ${}_3F_2 (a_1,a_2,a_3;b_1,b_2;z) $ is the generalized hypergeometric
function defined by
\begin{equation}
\label{hypeerdefh}
{}_3F_2 (a_1,a_2,a_3;b_1,b_2;z) =\sum_{k=0}^\infty
\frac{(a_1) _k (a_2) _k (a_3) _k\,z^k}{(b_1) _k (b_2) _k k!}.
\end{equation}
Observe that the moments scale as $\expec{L^m}_n\sim \Lambda^{-\frac{m}{2}}$
which indicates that the effective Hausdorff dimension is given by $d_H=2$,
just like in the pure CDT model \cite{Ambjorn:2000dv}.
\begin{figure}
\begin{center}
\includegraphics[width=4in]{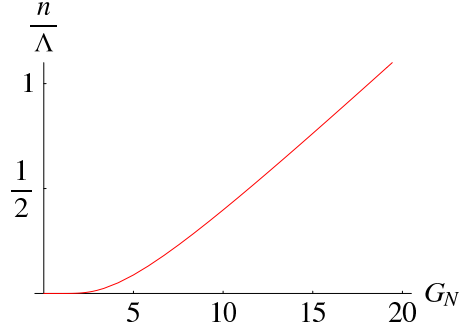}
\caption{The density of holes $n$ in units of $\Lambda$ as a
function of Newton's constant $G_N$.}\label{figure1}
\end{center}
\end{figure}

In addition to these well-known geometric observables,
the system possesses a new type of ``topological" observable
which involves the number of holes $N_\genus$, as already anticipated in
\cite{Loll:2003rn,Loll:2003yu}. As spelled out there, the presence of
holes in the quantum geometry and
their density can be determined from light scattering.
An interesting quantity to
calculate is the average number of holes in a piece of spacetime
of duration $T$, with initial and final spatial boundaries identified.
Because of the simple dependence of the action on the genus
this is easily computed by taking the derivative of the
partition function $Z_T$ with respect to the corresponding
coupling, namely,
\begin{equation}
\label{Nholedefpartfunc}
    \expec{N_\genus} = \frac{1}{Z_T}\frac{h_{ren}}{2}\frac{\partial\, Z_T}{\partial h_{ren}}.
\end{equation}
Upon inserting \myref{continuumpartfunc} this yields
\begin{equation}\label{Nholeresultpartfunc}
    \expec{N_\genus} =  T \, h_{ren}^2 \Lambda \frac{\coth \left(\sqrt{2\Lambda
    (1-h_{ren}^2) }\,T\right) }{ \sqrt{2\Lambda (1-h_{ren}^2) }}.
\end{equation}
In an analogous manner we can also calculate the average spacetime volume
\begin{equation}\label{averagevoldef}
    \expec{V} = -\frac{1}{Z_T}\frac{\partial\, Z_T}{\partial \Lambda},
\end{equation}
leading to
\begin{equation}\label{averagevolresult}
    \expec{V} = T \, \frac{\sqrt{(1-h_{ren}^2) }}{\sqrt{2 \Lambda}}
\coth\left(\sqrt{2\Lambda(1-h_{ren}^2) }\,T\right).
\end{equation}
Dividing \myref{Nholeresultpartfunc} by \myref{averagevolresult} we find
that the spacetime density $n$ of holes is constant,
\begin{equation}
\label{densityfinalresult}
n=\frac{\expec{N_\genus}}{\expec{V}}= \frac{h_{ren}^2}{1-h_{ren}^2} \,\Lambda.
\end{equation}
The density of holes in terms of the renormalized Newton's constant is given by
\begin{equation}
\label{densityonG}
n=\frac{1}{e^\frac{4\pi}{G_N}-1} \,\Lambda.
\end{equation}
The behaviour of $n$ in terms of the renormalized Newton's constant is shown
in fig.~\ref{figure1}. The density of holes vanishes as $G_N\rightarrow 0$ and
the model reduces to the case without topology change. -- An alternative calculation
of the density of holes from an infinitesimal strip, which leads to the same result,
is presented in Appendix B.

\begin{figure}
\begin{center}
\includegraphics[width=4in]{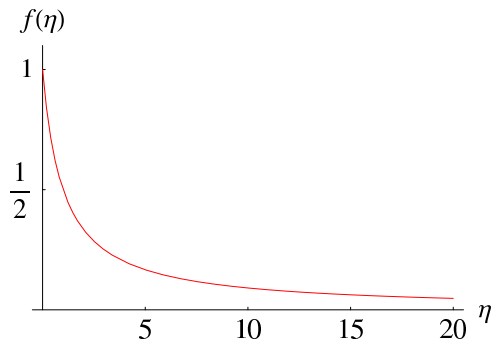}
\caption{The coefficient of the effective potential, $f(\eta) =1/(1+\eta) $,
as function of the density of holes in units of $\Lambda$,
$\eta=\frac{n}{\Lambda}$.}\label{figure2}
\end{center}
\end{figure}

We can now rewrite and interpret the Hamiltonian \myref{Hamiltonian1} in
terms of physical
quantities, namely, the cosmological scale $\Lambda$ and the density of holes in
units of $\Lambda$, i.e.~$\eta=\frac{n}{\Lambda}$, resulting in
\begin{equation}
\label{hamiltonianeta}
\hat{H} (L,\dL) =-L\ddL-\dL+  \frac{1}{1+\eta} \,2\,\Lambda\,L.
\end{equation}
One sees explicitly that the topology fluctuations affect the dynamics since the
effective potential depends on $\eta$, as illustrated by fig.~ \ref{figure2}.

It should be clear from expressions \myref{hamiltonianeta} and \myref{densityonG}
that the model has two scales instead of the single one of the pure
CDT model. As in the latter,
the cosmological constant defines the global length scale of the
two dimensional ``universe" through $\expec{L}\sim\frac{1}{\sqrt{\Lambda}}$.
The new scale in the model with topology change is the relative scale $\eta$
between the cosmological and topological fluctuations, which is parametrized
by Newton's constant $G_N$.

\subsection{Summary}\label{conclusions}

In this section, we have presented the complete analytic solution
of a previously proposed model \cite{Loll:2003rn} of two dimensional
Lorentzian quantum gravity including a sum over topologies.
The presence of causality constraints imposed on the
path-integral histories -- physically motivated in
\cite{Loll:2003rn,Loll:2003yu} --
enabled us to derive a new class of continuum theories by taking an
unambiguously defined double-scaling limit of a statistical
model of simplicially regularized spacetimes.
After computing the Laplace transform of the exact one-step
propagator of the discrete model, we investigated a
two-parameter family, defined by \rf{scalingh}  and \rf{d},  of
possible scalings for the gravitational (or Newton's)  coupling,
from which physical considerations singled out a unique one.
For this case, we computed the quantum Hamiltonian,
its spectrum and eigenfunctions, as well as the partition function and
propagator. Using these continuum results, we then calculated a
variety of physical observables, including the average spacetime
density of holes and the expectation values of the spatial volume
and all its moments.

This should be contrasted with the previous treatment in
\cite{Loll:2003rn,Loll:2003yu}, in which only the one-step
partition function with summed-over boundaries was evaluated.
Because of the lack of boundary information, no explicit Hamiltonian
was obtained there. Moreover, it turns out that the scaling of the
couplings which in the current work led to the essentially unique
Hamiltonian \rf{Hamiltonian1}  could {\it not} have been obtained
or even guessed
in the previous work. This is simply a consequence of the fact that
the dimensionful renormalized boundary cosmological constants
make an explicit appearance in the scaling relation \rf{hm1}  for
$h$, and thus for Newton's constant.
We conclude that -- unlike in the case of the original Lorentzian model -- for
two dimensional causal quantum gravity with topology changes
one cannot obtain the correct scalings for the ``bulk" coupling
constants from the one-step partition function with boundaries
summed over (which is easier to compute than the full one-step
propagator).

In contrast with what was extrapolated from the single-strip model in
\cite{Loll:2003yu}, the total number of holes in a finite patch of
spacetime turns out to be a finite
quantity determined by the cosmological and Newton's constants.
Note that this finiteness result has been obtained dynamically and
without invoking any fundamental discreteness.
Since the density of holes is finite and every hole in the model is
infinitesimal, this implies -- and is confirmed by explicit calculation -- that
the expectation value of the number of holes in a general spatial slice
of constant time is also
infinitesimal. The fact that physically sensible observables are
obtained in this toy model reiterates the earlier conclusion
\cite{Loll:2003rn} that causality-inspired
methods can be a useful tool in constructing gravitational path
integrals which include a sum over topologies.

From the effective potential displayed in fig.~\ref{figure2} one observes
that the presence of wormholes in our model leads to a decrease of the
``effective" cosmological constant $f(\eta) \Lambda$.
In Coleman's mechanism for
driving the cosmological constant $\Lambda$ to zero
\cite{Coleman:1988tj,Klebanov:1988eh}, an additional sum over different
baby universes is performed in the path integral, which leads to a
distribution of the cosmological constant that is peaked near zero.
We do not consider such an additional sum over baby universes,
but instead have an explicit expression for the effective potential which
shows that an increase in the number of wormholes is accompanied by a
decrease of the ``effective" cosmological constant.
A first step in establishing whether an analogue of our suppression
mechanism also exists in higher dimensions would be to try and understand
whether one can identify a class of causally preferred topology changes
which still leaves the sum over geometries exponentially bounded.

\chapter{Conclusions}

Despite many attempts, gravity has resiliently resisted a unification with the laws of quantum
mechanics. Besides a plethora of technical issues, one is also faced with many interesting
conceptual problems. The study of quantum gravity in lower dimensional models ameliorates the
technical difficulties while still preserving some of the conceptually fascinating characteristics
of quantum gravity.
\\
\\
In this thesis we analyze the very simple model of two dimensional quantum gravity. Although a
rather extreme simplification of four dimensional gravity, many of the most fundamental issues are
still relevant. Moreover, two dimensional gravity is interesting since it can be viewed as a
minimal version of string theory.
\\
\\
The first fundamental aspect where we make a contribution is the problem of topology change of
space. Particularly, we present an exactly solvable model which shows that it is possible to
incorporate spatial topology changes in the path integral rigorously. We show that if the change in
spatial topology is accompanied by a coupling constant it is possible to evaluate the path integral
to all orders in the coupling. Furthermore, the model can be viewed as a hybrid between causal and
Euclidean dynamical triangulation models. An interesting avenue for further research is the
question whether our model has an interpretation within string theory. In particular, is our new coupling
constant really equivalent to the string coupling?
\\
\\
The second conceptual topic that we cover is the emergence of geometry from a background
independent path integral. We show that from a path integral over noncompact manifolds a classical
geometry with constant negative curvature emerges. No initial singularity is present, so the model
naturally realizes the Hartle Hawking boundary condition. Furthermore, we demonstrate that under
certain conditions the superimposed quantum fluctuations are small! The model is an interesting
example where a classical background emerges from background independent quantum gravity in an
exactly solvable setting. How does the emergent geometry behave? Can we make contact with
effective descriptions of quantum geometry such as noncommutative geometry or doubly special
relativity? The answer to these questions is incomplete for now, but the exact solvability of the
model suggests that at least a detailed analysis could be possible.
\\
\\
To conclude, we tackle the problem of spacetime topology change. Although we are not able to
completely solve the path integral over all manifolds with arbitrary topology, we do obtain some
results that indicate that such a path integral might be consistent if suitable causality
restrictions are imposed. As a first step we extend the existing formalism of causal dynamical
triangulations by a perturbative computation of amplitudes that include manifolds up to genus two.
Further we present a toy model where we make the approximation that the holes in the manifold are
extremely small. This simplification allows us perform an explicit sum over all genera and analyze
the continuum limit exactly. Remarkably, the presence of the infinitesimal wormholes leads to a
decrease in the effective cosmological constant, reminiscent of the suppression mechanism
considered by Coleman and others in the four-dimensional Euclidean path integral.
\\
\\
The results of this thesis show that we still know very little about the ultimate configuration
space of quantum gravity. Even for the extremely simple case of two dimensional quantum gravity
various new models can be constructed that seem to lead to a well defined theory of quantum
geometry. To understand the situation even better it would be important to have a better continuum
understanding of the results of causal dynamical triangulations.

\renewcommand{\chaptermark}[1]{\markboth{\thechapter\ #1}{}}
\fancyhead{} \fancyhead[LE, RO]{\thepage}
\fancyhead[CO]{\slshape\leftmark}
\fancyhead[CE]{\slshape\leftmark}
\appendix

\chapter{Lorentzian triangles}\label{app:Lorentzian triangles}

\begin{figure}[t]
\begin{center}
\includegraphics[width=3.5in]{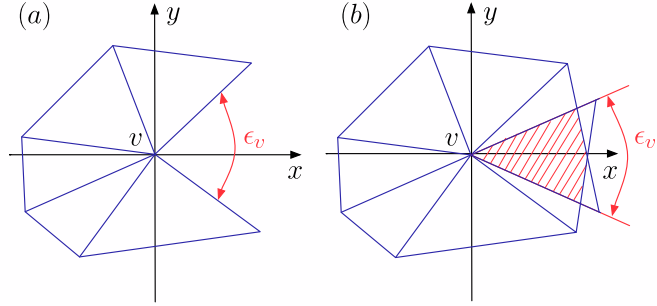}
\caption{Illustration of a positive (a)  and negative (b)  spacelike deficit angle $\epsilon_v$ at a vertex $v$.}
\label{fig:positivenegativecurvature}
\end{center}
\end{figure}

In this appendix, a brief summary of results on Lorentzian angles is presented, where we follow the
treatment and conventions of \cite{Sorkin:1975ah}.

\begin{figure}[t]
\begin{center}
\includegraphics[width=3.5in]{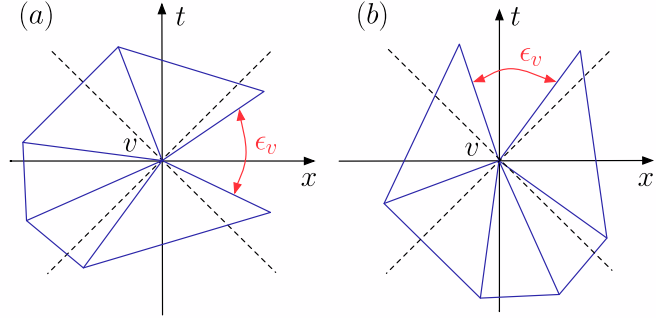}
\caption{Illustration of a spacelike (a)  and a timelike (b)  Lorentzian deficit angle $\epsilon_v$
at a vertex $v$.} \label{fig:appdeficit}
\end{center}
\end{figure}

Since in CDT one considers simplicial manifolds consisting of Minkowskian triangles, Lorentzian
angles or ``boosts'' naturally appear in the Regge action as rotations around vertices. Recall from
Section \rf{subsec:Quantum gravity from simplicial intrinsic geometry} that the definition of the
scalar curvature at a vertex $v$ is given by \eqref{eq:Reggecurvature},
\begin{equation}\label{eq:applor:K}
R_v=2 \frac{\epsilon_v}{V_v},
\end{equation}
where $\epsilon_v=2\pi-\sum_{i\supset v}\theta_i$ is the deficit angle at a vertex $v$ and $V_v$ is
the dual volume of the vertex $v$. In general, the spacelike deficit angle $\epsilon_v$ can be
positive or negative as illustrated in fig.~\ref{fig:positivenegativecurvature}. Furthermore, if the deficit
angle is timelike, as shown in fig.~\ref{fig:appdeficit}, it will be complex. The timelike
deficit angles are still additive, but contribute to the curvature \eqref{eq:applor:K} with the
opposite sign. Hence, both spacelike defect and timelike excess increase the curvature, whereas
spacelike excess and timelike defect decrease it.

The complex nature of the timelike deficit angles can be seen explicitly by noting that the angles
$\theta_i$ between two edges $\vec{a}_i$ and $\vec{b}_i$ (as vectors in Minkowski space)  are
calculated using
\begin{equation}
\label{eq:applor:theta}
\cos \theta_i
=\frac{\scprod{\vec{a}_i}{\vec{b}_i}}{\scprod{\vec{a}_i}{\vec{a}_i}^\frac{1}{2}\scprod{\vec{b}_i}{\vec{b}_i}^\frac{1}{2}},\quad
\sin \theta_i
=\frac{\sqrt{\scprod{\vec{a}_i}{\vec{a}_i}\scprod{\vec{b}_i}{\vec{b}_i}-\scprod{\vec{a}_i}{\vec{b}_i}^2}}{\scprod{\vec{a}_i}{\vec{a}_i}^\frac{1}{2}\scprod{\vec{b}_i}{\vec{b}_i}^\frac{1}{2}},
\end{equation}
where $\scprod{\cdot}{\cdot}$ denotes the flat Minkowskian scalar product and by definition, the
square roots of negative arguments are positive imaginary.

Having given a concrete meaning to  Lorentzian angles, we can now use \eqref{eq:applor:theta} to
calculate the volume of Minkowskian triangles which we will then use to explicitly compute the
volume terms of the Regge action.

The triangulations we are considering consist of Minkowskian triangles with one spacelike edge of
length squared $l_s^2=a^2$ and two timelike edges of length squared $l_t^2=-\alpha a^2$ with
$\alpha>0$. The general argument $\alpha>0$ is used to give a mathematically precise prescription
of the Wick rotation, but it can be set to $\alpha=1$ after the Wick rotation has been performed.
With the use of \eqref{eq:applor:theta} we can calculate the volume of such a Minkowskian triangle,
yielding
\begin{equation}
\Vol(\mathrm{triangle}) =\frac{a^2}{4}\sqrt{4\alpha+1}.
\end{equation}
Now one can define the Wick rotation $\mathcal{W}$ as the analytic continuation of
$\alpha\mapsto-\alpha$ through the lower-half plane. One then sees that for $\alpha>\frac{1}{2}$
under this prescription $\mbox{$i\,\Vol(\mathrm{triangle}) \mapsto -\Vol(\mathrm{triangle}) $}$ (up
to a $\mathcal{O}(1) $ constant which can be absorbed in the corresponding coupling constant in the
action). This ensures that
\begin{equation}
\mathcal{W}:\quad  e^{i\,S_{\mathrm{Regge}}(T^{lor}) }\mapsto  e^{-\,S_{\mathrm{Regge}}(T^{eu}) },
\quad \alpha>\frac{1}{2}.
\end{equation}
In the following we set $\alpha=1$ again. Generalizations of this treatment to dimension $d=3,4$
can be found in \cite{Ambjorn:2001cv}.

\chapter{Alternative scalings}\label{app2}

In this appendix we discuss the scalings with $\beta=1$ which we discarded as unphysical in Sec.\
\rf{continuumsection} above. We proceed as before by inserting the scaling relations
\myref{scalinga} and \myref{scalingbnew} into the composition law \myref{timeevolution1}. Instead
of using $\beta=\frac{3}{2}$ we set $\beta=1$, leading to the scaling
\begin{equation}
h=\frac{1}{4}\,h_{ren}\,a\,\sqrt{\Lambda}^\alpha (X+Y) ^{1-\alpha},
\end{equation}
where the normalization factor on the right-hand side has been chosen for later convenience. Up to
first order in $a$ one obtains
\begin{equation}
(1-a \hat{H}+\mathcal{O}(a^2) ) \psi(X)  =  \int^{i \infty}_{-i\infty} \frac{dZ}{2\pi i}
\left\{A(X,Z) +B(X,Z)  a + \mathcal{O}(a^2) \right\} \psi(Z),
\end{equation}
where the leading-order contribution is given by
\begin{equation}
A(X,Z) =\frac{2}{(Z-X) \left(1+C(X,Z) \right) }
\end{equation}
with
\begin{equation}
C(X,Z) =\sqrt{1-h_{ren}^2(X-Z) ^{-2\alpha}\Lambda^\alpha}.
\end{equation}
For the Laplace transform of $A(X,Z) $ to yield a delta-function, the scaling should be chosen such
that $\alpha\leqslant 0$. Considering now the terms of first order in $a$,
\begin{eqnarray}
B(X,Z) &=&\frac{  h_{ren}^2(X+Z-4 Z \gamma )
\Lambda ^{\alpha }}{(X-Z) ^{1+2\alpha} C(X,Z)  \left(1+C(X,Z) \right) ^2}\nonumber\\
&-&2\,\frac{X Z-2\Lambda + \gamma  (X-Z) ^2 } {(X-Z) ^2 C(X,Z) \left(1+C(X,Z) \right)
},\label{app_B}
\end{eqnarray}
one finds that for $\alpha\leqslant -1$ the continuum limit is independent of any ``hole
contribution" (i.e.~terms depending on $h_{ren}$)  and therefore leads to the usual Lorentzian
model. This becomes clear when one expands the last term of \myref{app_B} in $(X-Z) $, resulting in
\begin{equation}
\frac{X Z-2\Lambda }{(X-Z) ^2 C \left(1+C\right) } =\frac{1}{2}\frac{X Z-2\Lambda }{(X-Z)
^2}\left(1+\frac{3}{4}\,h_{ren}^2 \Lambda^\alpha (X-Z) ^{-2\alpha}+\mathcal{O}((X-Z) ^{-4\alpha})
\right).
\end{equation}
For $\alpha\leqslant -1$ the term depending on $h_{ren}$ does not have a pole and therefore does
not contribute to the Hamiltonian. Since we are only interested in non-fractional poles, this
leaves as possible $\alpha$-values only $\alpha=0$ and $\alpha=-\frac{1}{2}$.

\section{The case $\beta =1$, $\alpha=0$}

For $\alpha=0$ the Hamiltonian retains a $\gamma$-dependence contained in the first line of
\myref{app_B}. Since there is no immediate physical interpretation of $\gamma$ in our model, it
seems natural to choose $\gamma=0$, although strictly speaking this does not resolve the problem of
explaining the $\gamma$-dependence of the continuum limit. Setting this question aside, one may
simply look at the resulting model as an interesting integrable model in its own right. In order to
obtain a delta-function to leading order, one still needs to normalize the transfer matrix by a
constant factor $2/(1+s) $, with $s:=\sqrt{1-h_{ren}^2}$. After setting $\gamma=0$ and performing
an inverse Laplace transformation, the Hamiltonian reads
\begin{equation}
\label{haml} \hat{H}(L,\dL) =\frac{1}{s}\left(-L\ddL -s\dL +2\Lambda L \right).
\end{equation}
It is self-adjoint with respect to the measure $d\mu(L) =L^{s-1}dL$. Further setting
$L=\frac{\varphi^2}{2\,s}$ one encounters the one-dimensional Calogero Hamiltonian
\begin{equation}
\label{calo} \hat{H}(\varphi,\frac{\partial}{\partial\varphi})
=-\frac{1}{2}\frac{\partial^2}{\partial\varphi^2}+
\frac{1}{2}\omega^2\varphi^2-\frac{1}{8}\frac{A}{\varphi^2},
\end{equation}
with $\omega\equ\frac{\sqrt{2 \Lambda}}{s}$ and $A\equ 1-4(1-s) ^2$, which implies that the model
covers the parameter range $-3\leqslant A\leqslant 1$. The maximal range for which the Calogero
Hamiltonian is self-adjoint is $-\infty<A\leqslant 1$. The usual Lorentzian model without holes
corresponds to $A\equ 1$. The Hamiltonian (\rf{calo})  has already appeared in a causal dynamically
triangulated model where the two dimensional geometries were decorated with a certain type of
``outgrowth" or small ``baby universes" \cite{DiFrancesco:2000nn}. This model covered the parameter
range $0\leqslant A\leqslant 1$.

The eigenvectors of the Hamiltonian (\rf{haml})  are given by
\begin{equation}
\psi_n(L) =\mathcal{A}_n e^{- \sqrt{2\Lambda}L}\,{}_1F_1(-n,s,2\sqrt{2\Lambda} L), \quad d\mu(L)
=L^{s-1}dL,
\end{equation}
where ${}_1F_1(-n,a,b) $ is the Kummer confluent hypergeometric function. The eigenvectors form an
orthonormal basis with the normalization factors
\begin{equation}
\mathcal{A}_n=(8\Lambda) ^{\frac{s}{4}}\sqrt{\frac{\Gamma(n+s) }{\Gamma(n+1)  \Gamma(s) ^2}}
\end{equation}
and the corresponding eigenvalues
\begin{equation}
E_n=\frac{\sqrt{2\Lambda}}{s}(2n+s),\quad n=0,1,2,...\, .
\end{equation}
One sees explicitly that the case $s\equ 1$ or, equivalently, $A\equ 1$ corresponds to the pure two
dimensional CDT model.

\section{The case $\beta =1$, $\alpha=-\frac{1}{2}$}

For $\alpha=-\frac{1}{2}$ the result does not depend on $\gamma$ and therefore on the detailed
manner in which we approach the critical point. However, the Hamiltonian
\begin{equation}
\hat{H}(L,\dL) =-L\ddL-\dL+2\,\Lambda\,L-\frac{3}{4}\,h_{ren}^2\Lambda^{-1/2}\,\ddL
\end{equation}
cannot be made self-adjoint with respect to any measure $d\mu(L) $ because the boundary part of the
partial integration always gives a nonvanishing contribution. We therefore discard this
possibility.

\chapter{The density of holes of an infinitesimal strip} \label{app3}

In this appendix we give an alternative derivation of the spacetime density $n$ of holes and
explicitly show that the number of holes in a spacetime strip of infinitesimal time duration $a$ is
also infinitesimal. The operator in the $L$-representation of the number of holes per infinitesimal
strip with fixed initial boundary $L$ can be calculated by
\begin{equation}
\label{N1} \hat{N}_{\genus,a\rightarrow 0} =\hat{T}^{-1}\frac{h_{ren}}{2}\frac{\partial\,
\hat{T}}{\partial h_{ren}},
\end{equation}
where $\hat{T}$ is the transfer matrix defined in \myref{transfer}. Using
$\hat{T}=1-a\hat{H}+\mathcal{O}(a^2) $ and evaluating \myref{N1} to leading order in $a$ gives
\begin{equation}
\label{N2} \hat{N}_{\genus,a\rightarrow 0}=-a\,\frac{h_{ren}}{2}\frac{\partial \hat{H}}{\partial
h_{ren}}+ \mathcal{O}(a^2) =2\, \Lambda\,h_{ren}^2\,L\,a+\mathcal{O}(a^2).
\end{equation}
Similarly, the volume operator of the same infinitesimal spacetime strip in the $L$-representation
is given by
\begin{equation}
\label{volop} \hat{V}_{a\rightarrow 0} =-\hat{T}^{-1} \frac{\partial\, \hat{T}}{\partial \Lambda}=
a\,\frac{\partial \hat{H}}{\partial \Lambda}+\mathcal{O}(a^2) = 2\, (1-h_{ren}^2)
\,L\,a+\mathcal{O}(a^2).
\end{equation}
Although both expressions \rf{N2}  and \rf{volop}  vanish in the limit as $a\rightarrow 0$ (and
therefore the number of holes and the strip volume are both ``infinitesimal"), their quotient
evaluates to a finite number independent of $L$, namely,
\begin{equation}
n=\frac{N_{\genus,a\rightarrow 0}}{V_{a\rightarrow 0}}= \frac{h_{ren}^2}{1-h_{ren}^2} \,\Lambda.
\end{equation}
This is the exactly the same result for the spacetime density $n$ of holes as we obtained earlier
from the continuum partition function \rf{densityfinalresult}.

\backmatter
\renewcommand{\chaptermark}[1]{\markboth{ #1}{}}
\fancyhead{} \fancyhead[LE, RO]{\thepage}
\fancyhead[CO]{\slshape\nouppercase{\leftmark}}
\fancyhead[CE]{\slshape\nouppercase{\leftmark}}

\makeatletter
\def\thickhrulefill{\leavevmode \leaders \hrule height 1ex \hfill \kern \z@}
\def\@makechapterhead#1{%
  \vspace*{10\p@}%
  {\parindent \z@ \centering \reset@font
        \Huge \bfseries #1\par\nobreak
        \par
        \vspace*{10\p@}%
    \vskip 60\p@
  }}
\def\@makeschapterhead#1{%
  \vspace*{10\p@}%
  {\parindent \z@ \centering \reset@font
        \Huge \bfseries #1\par\nobreak
        \par
        \vspace*{10\p@}%
    \vskip 60\p@
  }}

\addcontentsline{toc}{chapter}{Bibliography}
\bibliographystyle{utphys}
\fontsize{11}{13pt}\selectfont
\bibliography{bib}

\newpage

\thispagestyle{empty}

\end{document}